%% file: lambdanotes.tex
\documentclass{article}
\input{lambdanotes.sty}

\title{Lecture Notes on the Lambda Calculus}
\author{Peter Selinger}
\date{Department of Mathematics and Statistics\\
  Dalhousie University, Halifax, Canada}

\usepackage[all]{xy}
\SelectTips{cm}{}

\begin{document}
\maketitle

\begin{abstract}
  This is a set of lecture notes that developed out of courses on the
  lambda calculus that I taught at the University of Ottawa in 2001
  and at Dalhousie University in 2007 and 2013. Topics covered in
  these notes include the untyped lambda calculus, the Church-Rosser
  theorem, combinatory algebras, the simply-typed lambda calculus, the
  Curry-Howard isomorphism, weak and strong normalization,
  polymorphism, type inference, denotational semantics, complete
  partial orders, and the language PCF.
\end{abstract}

\tableofcontents

\newpage

\section{Introduction}\label{sec-intro}

\subsection{Extensional vs. intensional view of functions}
\label{subsec-intro1}

What is a function? In modern mathematics, the prevalent notion is
that of ``functions as graphs'': each function $f$ has a fixed domain
$X$ and codomain $Y$, and a function $f:X\to Y$ is a set of pairs
$f\seq X\times Y$ such that for each $x\in X$, there exists exactly
one $y\in Y$ such that $(x,y)\in f$. Two functions $f,g:X\to Y$ are
considered equal if they yield the same output on each input, i.e.,
$f(x)=g(x)$ for all $x\in X$. This is called the {\em extensional}
view of functions, because it specifies that the only thing observable
about a function is how it maps inputs to outputs.

However, before the 20th century, functions were rarely looked at in
this way. An older notion of functions is that of ``functions as
rules''. In this view, to give a function means to give a rule for how
the function is to be calculated. Often, such a rule can be given by a
formula, for instance, the familiar $f(x)=x^2$ or $g(x)=\sin(e^x)$
from calculus. As before, two functions are {\em extensionally} equal
if they have the same input-output behavior; but now we can also speak
of another notion of equality: two functions are {\em
  intensionally}\footnote{Note that this word is intentionally spelled
  ``intensionally''.}  equal if they are given by (essentially) the
same formula.

When we think of functions as given by formulas, it is not always
necessary to know the domain and codomain of a function.  Consider for
instance the function $f(x)=x$. This is, of course, the identity
function. We may regard it as a function $f:X\to X$ for {\em any} set
$X$.

In most of mathematics, the ``functions as graphs'' paradigm is the
most elegant and appropriate way of dealing with functions. Graphs
define a more general class of functions, because it includes
functions that are not necessarily given by a rule. Thus, when we
prove a mathematical statement such as ``any differentiable function
is continuous'', we really mean this is true for {\em all} functions (in
the mathematical sense), not just those functions for which a rule can
be given.

On the other hand, in computer science, the ``functions as rules''
paradigm is often more appropriate. Think of a computer program as
defining a function that maps input to output. Most computer
programmers (and users) do not only care about the extensional
behavior of a program (which inputs are mapped to which outputs), but
also about {\em how} the output is calculated: How much time does it
take? How much memory and disk space is used in the process? How much
communication bandwidth is used? These are intensional questions
having to do with the particular way in which a function was defined.

\subsection{The lambda calculus}

The lambda calculus is a theory of {\em functions as formulas}. It is
a system for manipulating functions as {\em expressions}.  

Let us begin by looking at another well-known language of expressions,
namely arithmetic. Arithmetic expressions are made up from variables
($x,y,z\ldots$), numbers ($1,2,3,\ldots$), and operators (``$+$'',
``$-$'', ``$\times$'' etc.). An expression such as $x+y$ stands for
the {\em result} of an addition (as opposed to an {\em instruction} to
add, or the {\em statement} that something is being added). The great
advantage of this language is that expressions can be nested without
any need to mention the intermediate results explicitly. So for
instance, we write
\[   A = (x+y)\times z^2,   
\]
and not 
\[   \mbox{let $w=x+y$, then let $u=z^2$, then let $A=w\times u$.}    
\]
The latter notation would be tiring and cumbersome to manipulate. 

The lambda calculus extends the idea of an expression language to
include functions. Where we normally write
\[   \mbox{Let $f$ be the function $x\mapsto x^2$. Then consider $A=f(5)$,}
\]
in the lambda calculus we just write
\[   A = (\lam x.x^2)(5)  .
\]
The expression $\lam x.x^2$ stands for the function that maps $x$ to
$x^2$ (as opposed to the {\em statement} that $x$ is being mapped to
$x^2$). As in arithmetic, we use parentheses to group terms.

It is understood that the variable $x$ is a {\em local} variable in
the term $\lam x.x^2$. Thus, it does not make any difference if we
write $\lam y.y^2$ instead. A local variable is also called a {\em
  bound} variable.

One advantage of the lambda notation is that it allows us to easily
talk about {\em higher-order} functions, i.e., functions whose inputs
and/or outputs are themselves functions. An example is the operation
$f\mapsto f\cp f$ in mathematics, which takes a function $f$ and
maps it to $f\cp f$, the composition of $f$ with itself. In the lambda
calculus, $f\cp f$ is written as
\[         \lam x.f(f(x)),   
\]
and the operation that maps $f$ to $f\cp f$ is written as
\[         \lam f.\lam x.f(f(x)).
\]
The evaluation of higher-order functions can get somewhat complex; as
an example, consider the following expression:
\[         \left((\lam f.\lam x.f(f(x)))(\lam y.y^2)\right)(5)
\]
Convince yourself that this evaluates to 625. Another example is given
in the following exercise:

\begin{exercise}
Evaluate the lambda-expression
\[        \Big(
          \big(\left(\lam f.\lam x.f(f(f(x)))\right)
          \left(\lam g.\lam y.g(g(y))\right)\big)
          (\lam z.z+1)\Big)(0).
\]
\end{exercise}

We will soon introduce some conventions for reducing the number of
parentheses in such expressions.

\subsection{Untyped vs.\ typed lambda-calculi}

We have already mentioned that, when considering ``functions as
rules'', is not always necessary to know the domain and codomain of a
function ahead of time. The simplest example is the identity function
$f = \lam x.x$, which can have any set $X$ as its domain and codomain,
as long as domain and codomain are equal. We say that $f$ has the {\em
  type} $X\to X$.  Another example is the function $g = \lam f.\lam
x.f(f(x))$ that we encountered above. One can check that $g$ maps any
function $f:X\to X$ to a function $g(f):X\to X$. In this case, we say
that the type of $g$ is
\[          (X\to X)\to(X\to X).
\]
By being flexible about domains and codomains, we are able to
manipulate functions in ways that would not be possible in ordinary
mathematics. For instance, if $f=\lam x.x$ is the identity function,
then we have $f(x) = x$ for {\em any} $x$. In particular, we can take
$x=f$, and we get
\[     f(f) = (\lam x.x)(f) = f.
\]
Note that the equation $f(f)=f$ never makes sense in ordinary
mathematics, since it is not possible (for set-theoretic reasons) for
a function to be included in its own domain.

As another example, let $\omega = \lam x.x(x)$. 

\begin{exercise}
  What is $\omega(\omega)$?
\end{exercise}

We have several options regarding types in the lambda calculus.
\begin{itemize}
\item {\em Untyped lambda calculus.} In the untyped lambda calculus,
  we never specify the type of any expression. Thus we never specify
  the domain or codomain of any function. This gives us maximal
  flexibility. It is also very unsafe, because we might run into
  situations where we try to apply a function to an argument that it
  does not understand. 
\item {\em Simply-typed lambda calculus.} In the simply-typed lambda
  calculus, we always completely specify the type of every expression. 
  This is very similar to the situation in set theory. We never allow
  the application of a function to an argument unless the type of the
  argument is the same as the domain of the function. Thus, terms such
  as $f(f)$ are ruled out, even if $f$ is the identity function.
\item {\em Polymorphically typed lambda calculus.} This is an
  intermediate situation, where we may specify, for instance, that a
  term has a type of the form $X\to X$ for all $X$, without actually
  specifying $X$.
\end{itemize}

As we will see, each of these alternatives has dramatically different
properties from the others.

\subsection{Lambda calculus and computability}

In the 1930's, several people were interested in the question: what
does it mean for a function $f:\N\to\N$ to be {\em computable}? An
informal definition of computability is that there should be a
pencil-and-paper method allowing a trained person to calculate $f(n)$,
for any given $n$. The concept of a pencil-and-paper method is not so
easy to formalize. Three different researchers attempted to do so,
resulting in the following definitions of computability:

\begin{enumerate}
\item {\bf Turing} defined an idealized computer we now call a {\em
    Turing machine}, and postulated that a function is computable (in
  the intuitive sense) if and only if it can be computed by such a
  machine.
\item {\bf G\"odel} defined the class of {\em general recursive
    functions} as the smallest set of functions containing all the
  constant functions, the successor function, and closed under certain
  operations (such as compositions and recursion). He postulated that
    a function is computable (in the intuitive sense) if and only if
    it is general recursive.
\item {\bf Church} defined an idealized programming language called
  the {\em lambda calculus}, and postulated that a function is
  computable (in the intuitive sense) if and only if it can be
  written as a lambda term.
\end{enumerate}

It was proved by Church, Kleene, Rosser, and Turing that all three
computational models were equivalent to each other, i.e., each model
defines the same class of computable functions. Whether or not they
are equivalent to the ``intuitive'' notion of computability is a
question that cannot be answered, because there is no formal
definition of ``intuitive computability''. The assertion that they are
in fact equivalent to intuitive computability is known as the {\em
  Church-Turing thesis}.

\subsection{Connections to computer science}

The lambda calculus is a very idealized programming language;
arguably, it is the simplest possible programming language that is
Turing complete. Because of its simplicity, it is a useful tool for
defining and proving properties of programs.

Many real-world programming languages can be regarded as extensions of
the lambda calculus. This is true for all {\em functional programming
  languages}, a class that includes Lisp, Scheme, Haskell, and ML.
Such languages combine the lambda calculus with additional features,
such as data types, input/output, side effects, updatable memory,
object orientated features, etc. The lambda calculus provides a
vehicle for studying such extensions, in isolation and jointly, to see
how they will affect each other, and to prove properties of
programming language (such as: a well-formed program will not crash).

The lambda calculus is also a tool used in compiler construction, see
e.g. \cite{Pey87,App92}.

\subsection{Connections to logic}

In the 19th and early 20th centuries, there was a philosophical
dispute among mathematicians about what a proof is. The so-called {\em
  constructivists}, such as Brouwer and Heyting, believed that to prove
that a mathematical object exists, one must be able to construct it
explicitly. {\em Classical logicians}, such as Hilbert, held that it
is sufficient to derive a contradiction from the assumption that it
doesn't exist.

Ironically, one of the better-known examples of a proof that isn't
constructive is Brouwer's proof of his own fixed point theorem,
which states that every continuous function on the unit disk has a
fixed point. The proof is by contradiction and does not give any
information on the location of the fixed point.

The connection between lambda calculus and constructive logics is via
the ``proofs-as-programs'' paradigm. To a constructivist, a proof (of
an existence statement) must be a ``construction'', i.e., a program.
The lambda calculus is a notation for such programs, and it can also
be used as a notion for (constructive) proofs.

For the most part, constructivism has not prevailed as a philosophy in
mainstream mathematics. However, there has been renewed interest in
constructivism in the second half of the 20th century. The reason is
that constructive proofs give more information than classical ones,
and in particular, they allow one to compute solutions to problems (as
opposed to merely knowing the existence of a solution). The resulting
algorithms can be useful in computational mathematics, for instance in
computer algebra systems.

\subsection{Connections to mathematics}

One way to study the lambda calculus is to give mathematical models of
it, i.e., to provide spaces in which lambda terms can be given
meaning.  Such models are constructed using methods from algebra,
partially ordered sets, topology, category theory, and other areas of
mathematics.

\section{The untyped lambda calculus}

\subsection{Syntax}

The lambda calculus is a {\em formal language}. The expressions of the
language are called {\em lambda terms}, and we will give rules for
manipulating them.

\begin{definition}
  Assume given an infinite set $\Vars$ of {\em variables}, denoted by
  $x,y,z$ etc. The set of lambda terms is given by the following
  Backus-Naur Form:
  \[  \mbox{Lambda terms:}\ssep M,N \bnf x \bor (MN) \bor (\lam x.M)
  \]
\end{definition}

The above Backus-Naur Form (BNF) is a convenient abbreviation for the
following equivalent, more traditionally mathematical definition:

\begin{definition}
  Assume given an infinite set $\Vars$ of variables. Let $A$ be an
  alphabet consisting of the elements of $\Vars$, and the special
  symbols ``('', ``)'', ``$\lam$'', and ``.''. Let $A^*$ be the set of
  strings (finite sequences) over the alphabet $A$.  The set of lambda
  terms is the smallest subset $\Lambda\seq A^*$ such that:
  \begin{itemize}
  \item Whenever $x\in\Vars$ then $x\in\Lambda$.
  \item Whenever $M,N\in\Lambda$ then $(MN)\in\Lambda$.
  \item Whenever $x\in\Vars$ and $M\in\Lambda$ then $(\lam x.M)\in\Lambda$.
  \end{itemize}
\end{definition}

Comparing the two equivalent definitions, we see that the Backus-Naur
Form is a convenient notation because: (1) the definition of the
alphabet can be left implicit, (2) the use of distinct meta-symbols
for different syntactic classes ($x,y,z$ for variables and $M,N$ for
terms) eliminates the need to explicitly quantify over the sets
$\Vars$ and $\Lambda$. In the future, we will always present syntactic
definitions in the BNF style.

The following are some examples of lambda terms:
\[ (\lam x.x) \sep ((\lam x.(xx))(\lam y.(yy))) \sep (\lam f.(\lam x.(f(fx))))
\]
Note that in the definition of lambda terms, we have built in enough
mandatory parentheses to ensure that every term $M\in\Lambda$ can be
uniquely decomposed into subterms. This means, each term $M\in\Lambda$
is of precisely one of the forms $x$, $(MN)$, $(\lam x.M)$. Terms of
these three forms are called {\em variables}, {\em applications}, and
{\em lambda abstractions}, respectively. 

We use the notation $(MN)$, rather than $M(N)$, to denote the
application of a function $M$ to an argument $N$. Thus, in the lambda
calculus, we write $(fx)$ instead of the more traditional $f(x)$. This
allows us to economize more efficiently on the use of parentheses. To
avoid having to write an excessive number of parentheses, we establish
the following conventions for writing lambda terms:

\begin{convention}
  \begin{itemize}
  \item We omit outermost parentheses. For instance, we write $MN$
    instead of $(MN)$.
  \item Applications associate to the left; thus, $MNP$ means
    $(MN)P$. This is convenient when applying a function to a number
    of arguments, as in $fxyz$, which means $((fx)y)z$. 
  \item The body of a lambda abstraction (the part after the dot)
    extends as far to the right as possible. In particular, $\lam
    x.MN$ means $\lam x.(MN)$, and not $(\lam x.M)N$. 
  \item Multiple lambda abstractions can be contracted; thus $\lam
    xyz.M$ will abbreviate $\lam x.\lam y.\lam z.M$.
  \end{itemize}
\end{convention}  

It is important to note that this convention is only for notational
convenience; it does not affect the ``official'' definition of lambda
terms.

\begin{exercise}
  \begin{enumerate}\alphalabels
  \item Write the following terms with as few parenthesis as possible,
    without changing the meaning or structure of the terms: 
    \begin{enumerate}
    \item[(i)] $(\lam x.(\lam y.(\lam z.((xz)(yz)))))$, 
    \item[(ii)] $(((ab)(cd))((ef)(gh)))$, 
    \item[(iii)] $(\lam x.((\lam y.(yx))(\lam v.v)z)u)(\lam w.w)$.
    \end{enumerate}
  \item Restore all the dropped parentheses in the following terms,
    without changing the meaning or structure of the terms: 
    \begin{enumerate}
    \item[(i)] $xxxx$, 
    \item[(ii)] $\lam x.x\lam y.y$, 
    \item[(iii)] $\lam x.(x\lam y.yxx)x$.
    \end{enumerate}
  \end{enumerate}
\end{exercise}

\subsection{Free and bound variables, $\alpha$-equivalence}

In our informal discussion of lambda terms, we have already pointed
out that the terms $\lam x.x$ and $\lam y.y$, which differ only in the
name of their bound variable, are essentially the same. We will say
that such terms are $\alpha$-equivalent, and we write $M\eqa N$.  In
the rare event that we want to say that two terms are precisely equal,
symbol for symbol, we say that $M$ and $N$ are {\em identical} and we
write $M\equiv N$. We reserve ``$=$'' as a generic symbol used for
different purposes.

An occurrence of a variable $x$ inside a term of the form $\lam x.N$
is said to be {\em bound}. The corresponding $\lam x$ is called a {\em
  binder}, and we say that the subterm $N$ is the {\em scope} of the
binder. A variable occurrence that is not bound is {\em free}.  Thus,
for example, in the term
\[   M \equiv (\lam x.xy)(\lam y.yz), \]
$x$ is bound, but $z$ is free. The variable $y$ has both a free and a
bound occurrence. The set of free variables of $M$ is $\s{y,z}$. 

More generally, the set of free variables of a term $M$ is denoted
$\FV{M}$, and it is defined formally as follows:
\[ \begin{array}{lll}
  \FV{x} &=& \s{x}, \\
  \FV{MN} &=& \FV{M}\cup \FV{N}, \\
  \FV{\lam x.M} &=& \FV{M} \setminus \s{x}.
\end{array}
\]
This definition is an example of a definition by recursion on
terms. In other words, in defining $\FV{M}$, we assume that we have
already defined $\FV{N}$ for all subterms of $M$. We will often
encounter such recursive definitions, as well as inductive proofs. 

Before we can formally define $\alpha$-equivalence, we need to define
what it means to {\em rename} a variable in a term. If $x,y$ are
variables, and $M$ is a term, we write $\ren{M}{y}{x}$ for the result of
renaming $x$ as $y$ in $M$. Renaming is formally defined as follows:
\[ \begin{array}{llll}
  \ren{x}{y}{x} &\equiv& y, \\
  \ren{z}{y}{x} &\equiv& z, & \mbox{if $x\neq z$,} \\
  \ren{(MN)}{y}{x} &\equiv& (\ren{M}{y}{x})(\ren{N}{y}{x}), \\
  \ren{(\lam x.M)}{y}{x} &\equiv& \lam y.(\ren{M}{y}{x}), \\
  \ren{(\lam z.M)}{y}{x} &\equiv& \lam z.(\ren{M}{y}{x}), & \mbox{if
  $x\neq z$.}
\end{array}
\]
Note that this kind of renaming replaces all occurrences of $x$ by
$y$, whether free, bound, or binding. We will only apply it in cases
where $y$ does not already occur in $M$. 

Finally, we are in a position to formally define what it means for two
terms to be ``the same up to renaming of bound variables'':

\begin{definition}
  We define {\em $\alpha$-equivalence} to be the smallest congruence
  relation $\eqa$ on lambda terms, such that for all terms $M$ and all
  variables $y$ that do not occur in $M$,
  \[ \lam x.M \eqa \lam y.(\ren{M}{y}{x}). \]
\end{definition}

Recall that a relation on lambda terms is an equivalence relation if
it satisfies rules $\trule{refl}$, $\trule{symm}$, and
$\trule{trans}$. It is a congruence if it also satisfies rules
$\trule{cong}$ and $\nrule{\xi}$.  Thus, by definition,
$\alpha$-equivalence is the smallest relation on lambda terms
satisfying the six rules in Table~\ref{tab-alpha}.
\begin{table*}[tbp]
\[ \begin{array}{lc}
  \trule{refl} &
  \deriv{}{M=M} \nl
  \trule{symm} &
  \deriv{M=N}{N=M} \nl
  \trule{trans} &
  \deriv{M=N\sep N=P}{M=P}
\end{array} \sep
\begin{array}{lc}
  \trule{cong} &
  \deriv{M=M'\sep N=N'}{MN=M'N'} \nl
  \nrule{\xi} &
  \deriv{M=M'}{\lam x.M=\lam x.M'} \nl
  \nrule{\alpha} &
  \deriv{\mbox{$y\not\in M$}}{\lam x.M = \lam y.(\ren{M}{y}{x})}
\end{array}
\]
\caption{The rules for alpha-equivalence}
\label{tab-alpha}
\end{table*}

It is easy to prove by induction that any lambda term is
$\alpha$-equivalent to another term in which the names of all bound
variables are distinct from each other and from any free variables.
Thus, when we manipulate lambda terms in theory and in practice, we
can (and will) always assume {\wloss} that bound variables have been
renamed to be distinct. This convention is called {\em Barendregt's
  variable convention}.

As a remark, the notions of free and bound variables and
$\alpha$-equivalence are of course not particular to the lambda
calculus; they appear in many standard mathematical notations, as well
as in computer science. Here are four examples where the variable $x$
is bound.
\[ \begin{array}{l}
  \int_0^1 x^2\,dx \nl
  \sum_{x=1}^{10}\frac{1}{x} \nl
  \lim_{x\to\infty} e^{-x} \nl
  \verb!int succ(int x) { return x+1; }!
\end{array}
\]

\subsection{Substitution}\label{ssec-substitution}

In the previous section, we defined a renaming operation, which
allowed us to replace a variable by another variable in a lambda term.
Now we turn to a less trivial operation, called {\em substitution},
which allows us to replace a variable by a lambda term. We will write
$\subst{M}{N}{x}$ for the result of replacing $x$ by $N$ in $M$. The
definition of substitution is complicated by two circumstances:
\begin{enumerate}
\item[1.] We should only replace {\em free} variables. This is because
  the names of bound variables are considered immaterial, and should
  not affect the result of a substitution. Thus, $\subst{x(\lam
    xy.x)}{N}{x}$ is $N(\lam xy.x)$, and not $N(\lam xy.N)$.
\item[2.] We need to avoid unintended ``capture'' of free
  variables. Consider for example the term $M\equiv\lam x.yx$, and let
  $N\equiv\lam z.xz$. Note that $x$ is free in $N$ and bound in $M$.
  What should be the result of substituting $N$ for $y$ in $M$?
  If we do this naively, we get 
  \[ \subst{M}{N}{y}=\subst{(\lam x.yx)}{N}{y}=
     \lam x.Nx=\lam x.(\lam z.xz)x. 
  \]
  However, this is not what we intended, since the variable $x$ was
  free in $N$, and during the substitution, it got bound. We need to
  account for the fact that the $x$ that was bound in $M$ was not the
  ``same'' $x$ as the one that was free in $N$. The proper thing to do
  is to rename the bound variable {\em before} the substitution:
  \[ \subst{M}{N}{y}=\subst{(\lam x'.yx')}{N}{y}=
     \lam x'.Nx'=\lam x'.(\lam z.xz)x'. 
  \]
\end{enumerate}

Thus, the operation of substitution forces us to sometimes rename a
bound variable. In this case, it is best to pick a variable from
$\Vars$ that has not been used yet as the new name of the bound
variable. A variable that is currently unused is called {\em fresh}.
The reason we stipulated that the set $\Vars$ is infinite was to make
sure a fresh variable is always available when we need one.

\begin{definition}
  The (capture-avoiding) {\em substitution} of $N$ for free
  occurrences of $x$ in $M$, in symbols $\subst{M}{N}{x}$, is defined
  as follows:
  \[  \begin{array}{l@{~}l@{~}ll}
    \subst{x}{N}{x} &\equiv & N, \\
    \subst{y}{N}{x} &\equiv & y, &\mbox{if $x\neq y$,} \\
    \subst{(MP)}{N}{x} &\equiv & (\subst{M}{N}{x})(\subst{P}{N}{x}), \\
    \subst{(\lam x.M)}{N}{x} &\equiv & \lam x.M, \\
    \subst{(\lam y.M)}{N}{x} &\equiv & \lam y.(\subst{M}{N}{x}), 
    &\mbox{if $x\neq y$ and $y\not\in\FV{N}$,} \\
    \subst{(\lam y.M)}{N}{x} &\equiv & \lam y'.(\subst{\ren{M}{y'}{y}}{N}{x}), 
    &\mbox{if $x\neq y$, $y\in\FV{N}$, and $y'$ fresh.} \\
  \end{array}
  \]
\end{definition}

This definition has one technical flaw: in the last clause, we did not
specify which fresh variable to pick, and thus, technically,
substitution is not well-defined. One way to solve this problem is to
declare all lambda terms to be identified up to $\alpha$-equivalence,
and to prove that substitution is in fact well-defined modulo
$\alpha$-equivalence.  Another way would be to specify which variable
$y'$ to choose: for instance, assume that there is a well-ordering on
the set $\Vars$ of variables, and stipulate that $y'$ should be chosen
to be the least variable that does not occur in either $M$ or $N$.

\subsection{Introduction to $\beta$-reduction}

\begin{convention}
From now on, unless stated otherwise, we identify lambda terms up to
$\alpha$-equivalence. This means, when we speak of lambda terms being
``equal'', we mean that they are $\alpha$-equivalent. Formally, we
regard lambda terms as equivalence classes modulo $\alpha$-equivalence.
We will often use the ordinary equality symbol $M=N$ to denote
$\alpha$-equivalence. 
\end{convention}

The process of evaluating lambda terms by ``plugging arguments into
functions'' is called {\em $\beta$-reduction}. A term of the form
$(\lam x.M)N$, which consists of a lambda abstraction applied to
another term, is called a {\em $\beta$-redex}. We say that it {\em
  reduces} to $\subst{M}{N}{x}$, and we call the latter term the {\em
  reduct}. We reduce lambda terms by finding a subterm that is a
redex, and then replacing that redex by its reduct. We repeat this as
many times as we like, or until there are no more redexes left to
reduce. A lambda term without any $\beta$-redexes is said to be in {\em
  $\beta$-normal form}.

For example, the lambda term $(\lam x.y)((\lam z.zz)(\lam w.w))$ can
be reduced as follows. Here, we underline each redex just before
reducing it:
\[ \begin{array}{lll}
  (\lam x.y)(\ul{(\lam z.zz)(\lam w.w)}) 
  &\redb& (\lam x.y)(\ul{(\lam w.w)(\lam w.w)}) \\
  &\redb& \ul{(\lam x.y)(\lam w.w)} \\
  &\redb& y.
\end{array}
\]
The last term, $y$, has no redexes and is thus in normal form.
We could reduce the same term differently, by choosing the redexes in
a different order:
\[ \begin{array}{lll}
  \ul{(\lam x.y)((\lam z.zz)(\lam w.w))}
  &\redb& y.
\end{array}
\]
As we can see from this example:
\begin{itemize}
\item[-] reducing a redex can create new redexes,
\item[-] reducing a redex can delete some other redexes,
\item[-] the number of steps that it takes to reach a normal form
  can vary, depending on the order in which the redexes are reduced.
\end{itemize}
We can also see that the final result, $y$, does not seem to depend on
the order in which the redexes are reduced. In fact, this is true in
general, as we will prove later.

If $M$ and $M'$ are terms such that $M\redbs M'$, and if $M'$ is in
normal form, then we say that $M$ {\em evaluates} to $M'$. 

Not every term evaluates to something; some terms can be reduced
forever without reaching a normal form. The following is an example:
\[ \begin{array}{lll}
  (\lam x.xx)(\lam y.yyy) 
  &\redb& (\lam y.yyy)(\lam y.yyy) \\
  &\redb& (\lam y.yyy)(\lam y.yyy)(\lam y.yyy) \\
  &\redb& \ldots
\end{array}
\]
This example also shows that the size of a lambda term need not
decrease during reduction; it can increase, or remain the same. The
term $(\lam x.xx)(\lam x.xx)$, which we encountered in
Section~\ref{sec-intro}, is another example of a lambda term that
does not reach a normal form. 

\subsection{Formal definitions of $\beta$-reduction and $\beta$-equivalence}

The concept of $\beta$-reduction can be defined formally as follows:

\begin{definition}\label{page-def-beta}
  We define {\em single-step $\beta$-reduction} to be the smallest
  relation $\redb$ on terms satisfying:
  \[ \begin{array}{lc}
    \nrule{\beta} &
    \deriv{}{(\lam x.M)N\redb \subst{M}{N}{x}} \nl
    \trule{cong$_1$} &
    \deriv{M\redb M'}{MN\redb M'N} \nl
    \trule{cong$_2$} &
    \deriv{N\redb N'}{MN\redb MN'} \nl
    \nrule{\xi} &
    \deriv{M\redb M'}{\lam x.M\redb \lam x.M'}
  \end{array}
  \]
\end{definition}

Thus, $M\redb M'$ iff $M'$ is obtained from $M$ by reducing a {\em
  single} $\beta$-redex of $M$.

\begin{definition}
  We write $M\redbs M'$ if $M$ reduces to $M'$ in zero or more steps.
  Formally, $\redbs$ is defined to be the reflexive transitive
  closure of $\redb$, i.e., the smallest reflexive transitive
  relation containing $\redb$. 
\end{definition}
  
Finally, $\beta$-equivalence is obtained by allowing reduction steps
as well as inverse reduction steps, i.e., by making $\redb$
symmetric:

\begin{definition}
  We write $M\eqb M'$ if $M$ can be transformed into $M'$ by zero or
  more reduction steps and/or inverse reduction steps. Formally,
  $\eqb$ is defined to be the reflexive symmetric transitive closure
  of $\redb$, i.e., the smallest equivalence relation containing
  $\redb$. 
\end{definition}

\begin{exercise}
  This definition of $\beta$-equivalence is slightly different from
  the one given in class. Prove that they are in fact the same.
\end{exercise}

\section{Programming in the untyped lambda calculus}

One of the amazing facts about the untyped lambda calculus is that we
can use it to encode data, such as booleans and natural numbers, as well as
programs that operate on the data. This can be done purely within the
lambda calculus, without adding any additional syntax or axioms.

We will often have occasion to give names to particular lambda terms;
we will usually use boldface letters for such names.

\subsection{Booleans}\label{ssec-booleans}

We begin by defining two lambda terms to encode the truth values
``true'' and ``false'':
\[ \begin{array}{rcl}
  \truet &=& \lam xy.x \\
  \falset &=& \lam xy.y
\end{array}
\]
Let $\andt$ be the term $\lam ab.ab\falset$. Verify the following:
\[ \begin{array}{rcl}
  \andt \truet \truet &\redbs& \truet \\
  \andt \truet \falset &\redbs& \falset \\
  \andt \falset \truet &\redbs& \falset \\
  \andt \falset \falset &\redbs& \falset
\end{array}
\]
Note that $\truet$ and $\falset$ are normal forms, so we can really
say that a term such as $\andt \truet \truet$ {\em evaluates} to
$\truet$. We say that $\andt$ {\em encodes} the boolean function
``and''. It is understood that this coding is with respect to the
particular coding of ``true'' and ``false''. We don't claim that
$\andt MN$ evaluates to anything meaningful if $M$ or $N$ are terms
other than $\truet$ and $\falset$.

Incidentally, there is nothing unique about the term $\lam
ab.ab\falset$. It is one of many possible ways of encoding the ``and''
function. Another possibility is $\lam ab.bab$.

\begin{exercise}
  Find lambda terms $\ort$ and $\nott$ that encode the boolean
  functions ``or'' and ``not''. Can you find more than one term?
\end{exercise}

Moreover, we define the term $\ifthenelset=\lam x.x$. This term behaves
like an ``if-then-else'' function --- specifically, we have
\[ \begin{array}{rcl}
  \ifthenelset \truet M N &\redbs& M \\
  \ifthenelset \falset M N &\redbs& N \\
\end{array}
\]
for all lambda terms $M$, $N$.

\subsection{Natural numbers}\label{ssec-natural-numbers}

If $f$ and $x$ are lambda terms, and $n\geq 0$ a natural number, write
$f^nx$ for the term $f(f(\ldots(fx)\ldots))$, where $f$ occurs $n$
times.  For each natural number $n$, we define a lambda term
$\chnum{n}$, called the {\em $n$th Church numeral}, as $\chnum{n}=\lam
fx.f^nx$. Here are the first few Church numerals:
\[ \begin{array}{rcl}
  \chnum{0} &=& \lam fx.x \\
  \chnum{1} &=& \lam fx.fx \\
  \chnum{2} &=& \lam fx.f(fx) \\
  \chnum{3} &=& \lam fx.f(f(fx)) \\
  \ldots
\end{array}
\]
This particular way of encoding the natural numbers is due to Alonzo
Church, who was also the inventor of the lambda calculus.
Note that $\chnum{0}$ is in fact the same term as $\falset$; thus,
when interpreting a lambda term, we should know ahead of time whether
to interpret the result as a boolean or a numeral.

The successor function can be defined as follows: $\succt = \lam
nfx.f(nfx)$. What does this term compute when applied to a numeral?
\[ \begin{array}{rcl}
  \succt \chnum{n} &=& (\lam nfx.f(nfx))(\lam fx.f^nx) \\
  &\redb& \lam fx.f((\lam fx.f^nx)fx) \\
  &\redbs& \lam fx.f(f^nx) \\
  &=& \lam fx.f^{n+1}x \\
  &=& \chnum{n+1}
\end{array}
\]
Thus, we have proved that the term $\succt$ does indeed encode the
successor function, when applied to a numeral. Here are possible
definitions of addition and multiplication:
\[ \begin{array}{rcl}
  \addt &=& \lam nmfx.nf(mfx) \\
  \multt &=& \lam nmf.n(mf). 
\end{array}
\]

\begin{exercise}
  \begin{enumerate}
  \item[(a)] Manually evaluate the lambda terms
    $\addt\chnum{2}\chnum{3}$ and $\multt\chnum{2}\chnum{3}$.
  \item[(b)] Prove that $\addt\chnum{n}\chnum{m}\redbs\chnum{n+m}$,
    for all natural numbers $n$, $m$.
  \item[(c)] Prove that
    $\multt\chnum{n}\chnum{m}\redbs\chnum{n\cdot m}$, for all natural
    numbers $n$, $m$.
  \end{enumerate}
\end{exercise}

\begin{definition}
  Suppose $f:\N^k\to \N$ is a $k$-ary function on the natural numbers,
  and that $M$ is a lambda term. We say that $M$ {\em (numeralwise)
    represents} $f$ if for all $n_1,\ldots,n_k\in\N$,
  \[ M\chnum{n_1}\ldots\chnum{n_k} \redbs \chnum{f(n_1,\ldots,n_k)}.
  \]
\end{definition}

This definition makes explicit what it means to be an ``encoding''. We
can say, for instance, that the term $\addt = \lam nmfx.nf(mfx)$
represents the addition function.  The definition generalizes easily to
boolean functions, or functions of other data types.

Often handy is the function $\iszerot$ from natural numbers to booleans, which
is defined by
\[ \begin{array}{rcll}
  \iszerot(0) &=& \mbox{true} \\
  \iszerot(n) &=& \mbox{false,} & \mbox{if $n\neq 0$.}
\end{array}
\]
Convince yourself that the following term is a representation of
this function:
\[ \iszerot = \lam nxy.n(\lam z.y)x. \]

\begin{exercise}
  Find lambda terms that represent each of the following functions:
  \begin{enumerate}
  \item[(a)] 
    $ f(n)=(n+3)^2, $
  \item[(b)] 
    $  f(n)=\left\{\begin{array}{ll}\mbox{true}&\mbox{if
          $n$ is even,}\\\mbox{false}&\mbox{if $n$ is odd,}
      \end{array}\right. 
    $
  \item[(c)] 
    $ \expt(n,m)=n^m, $
  \item[(d)] 
    $ \predt(n)=n-1. $
  \end{enumerate}
  Note: part (d) is not easy. In fact, Church believed for a while
  that it was impossible, until his student Kleene found a solution.
  (In fact, Kleene said he found the solution while having his wisdom
  teeth pulled, so his trick for defining the predecessor function is
  sometimes referred to as the ``wisdom teeth trick''.)
\end{exercise}

We have seen how to encode some simple boolean and arithmetic
functions. However, we do not yet have a systematic method of
constructing such functions. What we need is a mechanism for defining
more complicated functions from simple ones. Consider for example the
factorial function, defined by:
\[ \begin{array}{rcll}
  0! &=& 1 \\
  n! &=& n\cdot (n-1)!,& \mbox{if $n\neq 0$}.
\end{array}
\]
The encoding of such functions in the lambda calculus is the subject
of the next section. It is related to the concept of a fixed point.

\subsection{Fixed points and recursive functions}\label{subsec-fixed-points}

Suppose $f$ is a function. We say that $x$ is a {\em fixed point} of $f$
if $f(x)=x$. In arithmetic and calculus, some functions have
fixed points, while others don't. For instance, $f(x)=x^2$ has two
fixed points $0$ and $1$, whereas $f(x)=x+1$ has no fixed points. Some
functions have infinitely many fixed points, notably $f(x)=x$. 

We apply the notion of fixed points to the lambda calculus. If $F$ and
$N$ are lambda terms, we say that $N$ is a fixed point of $F$ if $FN\eqb
N$.  The lambda calculus contrasts with arithmetic in that {\em every}
lambda term has a fixed point. This is perhaps the first surprising fact
about the lambda calculus we learn in this course.

\begin{theorem}\label{thm-fix}
In the untyped lambda calculus, every term $F$ has a fixed point.
\end{theorem}

\begin{proof}
  Let $A=\lam xy.y(xxy)$, and define $\Thetat=AA$. Now suppose $F$ is
  any lambda term, and let $N=\Thetat F$. We claim that $N$ is a
  fixed point of $F$.  This is shown by the following calculation:
\[ \begin{array}{rcl}
  N 
  &=& \Thetat F \\
  &=& AA F \\
  &=& (\lam xy.y(xxy))AF \\
  &\redbs& F(AAF) \\
  &=& F(\Thetat F) \\
  &=& FN.
\end{array}
\]
\eottwo
\end{proof}

The term $\Thetat$ used in the proof is called {\em Turing's fixed point
  combinator}. 

The importance of fixed points lies in the fact that they allow us to
solve {\em equations}. After all, finding a fixed point for $f$ is the
same thing as solving the equation $x = f(x)$. This covers equations
with an arbitrary right-hand side, whose left-hand side is $x$. From
the above theorem, we know that we can always solve such equations in
the lambda calculus.

To see how to apply this idea, consider the question from the last
section, namely, how to define the factorial function. The most
natural definition of the factorial function is recursive, and we can
write it in the lambda calculus as follows:
\[ \begin{array}{rcl} 
  \factt n &=& \ifthenelset (\iszerot n) (\chnum{1}) (\multt n (\factt
  (\predt n)))
 \end{array}
 \]
Here we have used various abbreviations for lambda terms that were
introduced in the previous section. The evident problem with a
recursive definition such as this one is that the term to be defined,
$\factt$, appears both on the left- and the right-hand side. In other
words, to find $\factt$ requires solving an equation!

We now apply our newfound knowledge of how to solve fixed point equations
in the lambda calculus. We start by rewriting the problem slightly:
\[ \begin{array}{rcl} 
  \factt &=& \lam n.\ifthenelset (\iszerot n) (\chnum{1}) (\multt n (\factt
  (\predt n))) \\
  \factt &=& (\lam f.\lam n.\ifthenelset (\iszerot n) (\chnum{1})
  (\multt n (f (\predt n)))) \factt
\end{array}
\]
Let us temporarily write $F$ for the term 
\[ \lam f.\lam n.\ifthenelset (\iszerot n) (\chnum{1}) (\multt n (f
   (\predt n))). 
\]
Then the last equation becomes $\factt = F \factt$, which is a
fixed point equation. We can solve it up to $\beta$-equivalence, by
letting
\[ \begin{array}{rcl} 
  \factt &=& \Thetat F \\
  &=& \Thetat (\lam f.\lam n.\ifthenelset (\iszerot n) (\chnum{1})
  (\multt n (f (\predt n))))
\end{array}
\]
Note that $\factt$ has disappeared from the right-hand side. The
right-hand side is a closed lambda term that represents the factorial
function. (A lambda term is called {\em closed} if it contains no free
variables).

To see how this definition works in practice, let us evaluate $\factt
\chnum{2}$. Recall from the proof of Theorem~\ref{thm-fix} that
$\Thetat F\redbs F(\Thetat F)$, therefore $\factt \redbs F\factt$.
\[ \begin{array}{@{}r@{~}c@{~}l@{}}
  \factt\chnum{2}
  &\redbs& F\factt\chnum{2} \\
  &\redbs& \ifthenelset(\iszerot \chnum{2}) (\chnum{1}) 
  (\multt \chnum{2} (\factt (\predt \chnum{2}))) \\
  &\redbs& \ifthenelset (\falset) (\chnum{1}) 
  (\multt \chnum{2} (\factt (\predt \chnum{2}))) \\
  &\redbs& \multt \chnum{2} (\factt (\predt \chnum{2})) \\
  &\redbs& \multt \chnum{2} (\factt \chnum{1}) \\
  &\redbs& \multt \chnum{2} (F\factt \chnum{1}) \\
  &\redbs& \ldots \\
  &\redbs& \multt \chnum{2} (\multt \chnum{1} (\factt\chnum{0})) \\
  &\redbs& \multt \chnum{2} (\multt \chnum{1} (F\factt\chnum{0})) \\
  &\redbs& \multt \chnum{2} (\multt \chnum{1} (\ifthenelset(\iszerot
  \chnum{0}) (\chnum{1}) (\multt \chnum{0} (\factt (\predt \chnum{0}))))) \\
  &\redbs& \multt \chnum{2} (\multt \chnum{1} (\ifthenelset(\truet)
  (\chnum{1}) (\multt \chnum{0} (\factt (\predt \chnum{0}))))) \\
\end{array}\]\[\begin{array}{rcl}
  &\redbs& \multt \chnum{2} (\multt \chnum{1} \chnum{1}) \\
  &\redbs& \chnum{2}
\end{array}
\]
Note that this calculation, while messy, is completely mechanical. You
can easily convince yourself that $\factt \chnum{3}$ reduces to
$\multt \chnum{3} (\factt \chnum{2})$, and therefore, by the above
calculation, to $\multt \chnum{3} \chnum{2}$, and finally to
$\chnum{6}$. It is now a matter of a simple induction to prove that
$\factt \chnum{n}\redbs \chnum{n!}$, for any $n$.

\begin{exercise}
  Write a lambda term that represents the Fibonacci
  function, defined by
  \[ f(0) = 1,\sep f(1) = 1,\sep f(n+2)=f(n+1)+f(n), \mbox{for $n\geq 2$}
  \]
\end{exercise}

\begin{exercise}
  Write a lambda term that represents the characteristic
  function of the prime numbers, i.e., $f(n)=\mbox{true}$ if $n$ is
  prime, and $\mbox{false}$ otherwise. 
\end{exercise}

\begin{exercise}
  We have remarked at the beginning of this section that
  the number-theoretic function $f(x)=x+1$ does not have a
  fixed point. On the other hand, the lambda term $F=\lam x.\succt x$,
  which represents the same function, does have a fixed point by
  Theorem~\ref{thm-fix}. How can you reconcile the two statements?
\end{exercise}

\begin{exercise}
  The first fixed point combinator for the lambda calculus was discovered
  by Curry.  Curry's fixed point combinator, which is also called the
  {\em paradoxical fixed point combinator}, is the term $\Y=\lam f.(\lam
  x.f(xx))(\lam x.f(xx))$.
  \begin{enumerate}
  \item[(a)] Prove that this is indeed a fixed point combinator, i.e.,
    that $\Y F$ is a fixed point of $F$, for any term $F$. 
  \item[(b)] Turing's fixed point combinator not only satisfies $\Thetat
    F\eqb F(\Thetat F)$, but also $\Thetat F\redbs F(\Thetat F)$. We
    used this fact in evaluating $\factt\chnum{2}$. Does an analogous
    property hold for $\Y$? Does this affect the outcome of the
    evaluation of $\factt\chnum{2}$?
  \item[(c)] Can you find another fixed point combinator, besides Curry's
    and Turing's?
  \end{enumerate}
\end{exercise}

\subsection{Other data types: pairs, tuples, lists, trees, etc.}

So far, we have discussed lambda terms that represented functions on
booleans and natural numbers. However, it is easily possible to encode more
general data structures in the untyped lambda calculus. Pairs and
tuples are of interest to everybody. The examples of lists and trees
are primarily interesting to people with experience in a
list-processing language such as LISP or PROLOG; you can safely ignore
these examples if you want to.

{\bf Pairs.} If $M$ and $N$ are lambda terms, we define the pair
$\pair{M,N}$ to be the lambda term $\lam z.zMN$. We also define two
terms $\leftt=\lam p.p(\lam xy.x)$ and $\rightt=\lam p.p(\lam xy.y)$.
We observe the following:
\[  \begin{array}{rcl}
  \leftt \pair{M,N} &\redbs& M \\
  \rightt \pair{M,N} &\redbs& N
\end{array}
\]
The terms $\leftt$ and $\rightt$ are called the left and right {\em
  projections}.

{\bf Tuples.} The encoding of pairs easily extends to arbitrary
$n$-tuples. If $M_1,\ldots,M_n$ are terms, we define the $n$-tuple
$\tuple{M_1,\ldots,M_n}$ as the lambda term $\lam z.zM_1\ldots M_n$,
and we define the $i$th projection $\pi^n_i=\lam p.p(\lam x_1\ldots
x_n.x_i)$. Then
\[ \begin{array}{rcl}
  \pi^n_i\tuple{M_1,\ldots,M_n} &\redbs& M_i, \mbox{for all $1\leq i\leq n$.}
\end{array}
\]

{\bf Lists.} A list is different from a tuple, because its length is
not necessarily fixed. A list is either empty (``nil''), or else it
consists of a first element (the ``head'') followed by another list
(the ``tail''). We write $\nilt$ for the empty list, and $H::T$ for
the list whose head is $H$ and whose tail is $T$. So, for instance,
the list of the first three numbers can be written as
$1::(2::(3::\nilt))$. We usually omit the parentheses, where it is
understood that ''$::$'' associates to the right. Note that every list
ends in $\nilt$.

In the lambda calculus, we can define $\nilt=\lam xy.y$ and $H::T =
\lam xy.xHT$. Here is a lambda term that adds a list of numbers:
\[ \nm{addlist} l = l(\lam h\,t.\addt h(\nm{addlist} t))(\chnum{0}).
\]
Of course, this is a recursive definition, and must be translated into
an actual lambda term by the method of Section~\ref{subsec-fixed-points}.
In the definition of $\nm{addlist}$, $l$ and $t$ are lists of numbers,
and $h$ is a number. If you are very diligent, you can calculate the sum of
last weekend's Canadian lottery results by evaluating the term
\[ \nm{addlist} (\chnum{4}::\chnum{22}::\chnum{24}::\chnum{32}::\chnum{42}::\chnum{43}::\nilt).
\]

Note that lists enable us to give an alternative encoding of the
natural numbers: We can encode a natural number as a list of booleans,
which we interpret as the binary digits 0 and 1. Of course, with this
encoding, we would have to carefully redesign our basic functions,
such as successor, addition, and multiplication. However, if done
properly, such an encoding would be a lot more efficient (in terms of
number of $\beta$-reductions to be performed) than the encoding by
Church numerals.

{\bf Trees.} A binary tree is a data structure that can be one of two
things: either a {\em leaf}, labeled by a natural number, or a {\em
  node}, which has a left and a right subtree. We write $\leaft(N)$ for
a leaf labeled $N$, and $\nodet(L,R)$ for a node with left subtree $L$
and right subtree $R$. We can encode trees as lambda terms, for
instance as follows:
\[ \leaft(n) = \lam xy.xn,\sep \nodet(L,R) = \lam xy.yLR \]
As an illustration, here is a program (i.e., a lambda term) that adds
all the numbers at the leafs of a given tree.
\[ \nm{addtree} t = t(\lam n.n)(\lam l\,
r.\addt(\nm{addtree}l)(\nm{addtree}r)).
\]

\begin{exercise}
  This is a voluntary programming exercise.
  \begin{enumerate}
  \item[(a)] Write a lambda term that calculates the length of a
    list.
  \item[(b)] Write a lambda term that calculates the depth (i.e., the
    nesting level) of a tree. You may need to define a function
    $\nm{max}$ that calculates the maximum of two numbers.
  \item[(c)] Write a lambda term that sorts a list of numbers. You
    may assume given a term $\nm{less}$ that compares two numbers. 
  \end{enumerate}
\end{exercise}

\section{The Church-Rosser Theorem}

\subsection{Extensionality, $\eta$-equivalence, and $\eta$-reduction}

In the untyped lambda calculus, any term can be applied to another
term. Therefore, any term can be regarded as a function.  Consider a
term $M$, not containing the variable $x$, and consider the term
$M'=\lam x.Mx$. Then for any argument $A$, we have $MA\eqb M'A$. So in
this sense, $M$ and $M'$ define ``the same function''.  Should $M$ and
$M'$ be considered equivalent as terms?

The answer depends on whether we want to accept the principle that
``if $M$ and $M'$ define the same function, then $M$ and $M'$ are
equal''. This is called the principle of {\em extensionality}, and we
have already encountered it in Section~\ref{subsec-intro1}. Formally, the
extensionality rule is the following:
\[ \begin{array}{lc}
  \trule{ext$_\forall$} & \deriv{\forall A.MA=M'A}{M=M'}.
\end{array}
\]
In the presence of the axioms $\nrule{\xi}$, $\trule{cong}$, and
$\nrule{\beta}$, it can be easily seen that $MA=M'A$ is true for {\em
  all} terms $A$ if and only if $Mx=M'x$, where $x$ is a fresh
variable. Therefore, we can replace the extensionality rule by the
following equivalent, but simpler rule:
\[ \begin{array}{lc}
  \trule{ext} & \deriv{Mx=M'x, \mbox{ where $x\not\in\FV{M,M'}$}}{M=M'}.
\end{array}
\]
Note that we can apply the extensionality rule in particular to the
case where $M'=\lam x.Mx$, where $x$ is not free in $M$. As we have
remarked above, $Mx\eqb M'x$, and thus extensionality implies that
$M=\lam x.Mx$. This last equation is called the $\eta$-law (eta-law):
\[ \begin{array}{lc}
  \nrule{\eta} & M=\lam x.Mx, \mbox{ where $x\not\in\FV{M}$}.
\end{array}
\]
In fact, $\nrule{\eta}$ and $\trule{ext}$ are equivalent in the
presence of the other axioms of the lambda calculus. We have already
seen that $\trule{ext}$ and $\nrule{\beta}$ imply $\nrule{\eta}$.
Conversely, assume $\nrule{\eta}$, and assume that $Mx=M'x$, for some
terms $M$ and $M'$ not containing $x$ freely.  Then by $\nrule{\xi}$,
we have $\lam x.Mx=\lam x.M'x$, hence by $\nrule{\eta}$ and
transitivity, $M=M'$. Thus $\trule{ext}$ holds.

We note that the $\eta$-law does not follow from the axioms and rules
of the lambda calculus that we have considered so far. In particular,
the terms $x$ and $\lam y.xy$ are not $\beta$-equivalent, although
they are clearly $\eta$-equivalent. We will prove that $x\not\eqb \lam
y.xy$ in Corollary~\ref{cor-beta-not-eta} below.

Single-step $\eta$-reduction is the smallest relation $\rede$
satisfying $\trule{cong$_1$}$, $\trule{cong$_2$}$, $\nrule{\xi}$, and
the following axiom (which is the same as the $\eta$-law, directed
right to left):
\[ \begin{array}{lc}\label{page-def-eta}
  \nrule{\eta} & \lam x.Mx\rede M, \mbox{ where $x\not\in\FV{M}$}.
\end{array}
\]
Single-step $\beta\eta$-reduction $\redbe$ is defined as the union of
the single-step $\beta$- and $\eta$-reductions, i.e., $M\redbe M'$ iff
$M\redb M'$ or $M\rede M'$. Multi-step $\eta$-reduction $\redes$,
multi-step $\beta\eta$-reduction $\redbes$, as well as
$\eta$-equivalence $\eqe$ and $\beta\eta$-equivalence $\eqbe$ are
defined in the obvious way as we did for $\beta$-reduction and
equivalence. We also get the evident notions of $\eta$-normal form,
$\beta\eta$-normal form, etc.

\subsection{Statement of the Church-Rosser Theorem, and some consequences}

\begin{un-theorem}[Church and Rosser, 1936]\label{thm-church-rosser}
  Let $\reds$ denote either $\redbs$ or $\redbes$.  Suppose $M$, $N$,
  and $P$ are lambda terms such that $M\reds N$ and $M\reds P$. Then
  there exists a lambda term $Z$ such that $N\reds Z$ and $P\reds Z$.
\end{un-theorem}

In pictures, the theorem states that the following diagram can always
be completed:
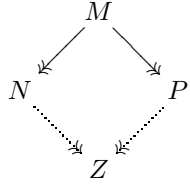
\[ \xymatrix@dr{M\ar@{->>}[r]\ar@{->>}[d] & P\ar@{.>>}[d]\\ N\ar@{.>>}[r] & Z}
\]

This property is called the {\em Church-Rosser property}, or {\em
  confluence}. Before we prove the Church-Rosser Theorem, let us
highlight some of its consequences.

\begin{corollary}\label{cor-cr-2}
  If $M\eqb N$ then there exists some $Z$ with $M,N\redbs
  Z$. Similarly for $\beta\eta$.
\end{corollary}

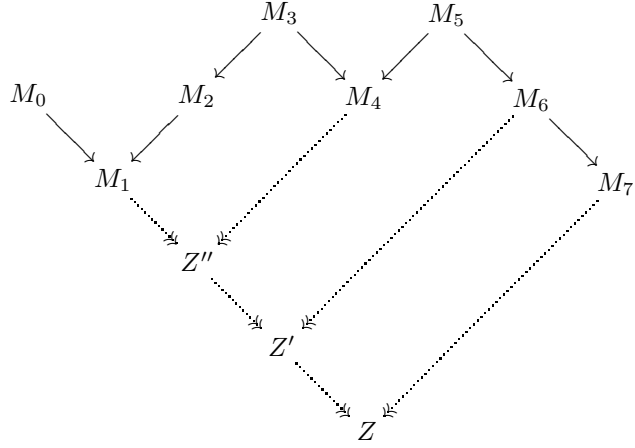
\begin{figure}
\vspace{-6em}
\[ \xymatrix@dr{
  && M_5\ar@{->}[r]\ar@{->}[d]
  & M_6\ar@{->}[r]\ar@{.>>}[ddd]
  & M_7\ar@{.>>}[ddd]
  \\& M_3\ar@{->}[r]\ar@{->}[d]
  & M_4\ar@{.>>}[dd]
  \\& M_2\ar@{->}[d]
  \\ M_0\ar@{->}[r]
  & M_1\ar@{.>>}[r]
  & Z''\ar@{.>>}[r]
  & Z'\ar@{.>>}[r]
  & Z
  }
\]
\caption{The proof of Corollary~\ref{cor-cr-2}}
\label{fig-cor-cr-2}
\end{figure}

\begin{proof}
  Please refer to Figure~\ref{fig-cor-cr-2} for an illustration of
  this proof. Recall that $\eqb$ is the reflexive symmetric transitive
  closure of $\redb$. Suppose that $M\eqb N$.  Then there exist $n\geq
  0$ and terms $M_0,\ldots,M_n$ such that $M=M_0$, $N=M_n$, and for
  all $i=1\ldots n$, either $M_{i-1}\redb M_{i}$ or $M_{i}\redb
  M_{i-1}$.  We prove the claim by induction on $n$. For $n=0$, we
  have $M=N$ and there is nothing to show. Suppose the claim has been
  proven for $n-1$. Then by induction hypothesis, there exists a term
  $Z'$ such that $M\redbs Z'$ and $M_{n-1}\redbs Z'$. Further, we know
  that either $N\redb M_{n-1}$ or $M_{n-1}\redb N$. In case $N\redb
  M_{n-1}$, then $N\redbs Z'$, and we are done. In case $M_{n-1}\redb
  N$, we apply the Church-Rosser Theorem to $M_{n-1}$, $Z'$, and $N$
  to obtain a term $Z$ such that $Z'\redbs Z$ and $N\redbs Z$. Since
  $M\redbs Z'\redbs Z$, we are done. The proof in the case of
  $\beta\eta$-reduction is identical.\eot
\end{proof}

\begin{corollary}\label{cor-cr-3}
  If $N$ is a $\beta$-normal form and $N\eqb M$, then $M\redbs N$, and
  similarly for $\beta\eta$.
\end{corollary}

\begin{proof}
  By Corollary~\ref{cor-cr-2}, there exists some $Z$ with $M,N\redbs
  Z$. But $N$ is a normal form, thus $N\eqa Z$. \eot
\end{proof}

\begin{corollary}\label{cor-cr-4}
  If $M$ and $N$ are $\beta$-normal forms such that $M\eqb N$, then
  $M\eqa N$, and similarly for $\beta\eta$.
\end{corollary}

\begin{proof}
  By Corollary~\ref{cor-cr-3}, we have $M\redbs N$, but since $M$ is a
  normal form, we have $M\eqa N$. \eot
\end{proof}

\begin{corollary}
  If $M\eqb N$, then neither or both have a $\beta$-normal
  form. Similarly for $\beta\eta$.
\end{corollary}

\begin{proof}
  Suppose that $M\eqb N$, and that one of them has a $\beta$-normal
  form. Say, for instance, that $M$ has a normal form $Z$. Then $N\eqb
  Z$, hence $N\redbs Z$ by Corollary~\ref{cor-cr-3}. \eot
\end{proof}

\begin{corollary}\label{cor-beta-not-eta}
  The terms $x$ and $\lam y.xy$ are not $\beta$-equivalent. In
  particular, the $\eta$-rule does not follow from the $\beta$-rule.
\end{corollary}

\begin{proof}
  The terms $x$ and $\lam y.xy$ are both $\beta$-normal forms, and
  they are not $\alpha$-equivalent. It follows by
  Corollary~\ref{cor-cr-4} that $x\not\eqb \lam y.xy$. \eot
\end{proof}

\subsection{Preliminary remarks on the proof of the Church-Rosser Theorem}
\label{subsec-prelim-cr}

Consider any binary relation $\red$ on a set, and let $\reds$ be its
reflexive transitive closure. Consider the following three
properties of such relations:
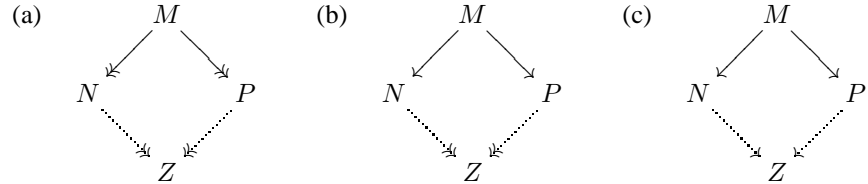
\[ \mbox{(a)} \ssep \xymatrix@dr{M\ar@{->>}[r]\ar@{->>}[d] & P\ar@{.>>}[d]\\ N\ar@{.>>}[r] & Z}
\sep
\mbox{(b)} \ssep \xymatrix@dr{M\ar@{->}[r]\ar@{->}[d] & P\ar@{.>>}[d]\\ N\ar@{.>>}[r] & Z}
\sep
\mbox{(c)} \ssep \xymatrix@dr{M\ar@{->}[r]\ar@{->}[d] & P\ar@{.>}[d]\\ N\ar@{.>}[r] & Z}
\]
Each of these properties states that for all $M,N,P$, if the solid
arrows exist, then there exists $Z$ such that the dotted arrows exist.
The only difference between (a), (b), and (c) is the difference
between where $\red$ and $\reds$ are used.

Property (a) is the Church-Rosser property. Property (c) is called the
diamond property (because the diagram is shaped like a diamond).

A naive attempt to prove the Church-Rosser Theorem might proceed as
follows: First, prove that the relation $\redb$ satisfies property (b)
(this is relatively easy to prove); then use an inductive argument to
conclude that it also satisfies property (a). 

\begin{figure}
\[ \xymatrix@-1em{
  \bullet\ar[dr]&&\bullet\ar[dl]\ar[dr]&&\bullet\ar[dr] \\
  &\bullet\ar[dl]&&\bullet\ar[dl]\ar[dr]&&\bullet\ar[dl] \\
  \bullet\ar[dr]&&\bullet\ar[dl]\ar[dr]&&\bullet\ar[dr] \\
  &\bullet\ar[dl]&&\bullet\ar[dl]\ar[dr]&&\bullet\ar[dl] \\
  \bullet\ar[dr]&&\bullet\ar[dl]\ar[dr]&&\bullet\ar[dr] \\
  &\bullet\ar[dl]&&\bullet\ar[dl]\ar[dr]&&\bullet\ar[dl] \\
  \bullet\ar[dr]&&\bullet\ar[dl]\ar[dr]&&\bullet\ar[dr] \\
  &\bullet\ar[dl]&&\bullet\ar[dl]\ar[dr]&&\bullet\ar[dl] \\
  \vdots &&   \vdots &&   \vdots & 
  }
\]
\caption{An example of a relation that satisfies property (b), but not
  property (a)}
\label{fig-b-not-a}
\end{figure}
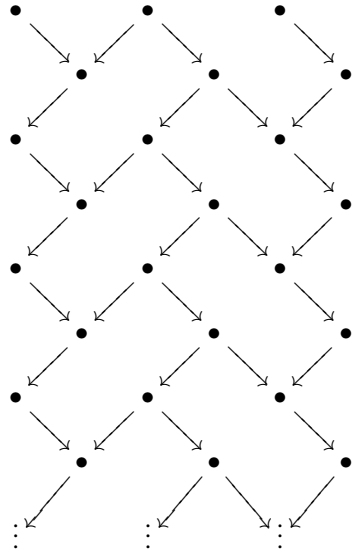

Unfortunately, this does not work: the reason is that in general,
property (b) does not imply property (a)! An example of a relation
that satisfies property (b) but not property (a) is shown in
Figure~\ref{fig-b-not-a}. In other words, a proof of property (b) is
not sufficient in order to prove property (a). 

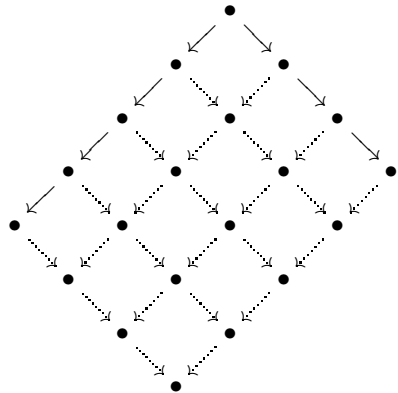
\begin{figure}
\[ \xymatrix@-1em@dr{
  \bullet\ar[r]\ar[d] & \bullet\ar[r]\ar@{.>}[d] & \bullet\ar[r]\ar@{.>}[d] &
  \bullet\ar@{.>}[d] \\
  \bullet\ar@{.>}[r]\ar[d] & \bullet\ar@{.>}[r]\ar@{.>}[d] & \bullet\ar@{.>}[r]\ar@{.>}[d] &
  \bullet\ar@{.>}[d] \\
  \bullet\ar@{.>}[r]\ar[d] & \bullet\ar@{.>}[r]\ar@{.>}[d] & \bullet\ar@{.>}[r]\ar@{.>}[d] &
  \bullet\ar@{.>}[d] \\
  \bullet\ar@{.>}[r]\ar[d] & \bullet\ar@{.>}[r]\ar@{.>}[d] & \bullet\ar@{.>}[r]\ar@{.>}[d] &
  \bullet\ar@{.>}[d] \\
  \bullet\ar@{.>}[r] & \bullet\ar@{.>}[r] & \bullet\ar@{.>}[r] &
  \bullet 
  }
\]
\caption{Proof that property (c) implies property (a)}
\label{fig-diamond-a}
\end{figure}

On the other hand, property (c), the diamond property, {\em does}
imply property (a). This is very easy to prove by induction, and the
proof is illustrated in Figure~\ref{fig-diamond-a}. But unfortunately,
$\beta$-reduction does not satisfy property (c), so again we are stuck.

To summarize, we are faced with the following dilemma:
\begin{itemize}
\item $\beta$-reduction satisfies property (b), but property (b) does
  not imply property (a).
\item Property (c) implies property (a), but $\beta$-reduction does
  not satisfy property (c).
\end{itemize}

On the other hand, it seems hopeless to prove property (a) directly.
In the next section, we will solve this dilemma by defining yet
another reduction relation $\tri$, with the following properties:
\begin{itemize}
\item $\tri$ satisfies property (c), and
\item the transitive closure of $\tri$ is the same as that of $\redb$
  (or $\redbe$).
\end{itemize}

\subsection{Proof of the Church-Rosser Theorem}
\label{subsec-proof-cr}

In this section, we will prove the Church-Rosser Theorem for
$\beta\eta$-reduction. The proof for $\beta$-reduction (without
$\eta$) is very similar, and in fact slightly simpler, so we omit it
here. The proof presented here is due to Tait and Martin-L\"of. We
begin by defining a new relation $M\tri M'$ on terms, called {\em
 parallel one-step reduction}. We define $\tri$ to be the smallest
relation satisfying
\[ \begin{array}{lc}
  (1) & \deriv{}{x\tri x} \nl
  (2) & \deriv{P \tri P' \sep N \tri N'}{PN \tri P'N'} \nl
  (3) & \deriv{N \tri N'}{\lam x.N \tri \lam x.N'} \nl
  (4) & \deriv{Q \tri Q' \sep N \tri N'}{(\lam x.Q)N \tri \subst{Q'}{N'}{x}}
  \nl
  (5) & \deriv{P \tri P',\mbox{ where $x\not\in\FV{P}$}}{\lam x.Px
    \tri P'}.
\end{array}
\]

\begin{lemma}\label{lem-tri-redbes}
  \begin{enumerate}
  \item[(a)] For all $M,M'$, if $M\redbe M'$ then $M\tri M'$.
  \item[(b)] For all $M,M'$, if $M\tri M'$ then $M\redbes M'$.
  \item[(c)] $\redbes$ is the reflexive, transitive closure of $\tri$.
  \end{enumerate}
\end{lemma}

\begin{proof}
  (a) First note that we have $P\tri P$, for any term $P$. This is
  easily shown by induction on $P$. We now prove the claim by
  induction on a derivation of $M\redbe M'$.  Please refer to
  pages~\pageref{page-def-beta} and {\pageref{page-def-eta}} for
  the rules that define $\redbe$. We make a case distinction based on
  the last rule used in the derivation of $M\redbe M'$.
  \begin{itemize}
  \item If the last rule was $\nrule{\beta}$, then $M=(\lam x.Q)N$ and
    $M'=\subst{Q}{N}{x}$, for some $Q$ and $N$. But then $M\tri M'$ by 
    (4), using the facts $Q\tri Q$ and $N\tri N$. 
  \item If the last rule was $\nrule{\eta}$, then $M=\lam x.Px$ and
    $M'=P$, for some $P$ such that $x\not\in\FV{P}$. Then $M\tri M'$
    follows from (5), using $P\tri P$.
  \item If the last rule was $\trule{cong$_1$}$, then $M=PN$ and
    $M'=P'N$, for some $P$, $P'$, and $N$ where $P\redbe P'$. By
    induction hypothesis, $P\tri P'$. From this and $N\tri N$, it
    follows immediately that $M\tri M'$ by (2).
  \item If the last rule was $\trule{cong$_2$}$, we proceed similarly
    to the last case.
  \item If the last rule was $\nrule{\xi}$, then $M=\lam x.N$ and
    $M'=\lam x.N'$ for some $N$ and $N'$ such that $N\redbe N'$. By
    induction hypothesis, $N\tri N'$, which implies $M\tri M'$ by (3).
  \end{itemize}
  
  (b) We prove this by induction on a derivation of $M\tri M'$. We
  distinguish several cases, depending on the last rule used in the
  derivation.
  \begin{itemize}
  \item If the last rule was (1), then $M=M'=x$, and we are done because
    $x\redbes x$.
  \item If the last rule was (2), then $M=PN$ and $M'=P'N'$, for some
    $P$, $P'$, $N$, $N'$ with $P\tri P'$ and $N\tri N'$. By induction
    hypothesis, $P\redbes P'$ and $N\redbes N'$. Since $\redbes$
    satisfies $\trule{cong}$, it follows that $PN\redbes P'N'$, hence
    $M\redbes M'$ as desired.
  \item If the last rule was (3), then $M=\lam x.N$ and $M'=\lam
    x.N'$, for some $N,N'$ with $N\tri N'$. By induction hypothesis,
    $N\redbes N'$, hence $M=\lam x.N\redbes \lam x.N'=M'$ by
    $\nrule{\xi}$. 
  \item If the last rule was (4), then $M=(\lam x.Q)N$ and
    $M'=\subst{Q'}{N'}{x}$, for some $Q,Q',N,N'$ with $Q\tri Q'$ and
    $N\tri N'$. By induction hypothesis, $Q\redbes Q'$ and $N\redbes
    N'$. Therefore $M=(\lam x.Q)N\redbes(\lam x.Q')N'\redbe
    \subst{Q'}{N'}{x}=M'$, as desired.
  \item If the last rule was (5), then $M=\lam x.Px$ and $M'=P'$, for
    some $P,P'$ with $P\tri P'$, and $x\not\in\FV{P}$. By induction
    hypothesis, $P\redbes P'$, hence $M=\lam x.Px\redbe P\redbes
    P'=M'$, as desired.
  \end{itemize}
  
  (c) This follows directly from (a) and (b). Let us write $R^*$ for
  the reflexive transitive closure of a relation $R$. By (a), we have
  ${\redbe}\seq{\tri}$, hence ${\redbes}={\redbe}^*\seq{\tri}^*$. By
  (b), we have ${\tri}\seq{\redbes}$, hence
  ${\tri}^*\seq{\redbes}^*={\redbes}$. It follows that
  ${\tri}^*={\redbes}$.\eot
\end{proof}

We will soon prove that $\tri$ satisfies the diamond property.  Note
that together with Lemma~\ref{lem-tri-redbes}(c), this will
immediately imply that $\redbes$ satisfies the Church-Rosser property.

\begin{lemma}[Substitution]
If $M \tri M'$ and $U \tri U'$, then
$\subst{M}{U}{y} \tri \subst{M'}{U'}{y}$.
\end{lemma}

\begin{proof}
  We assume without loss of generality that any bound variables of $M$
  are different from $y$ and from the free variables of $U$. The claim
  is now proved by induction on derivations of $M\tri M'$. We
  distinguish several cases, depending on the last rule used in the
  derivation:
  \begin{itemize}
  \item If the last rule was (1), then $M=M'=x$, for some variable
    $x$. If $x=y$, then $\subst{M}{U}{y}=U\tri U'=\subst{M'}{U'}{y}$.
    If $x\neq y$, then by (1), $\subst{M}{U}{y}=x\tri
    x=\subst{M'}{U'}{y}$.
  \item If the last rule was (2), then $M=PN$ and $M'=P'N'$, for some
    $P$, $P'$, $N$, $N'$ with $P\tri P'$ and $N\tri N'$. By induction
    hypothesis, $\subst{P}{U}{y}\tri \subst{P'}{U'}{y}$ and
    $\subst{N}{U}{y}\tri \subst{N'}{U'}{y}$, hence by (2),
    $\subst{M}{U}{y}=\subst{P}{U}{y}\subst{N}{U}{y} \tri 
    \subst{P'}{U'}{y}\subst{N'}{U'}{y} = \subst{M'}{U'}{y}$.
  \item If the last rule was (3), then $M=\lam x.N$ and $M'=\lam
    x.N'$, for some $N,N'$ with $N\tri N'$. By induction hypothesis,
    $\subst{N}{U}{y}\tri \subst{N'}{U'}{y}$, hence by (3)
    $\subst{M}{U}{y}=\lam x.\subst{N}{U}{y} \tri \lam
    x.\subst{N'}{U'}{y} = \subst{M'}{U'}{y}$.
  \item If the last rule was (4), then $M=(\lam x.Q)N$ and
    $M'=\subst{Q'}{N'}{x}$, for some $Q,Q',N,N'$ with $Q\tri Q'$ and
    $N\tri N'$.  By induction hypothesis, $\subst{Q}{U}{y}\tri
    \subst{Q'}{U'}{y}$ and $\subst{N}{U}{y}\tri \subst{N'}{U'}{y}$,
    hence by (4), $(\lam x.\subst{Q}{U}{y})\subst{N}{U}{y} \tri
    \subst{\subst{Q'}{U'}{y}}{\subst{N'}{U'}{y}}{x} =
    \subst{\subst{Q'}{N'}{x}}{U'}{y}$. Thus $\subst{M}{U}{y}
    \tri \subst{M'}{U'}{y}$.
  \item If the last rule was (5), then $M=\lam x.Px$ and $M'=P'$, for
    some $P,P'$ with $P\tri P'$, and $x\not\in\FV{P}$. By induction
    hypothesis, $\subst{P}{U}{y}\tri \subst{P'}{U'}{y}$, hence by (5),
    $\subst{M}{U}{y}=\lam x.\subst{P}{U}{y}x \tri
    \subst{P'}{U'}{y}=\subst{M'}{U'}{y}$.\eot
  \end{itemize}
\end{proof}
  
A more conceptual way of looking at this proof is the following:
consider any derivation of $M\tri M'$ from axioms (1)--(5). In this
derivation, replace any axiom $y\tri y$ by $U\tri U'$, and propagate
the changes (i.e., replace $y$ by $U$ on the left-hand-side, and by
$U'$ on the right-hand-side of any $\tri$). The result is a derivation
of $\subst{M}{U}{y}\tri\subst{M'}{U'}{y}$. (The formal proof that the
result of this replacement is indeed a valid derivation requires an
induction, and this is the reason why the proof of the substitution
lemma is so long).

Our next goal is to prove that $\tri$ satisfies the diamond
property. Before proving this, we first define the {\em maximal
  parallel one-step reduct} $M^*$ of a term $M$ as follows:

\begin{enumerate}
\item $x^* = x$, for a variable.
\item $(PN)^* = P^*N^*$, if $PN$ is not a $\beta$-redex. 
\item $((\lam x.Q)N)^* = \subst{Q^*}{N^*}{x}$.
\item $(\lam x.N)^* = \lam x.N^*$, if $\lam x.N$ is not an 
        $\eta$-redex.
\item $(\lam x.Px)^* = P^*$, if $x\not\in\FV{P}$.
\end{enumerate}

Note that $M^*$ depends only on $M$.  The following lemma implies the
diamond property for $\tri$.

\begin{lemma}[Maximal parallel one-step reductions]
\label{lem-max-par}
  Whenever $M \tri M'$, then $M' \tri M^*$.
\end{lemma}

\begin{proof}
  By induction on the size of $M$. We distinguish five cases,
  depending on the last rule used in the derivation of $M \tri M'$.
  As usual, we assume that all bound variables have been renamed to
  avoid clashes.
  \begin{itemize}
  \item If the last rule was (1), then $M=M'=x$, also $M^*=x$, and we
    are done.
  \item If the last rule was (2), then $M=PN$ and $M'=P'N'$, where
    $P\tri P'$ and $N\tri N'$. By
    induction hypothesis $P'\tri P^*$ and $N'\tri N^*$. Two cases:
    \begin{itemize}
    \item If $PN$ is not a $\beta$-redex, then $M^*=P^*N^*$. Thus
      $M'=P'N'\tri P^*N^*=M^*$ by (2), and we are done.
    \item If $PN$ is a $\beta$-redex, say $P=\lam x.Q$, then
      $M^*=\subst{Q^*}{N^*}{x}$. We distinguish two subcases,
      depending on the last rule used in the derivation of $P\tri P'$:
      \begin{itemize} 
      \item If the last rule was (3), then $P'=\lam x.Q'$, where
        $Q\tri Q'$.  By induction hypothesis $Q' \tri Q^*$, and with
        $N' \tri N^*$, it follows that $M' = (\lam x.Q')N' \tri
        \subst{Q^*}{N^*}{x} = M^*$ by (4).
      \item If the last rule was (5), then $P=\lam x.Rx$ and $P'=R'$,
        where $x\not\in\FV{R}$ and $R\tri R'$. Consider the term
        $Q=Rx$.  Since $Rx \tri R'x$, and $Rx$ is a subterm of $M$, by
        induction hypothesis $R'x \tri (Rx)^*$. By the substitution
        lemma, $M' = R'N' = \subst{(R'x)}{N'}{x} \tri
        \subst{(Rx)^*}{N^*}{x} = M^*$.
      \end{itemize}
    \end{itemize}
  \item If the last rule was (3), then $M = \lam x.N$ and $M' = \lam
    x.N'$, where $N\tri N'$. Two cases:
    \begin{itemize}
    \item If $M$ is not an $\eta$-redex, then $M^*=\lam x.N^*$. By
      induction hypothesis, $N'\tri N^*$, hence $M'\tri M^*$ by (3).
    \item If $M$ is an $\eta$-redex, then $N=Px$, where
      $x\not\in\FV{P}$. In this case, $M^*=P^*$. We distinguish two
      subcases, depending on the last rule used in the derivation of
      $N\tri N'$:
      \begin{itemize}
      \item If the last rule was (2), then $N' = P'x$, where $P\tri
        P'$. By induction hypothesis $P' \tri P^*$.  Hence $M' = \lam
        x.P'x \tri P^* = M^*$ by (5).
      \item If the last rule was (4), then $P = \lam y.Q$ and $N' =
        \subst{Q'}{x}{y}$, where $Q\tri Q'$. Then $M' = \lam
        x.\subst{Q'}{x}{y} = \lam y.Q'$ (note $x \not\in \FV{Q'}$).
        But $P \tri \lam y.Q'$, hence by induction hypothesis, $\lam
        y.Q' \tri P^* = M^*$.
      \end{itemize}
    \end{itemize}
  \item If the last rule was (4), then $M = (\lam x.Q)N$ and $M' =
    \subst{Q'}{N'}{x}$, where $Q\tri Q'$ and $N\tri N'$. Then $M^* =
    \subst{Q^*}{N^*}{x}$, and $M' \tri M^*$ by the substitution lemma.
  \item If the last rule was (5), then $M = \lam x.Px$ and $M' = P'$,
    where $P\tri P'$ and $x\not\in\FV{P}$. Then $M^*=P^*$. By
    induction hypothesis, $P'\tri P^*$, hence $M'\tri M^*$.  \eot
  \end{itemize}
\end{proof}

The previous lemma immediately implies the diamond property for
$\tri$:

\begin{lemma}[Diamond property for $\tri$]
  If $M\tri N$ and $M\tri P$, then there exists $Z$ such that $N\tri
  Z$ and $P\tri Z$. 
\end{lemma}

\begin{proof}
  Take $Z=M^*$.\eot
\end{proof}

Finally, we have a proof of the Church-Rosser Theorem:

\begin{proofof}{Theorem~\ref{thm-church-rosser}}
  Since $\tri$ satisfies the diamond property, it follows that its
  reflexive transitive closure $\tri^*$ also satisfies the diamond
  property, as shown in Figure~\ref{fig-diamond-a}. But $\tri^*$ is the same
  as $\redbes$ by Lemma~\ref{lem-tri-redbes}(c), and the diamond property for
  $\redbes$ is just the Church-Rosser property for $\redbe$.\eot
\end{proofof}

\subsection{Exercises}

\begin{exercise}
  Give a detailed proof that property (c) from
  Section~\ref{subsec-prelim-cr} implies property (a).
\end{exercise}

\begin{exercise}
  Prove that $M\tri M$, for all terms $M$.
\end{exercise}

\begin{exercise}
  Without using Lemma~\ref{lem-max-par}, prove that $M\tri M^*$ for
  all terms $M$.
\end{exercise}

\begin{exercise}
  Let $\Omega=(\lam x.xx)(\lam x.xx)$. Prove that
  $\Omega\not\eqbe\Omega\Omega$.
\end{exercise}

\begin{exercise}
  What changes have to be made to Section~\ref{subsec-proof-cr} to get
  a proof of the Church-Rosser Theorem for $\redb$, instead of
  $\redbe$?
\end{exercise}

\begin{exercise}
  Recall the properties (a)--(c) of binary relations $\red$ that were
  discussed in Section~\ref{subsec-prelim-cr}. Consider the following
  similar property, which is sometimes called the ``strip property'':
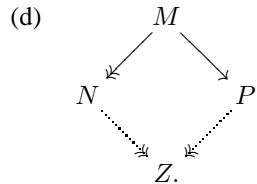
  \[ \mbox{(d)} \ssep \xymatrix@dr{M\ar@{->}[r]\ar@{->>}[d] & P\ar@{.>>}[d]\\ N\ar@{.>>}[r] & Z.}
  \]
  Does (d) imply (a)? Does (b) imply (d)? In each case, give either a
  proof or a counterexample.
\end{exercise}

\begin{exercise}
  To every lambda term $M$, we may associate a directed graph (with
  possibly multiple edges and loops) $\sG(M)$ as follows: (i) the
  vertices are terms $N$ such that $M\redbs N$, i.e., all the terms
  that $M$ can $\beta$-reduce to; (ii) the edges are given by a
  single-step $\beta$-reduction. Note that the same term may have two
  (or more) reductions coming from different redexes; each such
  reduction is a separate edge. For example, let $I=\lam x.x$. Let
  $M=I(Ix)$. Then
  \[ \sG(M)=\xymatrix{I(Ix)\ar@/^/[r]\ar@/_/[r]&Ix\ar[r]&x}.
  \]
  Note that there are two separate edges from $I(Ix)$ to $Ix$.  We
  also sometimes write bullets instead of terms, to get $
  \xymatrix{\bullet\ar@/^/[r]\ar@/_/[r]&\bullet\ar[r]&\bullet}$.  As
  another example, let $\Omega=(\lam x.xx)(\lam x.xx)$. Then
  \[ \sG(\Omega) = \xymatrix{\bullet\ar@(ur,dr)[]&}.
  \]
  \begin{enumerate}
  \item[(a)] Let $M=(\lam x.I(xx))(\lam x.xx)$. Find $\sG(M)$.
  \item[(b)] For each of the following graphs, find a term $M$ such
    that $\sG(M)$ is the given graph, or explain why no such term
    exists. (Note: the ``starting'' vertex need not always be the
    leftmost vertex in the picture). Warning: some of these terms are
    tricky to find!
    \begin{enumerate}
    \item[(i)] \[ \xymatrix{\bullet\ar[r]&\bullet\ar@(ur,dr)[]} \]
    \item[(ii)] \[ \xymatrix{\bullet&\bullet\ar@(ur,dr)[]\ar[l]} \]
    \item[(iii)] \[ \xymatrix{\bullet&\bullet\ar[r]\ar[l]&\bullet} \]
    \item[(iv)] 
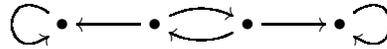
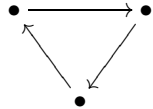
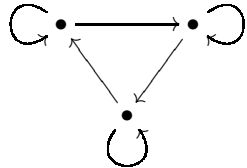
\[ \xymatrix{\bullet&\bullet\ar[l]\ar@/^/[r]&\bullet\ar@/^/[l]} \]
    \item[(v)] \[ \xymatrix{\bullet\ar@(ul,dl)[]&\bullet\ar@/^/[r]\ar[l]&\bullet\ar[r]\ar@/^/[l]&\bullet\ar@(ur,dr)[]} \]
    \item[(vi)] \[ \xymatrix@C-1em{\bullet\ar[rr]&&\bullet\ar[dl]\\&\bullet\ar[ul]}
      \]
    \item[(vii)] \[
      \xymatrix@C-1em{\bullet\ar@(ul,dl)[]\ar[rr]&&\bullet\ar@(ur,dr)[]\ar[dl]\\&\bullet\ar@(dl,dr)[]\ar[ul]}
      \]
    \end{enumerate}
  \end{enumerate}
\end{exercise}

\section{Combinatory algebras}

To give a model of the lambda calculus means to provide a mathematical
space in which the axioms of lambda calculus are satisfied. This
usually means that the elements of the space can be understood as
functions, and that certain functions can be understood as elements. 

Na\"ively, one might try to construct a model of lambda calculus by
finding a set $X$ such that $X$ is in bijective correspondence with
the set $X^X$ of {\em all} functions from $X$ to $X$. This, however,
is impossible: for cardinality reason, the equation $X\cong X^X$ has
no solutions except for a one-element set $X=1$. To see this, first
note that the empty set $\emptyset$ is not a solution. Also, suppose
$X$ is a solution with $|X|\geq 2$. Then $|X^X|\geq |2^X|$, but by
Cantor's argument, $|2^X|>|X|$, hence $X^X$ is of greater cardinality
than $X$, contradicting $X\cong X^X$.

There are two main strategies for constructing models of the lambda
calculus, and both involve a restriction on the class of functions to
make it smaller. The first approach, which will be discussed in this
section, uses {\em algebra}, and the essential idea is to replace the
set $X^X$ of all function by a smaller, and suitably defined set of
{\em polynomials}. The second approach is to equip the set $X$ with
additional structure (such as topology, ordered structure, etc), and
to replace $X^X$ by a set of structure-preserving functions (for
example, continuous functions, monotone functions, etc).

\subsection{Applicative structures}

\begin{definition}
  An {\em applicative structure} $(\Aa,\app)$ is a set $\Aa$ together
  with a binary operation ``$\app$''.
\end{definition}

Note that there are no further assumptions; in particular, we do {\em
  not} assume that application is an associative operation. We write
$ab$ for $a\app b$, and as in the lambda calculus, we follow the
convention of left associativity, i.e., we write $abc$ for $(ab)c$.

\begin{definition}
  Let $(\Aa,\app)$ be an applicative structure. A {\em polynomial} in
  a set of variables $x_1,\ldots,x_n$ and with coefficients in $\Aa$
  is a formal expression built from variables and elements of $\Aa$ by
  means of the application operation. In other words, the set of
  polynomials is given by the following grammar:
  \[ \begin{array}{lll}
    t,s&\bnf& x\bor a\bor ts,
  \end{array}
  \]
  where $x$ ranges over variables and $a$ ranges over the elements of
  $\Aa$. We write $\Aa\s{x_1,\ldots,x_n}$ for the set of polynomials
  in variables $x_1,\ldots,x_n$ with coefficients in $\Aa$.
\end{definition}

Here are some examples of polynomials in the variables $x,y,z$,
where $a,b\in\Aa$:
\[ x,\sep xy, \sep axx, \sep (x(y(zb)))(ax).
\]

If $t(x_1,\ldots,x_n)$ is a polynomial in the indicated variables,
and $b_1,\ldots,b_n$ are elements of $\Aa$, then we can evaluate the
polynomial at the given elements: the evaluation $t(b_1,\ldots,b_n)$
the element of $\Aa$ obtained by ``plugging'' $x_i=b_i$ into the
polynomial, for $i=1,\ldots,n$, and evaluating the resulting
expression in $\Aa$. Note that in this way, every polynomial $t$ in
$n$ variables can be understood as a {\em function} from $\Aa^n\to
\Aa$. This is very similar to the usual polynomials in algebra, which
can also either be understood as formal expressions or as functions.

If $t(x_1,\ldots,x_n)$ and $s(x_1,\ldots,x_n)$ are two polynomials
with coefficients in $\Aa$, we say that the equation
$t(x_1,\ldots,x_n) = s(x_1,\ldots,x_n)$ {\em holds} in $\Aa$ if for
all $b_1,\ldots,b_n\in\Aa$, $t(b_1,\ldots,b_n) = s(b_1,\ldots,b_n)$.

\subsection{Combinatory completeness}

\begin{definition}[Combinatory completeness]
  An applicative structure $(\Aa,\app)$ is {\em combinatorially
    complete} if for every polynomial $t(x_1,\ldots,x_n)$ of $n\geq 0$
  variables, there exists some element $a\in\Aa$ such that
  \[ ax_1\ldots x_n = t(x_1,\ldots,x_n)
  \]
  holds in $\Aa$.
\end{definition}

In other words, combinatory completeness means that every polynomial
{\em function} $t(x_1,\ldots,x_n)$ can be represented (in curried
form) by some {\em element} of $\Aa$. We are therefore setting up a
correspondence between functions and elements as discussed in the
introduction of this section.

Note that we do not require the element $a$ to be unique in the
definition of combinatory completeness. This means that we are dealing
with an intensional view of functions, where a given function might in
general have several different names (but see the discussion of
extensionality in Section~\ref{subsec-extensional-combinatory}).

The following theorem characterizes combinatory completeness in terms
of a much simpler algebraic condition.

\begin{theorem}\label{thm-combinatory-completeness}
  An applicative structure $(\Aa,\app)$ is combinatorially complete if
  and only if there exist two elements $s,k\in \Aa$, such that the
  following equations are satisfied for all $x,y,z\in\Aa$:
  \[ \begin{array}{ll}
    (1) & sxyz = (xz)(yz) \\
    (2) & kxy = x \\
  \end{array}
  \]
\end{theorem}

\begin{example}\label{exa-combinatory-completeness}
  Before we prove this theorem, let us look at a few examples.
  \begin{enumerate}\alphalabels
  \item The identity function. Can we find an element $i\in\Aa$ such
    that $ix=x$ for all $x$? Yes, indeed, we can let $i=skk$. We check
    that for all $x$, $skkx = (kx)(kx) = x$.
  \item The boolean ``true''. Can we find an element $\truet$ such
    that for all $x,y$, $\truet xy = x$? Yes, this is easy: $\truet = k$.
  \item The boolean ``false''. Can we find $\falset$ such that
    $\falset xy = y$?  Yes, what we need is $\falset x = i$. Therefore
    a solution is $\falset = ki$.  And indeed, for all $y$, we have
    $kixy = iy = y$.
  \item Find a function $f$ such that $fx = xx$ for all $x$. Solution:
    let $f=sii$. Then $siix = (ix)(ix) = xx$. 
  \end{enumerate}
\end{example}

\begin{proofof}{Theorem~\ref{thm-combinatory-completeness}}
  The ``only if'' direction is trivial. If $\Aa$ is combinatorially
  complete, then consider the polynomial $t(x,y,z)=(xz)(yz)$. By
  combinatory completeness, there exists some $s\in\Aa$ with
  $sxyz=t(x,y,z)$, and similarly for $k$.
  
  We thus have to prove the ``if'' direction.  Recall that
  $\Aa\s{x_1,\ldots,x_n}$ is the set of polynomials with variables
  $x_1,\ldots,x_n$. For each polynomial $t\in\Aa\s{x,y_1,\ldots,y_n}$ in
  $n+1$ variables, we will define a new polynomial $\lam^*x.t \in
  \Aa\s{y_1,\ldots,y_n}$ in $n$ variables, as follows by recursion on
  $t$:
  \[ \begin{array}{llll}
    \lam^*x.x &:=& i,\\
    \lam^*x.y_i &:=& ky_i & \mbox{where $y_i\neq x$ is a variable,} \\
    \lam^*x.a &:=& ka     & \mbox{where $a\in \Aa$,}\\
    \lam^*x.pq &:=& s(\lam^*x.p)(\lam^*x.q). \\
  \end{array}
  \]
  We claim that for all $t$, the equation $(\lam^* x.t)x = t$ holds
  in $\Aa$. Indeed, this is easily proved by induction on $t$, using
  the definition of $\lam^*$:
  \[ \begin{array}{llll}
    (\lam^*x.x)x &=& ix = x,\\
    (\lam^*x.y_i)x &=& ky_ix = y_i,\\
    (\lam^*x.a)x &=& kax = a,\\
    (\lam^*x.pq)x &=& s(\lam^*x.p)(\lam^*x.q)x = ((\lam^*x.p)x)((\lam^*x.q)x) = pq. \\
  \end{array}
  \]
  Note that the last case uses the induction hypothesis for $p$ and $q$. 
  
  Finally, to prove the theorem, assume that $\Aa$ has elements $s,k$
  satisfying equations (1) and (2), and consider a polynomial
  $t\in\Aa\s{x_1,\ldots,x_n}$. We must show that there exists $a\in\Aa$
  such that $ax_1\ldots x_n = t$ holds in $\Aa$. We let
  \[ a = \lam^*x_1.\ldots.\lam^*x_n.t.
  \]
  Note that $a$ is a polynomial in $0$ variables, which we may
  consider as an element of $\Aa$. Then from the previous claim, it
  follows that
  \[ \begin{array}{lll}
    ax_1\ldots x_n &=& (\lam^*x_1.\lam^*x_2.\ldots.\lam^*x_n.t)x_1x_2\ldots x_n\\
    &=& (\lam^*x_2.\ldots.\lam^*x_n.t)x_2\ldots x_n\\
    &=& \ldots\\
    &=& (\lam^*x_n.t)x_n\\
    &=& t\\
  \end{array}
  \]
  holds in $\Aa$. \eot
\end{proofof}

\subsection{Combinatory algebras}\label{ssec-comb-alg}

By Theorem~\ref{thm-combinatory-completeness}, combinatory
completeness is equivalent to the existence of the $s$ and $k$
operators. We enshrine this in the following definition:

\begin{definition}[Combinatory algebra]
  A {\em combinatory algebra} $(\Aa,\app,s,k)$ is an applicative
  structure $(\Aa,\app)$ together with elements $s,k\in\Aa$,
  satisfying the following two axioms:
    \[ \begin{array}{ll}
    (1) & sxyz = (xz)(yz) \\
    (2) & kxy = x \\
  \end{array}
  \]
\end{definition}

\begin{remark}\label{rem-derived-lambda}
  The operation $\lam^*$, defined in the proof of
  Theorem~\ref{thm-combinatory-completeness}, is defined on the
  polynomials of any combinatory algebra. It is called the {\em derived
    lambda abstractor}, and it satisfies the law of $\beta$-equivalence,
  i.e., $(\lam^*x.t)b = t[b/x]$, for all $b\in\Aa$.
\end{remark}

Finding actual examples of combinatory algebras is not so easy. Here
are some examples:

\begin{example}
  The one-element set $\Aa=\s{*}$, with $*\app *=*$, $s=*$, and $k=*$,
  is a combinatory algebra. It is called the {\em trivial} combinatory
  algebra. 
\end{example}

\begin{example}
  Recall that $\Lambda$ is the set of lambda terms. Let
  $\Aa=\Lambda/{\eqb}$, the set of lambda terms modulo
  $\beta$-equivalence. Define $M\app N=MN$, $S=\lam xyz.(xz)(yz)$,
  and $K=\lam xy.x$. Then $(\Lambda,\app,S,K)$ is a combinatory
  algebra. Also note that, by Corollary~\ref{cor-beta-not-eta}, this
  algebra is non-trivial, i.e., it has more than one element.
  
  Similar examples are obtained by replacing $\eqb$ by $\eqbe$, and/or
  replacing $\Lambda$ by the set $\Lambda_0$ of closed terms.
\end{example}

\begin{example}\label{exa-sk-term-alg}
  We construct a combinatory algebra of $SK$-terms as follows.  Let
  $V$ be a given set of variables. The set $\CTerm$ of {\em terms}
  of combinatory logic is given by the grammar:
  \[   \begin{array}{lll}
    A,B &\bnf& x\bor \S\bor \K\bor AB,
  \end{array}
  \]
  where $x$ ranges over the elements of $V$.
  
  On $\CTerm$, we define combinatory equivalence $\eqc$ as the
  smallest equivalence relation satisfying $\S ABC \eqc (AC)(BC)$, $\K
  AB \eqc A$, and the rules $\trule{cong$_1$}$ and $\trule{cong$_2$}$
  (see page~\pageref{page-def-beta}). Then the set $\CTerm/{\eqc}$ is a
  combinatory algebra (called the {\em free} combinatory algebra
  generated by $V$, or the {\em term algebra}). You will prove in
  Exercise~\ref{exe-combinatory-cr} that it is non-trivial.
\end{example}

\void{
Note that all of the above examples of combinatory algebras are either
trivial or syntactic.  It is not easy to find a true ``mathematical''
example of a combinatory algebra. We will see how to find such models
later.
}

\begin{exercise}\label{exe-combinatory-cr}
  On the set $\CTerm$ of combinatory terms, define a notion of {\em
    single-step reduction} by the following laws:
  \[ \begin{array}{l}
    \S ABC \redc (AC)(BC), \\
    \K AB \redc A,\\
  \end{array}
  \]
  together with the usual rules $\trule{cong$_1$}$ and
  $\trule{cong$_2$}$ (see page~\pageref{page-def-beta}). As in lambda
  calculus, we call a term a {\em normal form} if it cannot be
  reduced. Prove that the reduction $\redc$ satisfies the
  Church-Rosser property. (Hint: similarly to the lambda calculus,
  first define a suitable parallel one-step reduction $\tri$ whose
  reflexive transitive closure is that of $\redc$. Then show that it
  satisfies the diamond property.)
\end{exercise}

\begin{corollary}\label{cor-sk-nf}
  It immediately follows from the Church-Rosser Theorem for
  combinatory logic (Exercise~\ref{exe-combinatory-cr}) that two
  normal forms are $\eqc$-equivalent if and only if they are equal.
\end{corollary}

\subsection{The failure of soundness for combinatory algebras}

A combinatory algebra is almost a model of the lambda calculus.
Indeed, given a combinatory algebra $\Aa$, we can interpret any lambda
term as follows. To each term $M$ with free variables among
$x_1,\ldots,x_n$, we recursively associate a polynomial
$\semm{M}\in\Aa\s{x_1,\ldots,x_n}$:
\[ \begin{array}{lll}
  \semm{x} := x, \\
  \semm{NP} := \semm{N}\semm{P}, \\
  \semm{\lam x.M} := \lam^*x.\semm{M}.\\
\end{array}
\]
Notice that this definition is almost the identity function, except
that we have replaced the ordinary lambda abstractor of lambda
calculus by the derived lambda abstractor of combinatory logic. The
result is a polynomial in $\Aa\s{x_1,\ldots,x_n}$. In the particular
case where $M$ is a closed term, we can regard $\semm{M}$ as an
element of $\Aa$.

To be able to say that $\Aa$ is a ``model'' of the lambda calculus, we
would like the following property to be true:
\[ M\eqb N \imp \semm{M}=\semm{N}\mbox{ holds in $\Aa$}.
\]
This property is called {\em soundness} of the interpretation.
Unfortunately, it is in general false for combinatory algebras, as the
following example shows.

\begin{example}\label{exa-failure-sound}
  Let $M=\lam x.x$ and $N=\lam x.(\lam y.y)x$. Then clearly
  $M\eqb N$. On the other hand, 
  \[ \begin{array}{l}
    \semm{M} = \lam^*x.x=i, \\
    \semm{N} = \lam^*x.(\lam^*y.y)x = \lam^*x.ix
    = s(ki)i.
  \end{array}
  \]
  It follows from Exercise~\ref{exe-combinatory-cr} and
  Corollary~\ref{cor-sk-nf} that the equation $i=s(ki)i$ does not hold
  in the combinatory algebra $\CTerm/{\eqc}$. In other words, the
  interpretation is not sound.
\end{example}

Let us analyze the failure of the soundness property further. Recall
that $\beta$-equi\-va\-lence is the smallest equivalence relation on
lambda terms satisfying the six rules in Table~\ref{tab-beta}. 
\begin{table*}[tbp]
  \[ \begin{array}{lc}
    \trule{refl} &
    \deriv{}{M=M} \nl
    \trule{symm} &
    \deriv{M=N}{N=M} \nl
    \trule{trans} &
    \deriv{M=N\sep N=P}{M=P}
  \end{array} \sep
  \begin{array}{lc}
    \trule{cong} &
    \deriv{M=M'\sep N=N'}{MN=M'N'} \nl
    \nrule{\xi} &
    \deriv{M=M'}{\lam x.M=\lam x.M'} \nl
    \nrule{\beta} &
    \deriv{}{(\lam x.M)N= \subst{M}{N}{x}} \nl
  \end{array}
  \]
\caption{The rules for $\beta$-equivalence}
\label{tab-beta}
\end{table*}

If we define a relation $\sim$ on lambda terms by
\[   M\sim N\sep\iff\sep \semm{M}=\semm{N}\mbox{ holds in $\Aa$},
\]
then we may ask which of the six rules of Table~\ref{tab-beta} the
relation $\sim$ satisfies. Clearly, not all six rules can be
satisfied, or else we would have $M\eqb N\imp M\sim N\imp
\semm{M}=\semm{N}$, i.e., the model would be sound.

Clearly, $\sim$ is an equivalence relation, and therefore satisfies
$\trule{refl}$, $\trule{symm}$, and $\trule{trans}$. Also,
$\trule{cong}$ is satisfied, because whenever $p,q,p',q'$ are
polynomials such that $p=p'$ and $q=q'$ holds in $\Aa$, then clearly
$pq=p'q'$ holds in $\Aa$ as well. Finally, we know from
Remark~\ref{rem-derived-lambda} that the rule $\nrule{\beta}$ is
satisfied.

So the rule that fails is the $\nrule{\xi}$ rule. Indeed,
Example~\ref{exa-failure-sound} illustrates this. Note that $x\sim
(\lam y.y)x$ (from the proof of
Theorem~\ref{thm-combinatory-completeness}), but $\lam x.x \not\sim
\lam x.(\lam y.y)x$, and therefore the $\nrule{\xi}$ rule is
violated.

\subsection{Lambda algebras}

A lambda algebra is, by definition, a combinatory algebra that is a
sound model of lambda calculus, and in which $s$ and $k$ have their
expected meanings.

\begin{definition}[Lambda algebra]
  A {\em lambda algebra} is a combinatory algebra $\Aa$ satisfying the
  following properties: 
  \[\begin{array}{l@{~~~}l}
    (\forall M,N\in\Lambda)\sssep M\eqb N \sssep\imp\sssep \semm{M}=\semm{N}& \trule{soundness},\\
    s = \lam^*x.\lam^*y.\lam^*z.(xz)(yz) &\trule{s-derived},\\
    k = \lam^*x.\lam^*y.x &\trule{k-derived}.\\
  \end{array}
  \]
\end{definition}

The purpose of the remainder of this section is to give an axiomatic
description of lambda algebras.

\begin{lemma}
  Recall that $\Lambda_0$ is the set of closed lambda terms, i.e.,
  lambda terms without free variables. Soundness is equivalent to the
  following:
  \[
  (\forall M,N\in\Lambda_0)\sssep M\eqb N \sssep\imp\sssep \semm{M}=\semm{N}
  \ssep\trule{closed soundness}
  \]
\end{lemma}

\begin{proof}
  Clearly soundness implies closed soundness. For the converse, assume
  closed soundness and let $M,N\in\Lambda$ with $M\eqb N$. Let
  $\FV{M}\cup\FV{N}=\s{x_1,\ldots,x_n}$.  Then
  \[ \begin{array}{llll}
    \multicolumn{3}{l}{M\eqb N} \\
    &\imp& \lam x_1\ldots x_n.M\eqb \lam x_1\ldots x_n.N & \mbox{by $\nrule{\xi}$} \\
    &\imp& \semm{\lam x_1\ldots x_n.M}=\semm{\lam x_1\ldots x_n.N} & \mbox{by closed soundness} \\
    &\imp& \lam^* x_1\ldots x_n.\semm{M} = \lam^* x_1\ldots x_n.\semm{N} & \mbox{by def. of $\semm{-}$} \\
    &\imp& (\lam^* x_1\ldots x_n.\semm{M})x_1\ldots x_n \\
    &&  \ssep= (\lam^* x_1\ldots x_n.\semm{N})x_1\ldots x_n \\
    &\imp& \semm{M}=\semm{N} &\mbox{by proof of Thm~\ref{thm-combinatory-completeness}}
  \end{array}
  \]
  This proves soundness.\eot
\end{proof}

\begin{definition}[Translations between combinatory logic and lambda calculus]
  Let $A\in\CTerm$ be a combinatory term (see
  Example~\ref{exa-sk-term-alg}). We define its translation to lambda
  calculus in the obvious way: the translation $A_\lam$ is given
  recursively by:
  \[\begin{array}{lll}
    \S_\lam &=& \lam xyz.(xz)(yz), \\
    \K_\lam &=& \lam xy.x, \\
    x_\lam &=& x, \\
    (AB)_\lam &=& A_\lam B_\lam.
  \end{array}
  \]
  Conversely, given a lambda term $M\in\Lambda$, we recursively define
  its translation $M_c$ to combinatory logic like this:
  \[\begin{array}{lll}
    x_c &=& x,\\
    (MN)_c &=& M_c N_c,\\
    (\lam x.M)_c &=& \lam^*x.(M_c).\\
  \end{array}
  \]
\end{definition}

\begin{lemma}\label{lem-c-lam}
  For all lambda terms $M$, $(M_c)_\lam \eqb M$.
\end{lemma}

\begin{lemma}\label{lem-lam-c}
  Let $\Aa$ be a combinatory algebra satisfying $k =
  \lam^*x.\lam^*y.x$ and $s = \lam^*x.\lam^*y.\lam^*z.(xz)(yz)$.  Then
  for all combinatory terms $A$, $(A_\lam)_c=A$ holds in $\Aa$.
\end{lemma}

\begin{exercise}
  Prove Lemmas~\ref{lem-c-lam} and {\ref{lem-lam-c}}.
\end{exercise}

Let $\CTerm_0$ be the set of {\em closed} combinatory terms.
The following is our first useful characterization of lambda calculus.

\begin{lemma}\label{lem-alt-soundness}
  Let $\Aa$ be a combinatory algebra. Then $\Aa$ is a lambda algebra
  if and only if it satisfies the following property:
  \[ (\forall A,B\in\CTerm_0)\sssep A_\lam \eqb B_\lam 
  \sssep\imp\sssep A=B\mbox{ holds in $\Aa$}.
  \ssep \trule{alt-soundness}
  \]
\end{lemma}

\begin{proof}
  First, assume that $\Aa$ satisfies $\trule{alt-soundness}$. To prove
  $\trule{closed soundness}$, let $M,N$ be lambda terms with $M\eqb
  N$. Then $(M_c)_\lam \eqb M\eqb N\eqb (N_c)_\lam$, hence by
  $\trule{alt-soundness}$, $M_c=N_c$ holds in $\Aa$. But this is the
  definition of $\semm{M}=\semm{N}$. 
  
  To prove $\trule{k-derived}$, note that 
  \[\begin{array}{llll}
    k_\lam &=& (\lam x.\lam y.x) &\mbox{by definition of $(-)_\lam$} \\
    &=& ((\lam x.\lam y.x)_c)_\lam &\mbox{by Lemma~\ref{lem-c-lam}}\\
    &=& (\lam^* x.\lam^* y.x)_\lam &\mbox{by definition of $(-)_c$}.
  \end{array}
  \]
  Hence, by $\trule{alt-soundness}$, it follows that $k=(\lam^*
  x.\lam^* y.x)$ holds in $\Aa$. Similarly for $\trule{s-derived}$.
  
  Conversely, assume that $\Aa$ is a lambda algebra. Let
  $A,B\in\CTerm_0$ and assume $A_\lam\eqb B_\lam$. By soundness,
  $\semm{A_\lam}=\semm{B_\lam}$. By definition of the interpretation,
  $(A_\lam)_c=(B_\lam)_c$ holds in $\Aa$. But by $\trule{s-derived}$,
  $\trule{k-derived}$, and Lemma~\ref{lem-lam-c},
  $A=(A_\lam)_c=(B_\lam)_c=B$ holds in $\Aa$, proving
  $\trule{alt-soundness}$.\eot
\end{proof}

\begin{definition}[Homomorphism]
  Let $(\Aa,\app_\Aa,s_\Aa,k_\Aa)$, $(\Bb,\app_\Bb,s_\Bb,k_\Bb)$ be
  combinatory algebras.  A {\em homomorphism} of combinatory algebras
  is a function $\phi:\Aa\to\Bb$ such that $\phi(s_\Aa)=s_\Bb$,
  $\phi(k_\Aa)=k_\Bb$, and for all $a,b\in\Aa$, $\phi(a\app_\Aa
  b)=\phi(a)\app_\Bb \phi(b)$.
\end{definition}

Any given homomorphism $\phi:\Aa\to\Bb$ can be extended to polynomials
in the obvious way: we define
$\phihat:\Aa\s{x_1,\ldots,x_n}\to\Bb\s{x_1,\ldots,x_n}$ by
\[\begin{array}{ll}
  \phihat(a)=\phi(a) & \mbox{for $a\in\Aa$,}\\
  \phihat(x)=x       & \mbox{if $x\in\s{x_1,\ldots,x_n}$,}\\
  \phihat(pq)=\phihat(p)\phihat(q). \\
\end{array}
\]

\begin{example}
  If $\phi(a)=a'$ and $\phi(b)=b'$, then $\phihat((ax)(by)) = (a'x)(b'y)$. 
\end{example}

The following is the main technical concept needed in the
characterization of lambda algebras. We say that an equation {\em
  holds absolutely} if it holds in $\Aa$ and in any homomorphic image
of $\Aa$. If an equation holds only in the previous sense, then we
sometimes say it holds {\em locally}.

\begin{definition}[Absolute equation]
  Let $p,q\in\Aa\s{x_1,\ldots,x_n}$ be two polynomials with
  coefficients in $\Aa$. We say that the equation $p=q$ {\em holds
    absolutely} in $\Aa$ if for all combinatory algebras $\Bb$ and all
  homomorphisms $\phi:\Aa\to\Bb$, $\phihat(p)=\phihat(q)$ holds in $\Bb$.
  If an equation holds absolutely, we write $p\eqabs q$.
\end{definition}

We can now state the main theorem characterizing lambda algebras. Let
$\one=s(ki)$.

\begin{theorem}\label{thm-tfae-lambda-algebra}
  Let $\Aa$ be a combinatory algebra. Then the following are equivalent:
  \begin{enumerate}
  \item $\Aa$ is a lambda algebra,
  \item $\Aa$ satisfies $\trule{alt-soundness}$,
  \item for all $A,B\in\CTerm$ such that $ A_\lam \eqb B_\lam$, 
    the equation $A=B$ holds absolutely in $\Aa$,
  \item $\Aa$ absolutely satisfies the nine axioms in
    Table~\ref{tab-lambda-algebra-axioms},
  \item $\Aa$ satisfies $\trule{s-derived}$ and $\trule{k-derived}$,
    and for all $p,q\in\Aa\s{y_1,\ldots,y_n}$, if $px\eqabs qx$ then
    $\one p\eqabs \one q$,
  \item\label{thm-tfae-lambda-algebra6} $\Aa$ satisfies
    $\trule{s-derived}$ and $\trule{k-derived}$, and for all
    $p,q\in\Aa\s{x,y_1,\ldots,y_n}$, if $p\eqabs q$ then
    $\lam^*x.p\eqabs \lam^*y.q$.
  \end{enumerate}
\end{theorem}

\begin{table}
\[
\begin{array}{crcl}     
  \eqna &    \one k &\eqabs& k,
  \\      \eqnb &    \one s &\eqabs& s,
  \\      \eqnc &    \one(kx) &\eqabs& kx,
  \\      \eqnd &    \one(sx) &\eqabs& sx,
  \\      \eqne &    \one(sxy) &\eqabs& sxy,
  \\      \eqnf &    s(s(kk)x)y  &\eqabs&\one x,
  \\      \eqng &    s(s(s(ks)x)y)z  &\eqabs&  s(sxz)(syz),
  \\      \eqnh &    k(xy)  &\eqabs&  s(kx)(ky),
  \\      \eqni &    s(kx)i  &\eqabs&\one x.
\end{array}
\]
  \caption{An axiomatization of lambda algebras. Here $\one=s(ki)$.}\label{tab-lambda-algebra-axioms}
\end{table}

The proof proceeds via $1\imp 2\imp 3\imp 4\imp 5\imp 6\imp 1$. 

We have already proven $1\imp 2$ in Lemma~\ref{lem-alt-soundness}.

To prove $2\imp 3$, let $\FV{A}\cup\FV{B}\seq\s{x_1,\ldots,x_n}$, and
assume $A_\lam \eqb B_\lam$. Then $\lam x_1\ldots x_n.(A_\lam)\eqb
\lam x_1\ldots x_n.(B_\lam)$, hence $(\lam^* x_1\ldots x_n.A)_\lam
\eqb (\lam^* x_1\ldots x_n.B)_\lam$ (why?). Since the latter terms are
closed, it follows by the rule $\trule{alt-soundness}$ that $\lam^* x_1\ldots
x_n.A=\lam^* x_1\ldots x_n.B$ holds in $\Aa$. Since closed equations
are preserved by homomorphisms, the latter also holds in $\Bb$ for any
homomorphism $\phi:\Aa\to\Bb$. Finally, this implies that $A=B$ holds
for any such $\Bb$, proving that $A=B$ holds absolutely in $\Aa$.

\begin{exercise}
  Prove the implication $3\imp 4$.
\end{exercise}

The implication $4\imp 5$ is the most difficult part of the theorem.
We first dispense with the easier part:

\begin{exercise}
  Prove that the axioms from Table~\ref{tab-lambda-algebra-axioms}
  imply $\trule{s-derived}$ and $\trule{k-derived}$.
\end{exercise}

The last part of $4\imp 5$ needs the following lemma:

\begin{lemma}\label{lem-a-b}
  Suppose $\Aa$ satisfies the nine axioms from
  Table~\ref{tab-lambda-algebra-axioms}. Define a structure
  $(\Bb,\bullet,S,K)$ by:
  \[ \begin{array}{lll}
    \Bb = \s{a\in\Aa\such a=\one a},\\
    a\bullet b = sab,\\
    S=ks,\\
    K=kk.
  \end{array}
  \]
  Then $\Bb$ is a well-defined combinatory algebra. Moreover, the
  function $\phi:\Aa\to\Bb$ defined by $\phi(a)=ka$ defines a
  homomorphism.
\end{lemma}

\begin{exercise}
  Prove Lemma~\ref{lem-a-b}.
\end{exercise}

To prove the implication $4\imp 5$, assume $ax=bx$ holds absolutely in
$\Aa$. Then $\phihat(ax)=\phihat(bx)$ holds in $\Bb$ by definition of
``absolute''. But $\phihat(ax) = (\phi a)x = s(ka)x$ and $\phihat(bx)
= (\phi b)x = s(kb)x$. Therefore $s(ka)x=s(kb)x$ holds in $\Aa$. We
plug in $x=i$ to get $s(ka)i=s(kb)i$. By axiom $\eqni$, $\one a=\one b$.

To prove $5\imp 6$, assume $p\eqabs q$. Then $(\lam^*x.p)x \eqabs
p\eqabs q \eqabs (\lam^*x.q)x$ by the proof of
Theorem~\ref{thm-combinatory-completeness}. Then by 5.,
$(\lam^*x.p)\eqabs (\lam^*x.q)$.

Finally, to prove $6\imp 1$, note that if $6$ holds, then the absolute
interpretation satisfies the $\xi$-rule, and therefore satisfies all
the axioms of lambda calculus.

\begin{exercise}
  Prove $6\imp 1$.
\end{exercise}

\begin{remark}
  The axioms in Table~\ref{tab-lambda-algebra-axioms} are required to
  hold {\em absolutely}. They can be replaced by local axioms by
  prefacing each axiom with $\lam^*xyz$. Note that this makes the
  axioms much longer.
\end{remark}

\subsection{Extensional combinatory algebras}
\label{subsec-extensional-combinatory}

\begin{definition}
  An applicative structure $(\Aa,\cdot)$ is {\em extensional} if for all
  $a,b\in \Aa$, if $ac=bc$ holds for all $c\in \Aa$, then $a=b$.
\end{definition}

\begin{proposition}
  In an extensional combinatory algebra, the $\nrule{\eta}$ axioms is
  valid.
\end{proposition}

\begin{proof}
  By $\nrule{\beta}$, $(\lam^*x.Mx)c = Mc$ for all $c\in\Aa$. Therefore, by
  extensionality, $(\lam^*x.Mx)=M$.\eot
\end{proof}

\begin{proposition}
  In an extensional combinatory algebra, an equation holds locally if
  and only if it holds absolutely.
\end{proposition}

\begin{proof}
  Clearly, if an equation holds absolutely, then it holds locally.
  Conversely, assume the equation $p=q$ holds locally in $\Aa$. Let
  $x_1,\ldots,x_n$ be the variables occurring in the equation. By
  $\nrule{\beta}$, 
  \[(\lam^*x_1\ldots x_n.p)x_1\ldots
  x_n=(\lam^*x_1\ldots x_n.q)x_1\ldots x_n\] holds locally. By
  extensionality, \[\lam^*x_1\ldots x_n.p=\lam^*x_1\ldots x_n.q\]
  holds. Since this is a closed equation (no free variables), it
  automatically holds absolutely. This implies that $(\lam^*x_1\ldots
  x_n.p)x_1\ldots
  x_n=(\lam^*x_1\ldots x_n.q)x_1\ldots x_n$ holds absolutely, and
  finally, by $\nrule{\beta}$ again, that $p=q$ holds absolutely. \eot
\end{proof}

\begin{proposition}
  Every extensional combinatory algebra is a lambda
  algebra.
\end{proposition}

\begin{proof}
  By
  Theorem~\ref{thm-tfae-lambda-algebra}(\ref{thm-tfae-lambda-algebra6}),
  it suffices to prove $\trule{s-derived}$, $\trule{k-derived}$ and
  the $\nrule{\xi}$-rule. Let $a,b,c\in\Aa$ be arbitrary. Then 
  \[ (\lam^*x.\lam^*y.\lam^*z.(xz)(yz))abc = (ac)(bc) = sabc 
  \]
  by $\nrule{\beta}$ and definition of $s$. Applying extensionality three
  times (with respect to $c$, $b$, and $a$), we get
  \[ \lam^*x.\lam^*y.\lam^*z.(xz)(yz) = s.
  \]
  This proves $\trule{s-derived}$. The proof of $\trule{k-derived}$ is
  similar. Finally, to prove $\nrule{\xi}$, assume that $p\eqabs q$.
  Then by $\nrule{\beta}$, $(\lam^*x.p)c=(\lam^*x.q)c$ for all
  $c\in\Aa$. By extensionality, $\lam^*x.p=\lam^*x.q$ holds.\eot
\end{proof}

\section{Simply-typed lambda calculus, propositional logic, and the Curry-Howard isomorphism}\label{sec-simply-typed-lc}

In the untyped lambda calculus, we spoke about functions without
speaking about their domains and codomains. The domain and codomain of
any function was the set of all lambda terms. We now introduce types
into the lambda calculus, and thus a notion of domain and codomain for
functions. The difference between types and sets is that types are
{\em syntactic} objects, i.e., we can speak of types without having to
speak of their elements. We can think of types as {\em names} for
sets.

\subsection{Simple types and simply-typed terms}

We assume a set of basic types. We usually use the Greek letter
$\iota$ (``iota'') to denote a basic type. The set of simple types is
given by the following BNF:
\[ \mbox{Simple types:}\ssep A,B \bnf \iota \bor A\to B\bor A\times
B\bor 1
\]
The intended meaning of these types is as follows: base types are
things like the type of integers or the type of booleans. The type $A\to
B$ is the type of functions from $A$ to $B$. The type $A\times B$ is
the type of pairs $\pair{x,y}$, where $x$ has type $A$ and $y$ has
type $B$. The type $1$ is a one-element type. You can think of $1$ as
an abridged version of the booleans, in which there is only one
boolean instead of two. Or you can think of $1$ as the ``void'' or
``unit'' type in many programming languages: the result type of a
function that has no real result.

When we write types, we adopt the convention that $\times$ binds
stronger than $\to$, and $\to$ associates to the right. Thus, $A\times
B\to C$ is $(A\times B)\to C$, and $A\to B\to C$ is $A\to (B\to C)$.

The set of {\em raw typed lambda terms} is given by the following BNF:
\[ \mbox{Raw terms:}\ssep M,N \bnf x \bor MN \bor \lamabs{x}{A}.M
\bor \pair{M,N} \bor \proj1 M \bor \proj2 M \bor \unit \] Unlike what
we did in the untyped lambda calculus, we have added special syntax
here for pairs. Specifically, $\pair{M,N}$ is a pair of terms,
$\proj{i} M$ is a projection, with the intention that
$\proj{i}\pair{M_1,M_2}=M_i$. Also, we have added a term $\unit$,
which is the unique element of the type $1$. One other change from the
untyped lambda calculus is that we now write $\lamabs{x}{A}.M$ for a
lambda abstraction to indicate that $x$ has type $A$. However, we will
sometimes omit the superscripts and write $\lam x.M$ as before. The
notions of free and bound variables and $\alpha$-conversion are
defined as for the untyped lambda calculus; again we identify
$\alpha$-equivalent terms.

We call the above terms the {\em raw} terms, because we have not yet
imposed any typing discipline on these terms. To avoid meaningless
terms such as $\pair{M,N}(P)$ or $\proj1(\lam x.M)$, we introduce {\em
  typing rules}. 

We use the colon notation $M:A$ to mean ``$M$ is of type $A$''.
(Similar to the element notation in set theory).  The typing rules are
expressed in terms of {\em typing judgments}. A typing judgment is an
expression of the form
\[  \typ{x_1}{A_1},\typ{x_2}{A_2},\ldots,\typ{x_n}{A_n} \tj M:A.
\]
Its meaning is: ``under the assumption that $x_i$ is of type $A_i$,
for $i=1\ldots n$, the term $M$ is a well-typed term of type $A$.'' 
The free variables of $M$ must be contained in $x_1,\ldots,x_n$. The
idea is that in order to determine the type of $M$, we must make some
assumptions about the type of its free variables. For instance, the
term $xy$ will have type $B$ if $\typ{x}{A\to B}$ and $\typ{y}{A}$. 
Clearly, the type of $xy$ depends on the type of its free variables. 

A sequence of assumptions of the form
$\typ{x_1}{A_1},\ldots,\typ{x_n}{A_n}$, as in the left-hand-side of a
typing judgment, is called a {\em typing context}. We always assume
that no variable appears more than once in a typing context, and we
allow typing contexts to be re-ordered implicitly. We often use the
Greek letter $\Gamma$ to stand for an arbitrary typing context, and we
use the notations $\Gamma,\Gamma'$ and $\Gamma,\typ{x}{A}$ to denote
the concatenation of typing contexts, where it is always assumed that
the sets of variables are disjoint.

The symbol $\tj$, which appears in a typing judgment, is called the
{\em turnstile} symbol. Its purpose is to separate the left-hand side
from the right-hand side. 

The typing rules for the simply-typed lambda calculus are shown in
Table~\ref{tab-simple-typing-rules}.
\begin{table*}[tbp]
\[
\begin{array}{rc}
        \tjvar
&       \deriv{}{\Gamma,\typ{x}{ A}\tj x: A}
\nl     \tjapp
&       \deriv{\Gamma\tj M: A\to B\sep\Gamma\tj N: A}
                {\Gamma\tj MN: B}
\nl     \tjlam
&       \deriv{\Gamma,\typ{x}{ A}\tj M: B}
                {\Gamma\tj\lamabs{x}{A}.M: A\to B}
\nl     \tjpair
&       \deriv{\Gamma\tj M: A\sep\Gamma\tj N: B}
                {\Gamma\tj\pair{M,N}: A\times B}
\end{array}
\begin{array}{rc}
\nl     \tjproja
&       \deriv{\Gamma\tj M: A\times B}{\Gamma\tj\proj1 M: A}
\nl     \tjprojb
&       \deriv{\Gamma\tj M: A\times B}{\Gamma\tj\proj2 M: B}
\nl     \tjunit
&       \deriv{}{\Gamma\tj\unit:1}
\end{array}
\]
\caption{Typing rules for the simply-typed lambda calculus}
\label{tab-simple-typing-rules}
\end{table*}
The rule $\tjvar$ is a tautology: under the assumption that $x$ has
type $A$, $x$ has type $A$. The rule $\tjapp$ states that a function
of type $A\to B$ can be applied to an argument of type $A$ to produce
a result of type $B$. The rule $\tjlam$ states that if $M$ is a term
of type $B$ with a free variable $x$ of type $A$, then
$\lamabs{x}{A}.M$ is a function of type $A\to B$. The other rules
have similar interpretations.

Here is an example of a valid typing derivation:
{
\footnotesize
\[
\deriv{
  \deriv{
    }{
    \typ{x}{A\to A},\typ{y}{A}\tj x:A\to A
    }
  \sep
  \deriv{
    \deriv{
      }{
      \typ{x}{A\to A},\typ{y}{A}\tj x:A\to A
      }
    \sep
    \deriv{
      }{
      \typ{x}{A\to A},\typ{y}{A}\tj y:A
      }
    }{
    \typ{x}{A\to A},\typ{y}{A}\tj xy:A
    }
  }{
  \deriv{
    \typ{x}{A\to A},\typ{y}{A}\tj x(xy):A
    }{
    \deriv{
      \typ{x}{A\to A}\tj\lamabs{y}{A}.x(xy):A\to A
      }{
      \tj\lamabs{x}{A\to A}.\lamabs{y}{A}.x(xy):(A\to A)\to A\to A
      }
    }
  }
\]
}

One important property of these typing rules is that there is
precisely one rule for each kind of lambda term. Thus, when we
construct typing derivations in a bottom-up fashion, there is always a
unique choice of which rule to apply next. The only real choice we
have is about which types to assign to variables. 

\begin{exercise}
  Give a typing derivation of each of the following typing judgments:
  \begin{enumerate}
  \item[(a)] $\tj\lamabs{x}{(A\to A)\to B}.x(\lamabs{y}{A}.y):((A\to
    A)\to B)\to B$
  \item[(b)] $\tj\lamabs{x}{A\times B}.\pair{\proj2 x,\proj1
      x}:(A\times B)\to (B\times A)$
  \end{enumerate}
\end{exercise}

Not all terms are typable. For instance, the terms $\proj1(\lam x.M)$
and $\pair{M,N}(P)$ cannot be assigned a type, and neither can the
term $\lam x.xx$. Here, by ``assigning a type'' we mean, assigning
types to the free and bound variables such that the corresponding
typing judgment is derivable. We say that a term is typable if it can
be assigned a type.

\begin{exercise}
  Show that neither of the three terms mentioned in the previous paragraph
  is typable. 
\end{exercise}

\begin{exercise}
  We said that we will identify $\alpha$-equivalent terms. Show that
  this is actually necessary. In particular, show that if we didn't
  identify $\alpha$-equivalent terms, there would be no valid
  derivation of the typing judgment
  \[ \tj\lamabs{x}{A}.\lamabs{x}{B}.x:A\to B\to B. \]
  Give a derivation of this typing judgment using the bound variable
  convention.
\end{exercise}

\subsection{Connections to propositional logic}
\label{subsec-connprop}

Consider the following types:
\[ \begin{array}{ll}
  (1)& (A\times B)\to A \\
  (2)& A\to B\to (A\times B) \\
  (3)& (A\to B)\to(B\to C)\to(A\to C) \\
  (4)& A\to A\to A \\
  (5)& ((A\to A)\to B)\to B \\
  (6)& A\to (A\times B) \\
  (7)& (A\to C)\to C
\end{array}
\]
Let us ask, in each case, whether it is possible to find a closed term
of the given type. We find the following terms:
\[ \begin{array}{ll}
  (1)& \lamabs{x}{A\times B}.\proj1 x \\
  (2)& \lamabs{x}{A}.\lamabs{y}{B}.\pair{x,y} \\
  (3)& \lamabs{x}{A\to B}.\lamabs{y}{B\to C}.\lamabs{z}{A}.y(xz) \\
  (4)& \lamabs{x}{A}.\lamabs{y}{A}.x 
  \sep\mbox{and}\sep
  \lamabs{x}{A}.\lamabs{y}{A}.y \\
  (5)& \lamabs{x}{(A\to A)\to B}.x(\lamabs{y}{A}.y) \\
  (6)& \mbox{can't find a closed term} \\
  (7)& \mbox{can't find a closed term}
\end{array}
\]
Can we answer the general question, given a type, whether there exists
a closed term for it?

For a new way to look at the problem, take the types (1)--(7) and
make the following replacement of symbols: replace ``$\to$'' by
``$\imp$'' and replace ``$\times$'' by ``$\cand$''. We obtain the
following formulas:
\[ \begin{array}{ll}
  (1)& (A\cand B)\imp A \\
  (2)& A\imp B\imp (A\cand B) \\
  (3)& (A\imp B)\imp(B\imp C)\imp(A\imp C) \\
  (4)& A\imp A\imp A \\
  (5)& ((A\imp A)\imp B)\imp B \\
  (6)& A\imp (A\cand B) \\
  (7)& (A\imp C)\imp C
\end{array}
\]
Note that these are formulas of propositional logic, where ``$\imp$''
is implication, and ``$\cand$'' is conjunction (``and''). What can we
say about the validity of these formulas? It turns out that (1)--(5)
are tautologies, whereas (6)--(7) are not. Thus, the types for which we
could find a lambda term turn out to be the ones that are valid
when considered as formulas in propositional logic! This is not
entirely coincidental.

Let us consider, for example, how to prove $(A\cand B)\imp A$. The
proof is very short. It goes as follows: ``Assume $A\cand B$. Then, by
the first part of that assumption, $A$ holds. Thus $(A\cand B)\imp
A$.'' On the other hand, the lambda term of the corresponding type is
$\lamabs{x}{A\times B}.\proj1 x$. You can see that there is a close
connection between the proof and the lambda term. Namely, if one reads
$\lamabs{x}{A\times B}$ as ``assume $A\cand B$ (call the assumption
`$x$')'', and if one reads $\proj1 x$ as ``by the first part of
assumption $x$'', then this lambda term can be read as a proof of 
the proposition $(A\cand B)\imp A$.

This connection between simply-typed lambda calculus and propositional
logic is known as the ``Curry-Howard isomorphism''. Since types of the
lambda calculus correspond to formulas in propositional logic, and
terms correspond to proofs, the concept is also known as the
``proofs-as-programs'' paradigm, or the ``formulas-as-types''
correspondence. We will make the actual correspondence more precise in
the next two sections.

Before we go any further, we must make one important point. When we
are going to make precise the connection between simply-typed lambda
calculus and propositional logic, we will see that the appropriate
logic is {\em intuitionistic logic}, and not the ordinary {\em
  classical logic} that we are used to from mathematical practice. The
main difference between intuitionistic and classical logic is that the
former misses the principles of ``proof by contradiction'' and
``excluded middle''. The principle of proof by contradiction states
that if the assumption ``not $A$'' leads to a contradiction then we
have proved $A$. The principle of excluded middle states that either
``$A$'' or ``not $A$'' must be true. 

Intuitionistic logic is also known as {\em constructive logic},
because all proofs in it are by construction. Thus, in intuitionistic
logic, the only way to prove the existence of some object is by
actually constructing the object. This is in contrast with classical
logic, where we may prove the existence of an object simply by
deriving a contradiction from the assumption that the object doesn't
exist. The disadvantage of constructive logic is that it is generally
more difficult to prove things. The advantage is that once one has a
proof, the proof can be transformed into an algorithm. 

\subsection{Propositional intuitionistic logic}
\label{subsec-proplogic}

We start by introducing a system for intuitionistic logic that uses
only three connectives: ``$\cand$'', ``$\to$'', and ``$\ctrue$''.
{\em Formulas} $A,B\ldots$ are built from atomic formulas
$\alpha,\beta,\ldots$ via the BNF
\[ \mbox{Formulas:}\ssep A,B \bnf \alpha \bor A\to B\bor A\cand
B\bor\ctrue.
\]
We now need to formalize proofs. The formalized proofs will be called
``derivations''. The system we introduce here is known as {\em natural
  deduction}, and is due to Gentzen (1935).

In natural deduction, derivations are certain kinds of trees. In
general, we will deal with derivations of a formula $A$ from a
set of assumptions $\Gamma=\s{A_1,\ldots,A_n}$. Such a derivation will
be written schematically as
\[  \ideriv{\typ{x_1}{A_1},\ldots,\typ{x_n}{A_n}}{A}.
\]
We simplify the bookkeeping by giving a name to each assumption, and
we will use lower-case letters such as $x,y,z$ for such names.
In using the above notation for schematically writing a derivation of
$A$ from assumptions $\Gamma$, it is understood that the derivation
may in fact use a given assumption more than once, or zero times.
The rules for constructing derivations are as follows:

\begin{enumerate}
\item (Axiom) 
\[ \trule{ax}\ \deriv{\typ{x}{A}}{A} x
\]
is a derivation of $A$ from assumption $A$ (and possibly other
assumptions that were used zero times). We have written the letter
``$x$'' next to the rule, to indicate precisely which assumption we
have used here.
\item ($\cand$-introduction)
If 
\[ \ideriv{\Gamma}{A} \sep\mbox{and}\sep
   \ideriv{\Gamma}{B}
\]
are derivations of $A$ and $B$, respectively, then
\[
\trule{$\cand$-I}\ \deriv{\ideriv{\Gamma}{A}\sep\ideriv{\Gamma}{B}}{A\cand
  B}
\]
is a derivation of $A\cand B$. In other words, a proof of $A\cand B$
is a proof of $A$ and a proof of $B$.
\item ($\cand$-elimination)
If 
\[ \ideriv{\Gamma}{A\cand B}
\]
is a derivation of $A\cand B$, then
\[
\trule{$\cand$-E$_1$}\ \deriv{\ideriv{\Gamma}{A\cand
    B}}{A} 
\sep\mbox{and}\sep
\trule{$\cand$-E$_2$}\ \deriv{\ideriv{\Gamma}{A\cand
    B}}{B} 
\]
are derivations of $A$ and $B$, respectively. In other words, from
$A\cand B$, we are allowed to conclude both $A$ and $B$. 
\item ($\ctrue$-introduction)
\[ \trule{$\ctrue$-I}\ \deriv{\hspace{.5in}}{\ctrue}
\]
is a derivation of $\ctrue$ (possibly from some assumptions, which
were not used). In other words, $\ctrue$ is always true.
\item ($\to$-introduction)
If
\[ \ideriv{\Gamma,\typ{x}{A}}{B}
\]
is a derivation of $B$ from assumptions $\Gamma$ and $A$, then 
\[ \trule{$\to$-I}\ \deriv{\ideriv{\Gamma,[\typ{x}{A}]}{B}}{A\to B} x
\]
is a derivation of $A\to B$ from $\Gamma$ alone. Here, the assumption
$\typ{x}{A}$ is no longer an assumption of the new derivation --- we
say that it has been ``canceled''. We indicate canceled assumptions
by enclosing them in brackets $[\,]$, and we indicate the place where
the assumption was canceled by writing the letter $x$ next to the rule
where it was canceled.
\item ($\to$-elimination)
If 
\[  \ideriv{\Gamma}{A\to B} \sep\mbox{and}\sep
   \ideriv{\Gamma}{A}
\]
are derivations of $A\to B$ and $A$, respectively, then
\[ \trule{$\to$-E}\ \deriv{\ideriv{\Gamma}{A\to B} \sep
  \ideriv{\Gamma}{A}}{B}
\]
is a derivation of $B$. In other words, from $A\to B$ and $A$, we are
allowed to conclude $B$. This rule is sometimes called by its Latin
name, ``modus ponens''.
\suspendenumerate
\end{enumerate}

This finishes the definition of derivations in natural deduction. Note
that, with the exception of the axiom, each rule belongs to some
specific logical connective, and there are introduction and
elimination rules. ``$\cand$'' and ``$\to$'' have both introduction
and elimination rules, whereas ``$\ctrue$'' only has an introduction
rule.

In natural deduction, like in real mathematical life, assumptions can
be made at any time. The challenge is to get rid of assumptions once
they are made. In the end, we would like to have a derivation of a
given formula that depends on as few assumptions as possible --- in
fact, we don't regard the formula as proven unless we can derive it
from {\em no} assumptions. The rule {\trule{$\to$-I}} allows us to
discard temporary assumptions that we might have made during the
proof.

\begin{exercise}
  Give a derivation, in natural deduction, for each of the formulas
  (1)--(5) from Section~\ref{subsec-connprop}.
\end{exercise}

\subsection{An alternative presentation of natural deduction}

The above notation for natural deduction derivations suffers from a
problem of presentation: since assumptions are first written down,
later canceled dynamically, it is not easy to see when each assumption
in a finished derivation was canceled. 

The following alternate presentation of natural deduction works by
deriving entire {\em judgments}, rather than {\em formulas}. Rather
than keeping track of assumptions as the leaves of a proof tree, we
annotate each formula in a derivation with the entire set of
assumptions that were used in deriving it. In practice, this makes
derivations more verbose, by repeating most assumptions on each line.
In theory, however, such derivations are easier to reason about.

A {\em judgment} is a statement of the form
$\typ{x_1}{A_1},\ldots,\typ{x_n}{A_n} \tj B$. It states that the
formula $B$ is a consequence of the (labeled) assumptions
$A_1,\ldots,A_n$. The rules of natural deduction can now be
reformulated as rules for deriving judgments:

\begin{enumerate}
\item (Axiom)
\[ \trule{ax$_x$}\ \deriv{}{\Gamma,\typ{x}{A}\tj A}
\]
\item ($\cand$-introduction)
\[ \trule{$\cand$-I}\ \deriv{\Gamma\tj A\sep\Gamma\tj B}{\Gamma\tj A\cand B}
\]
\item ($\cand$-elimination)
\[ \trule{$\cand$-E$_1$}\ \deriv{\Gamma\tj A\cand B}{\Gamma\tj A}
\sep
\trule{$\cand$-E$_2$}\ \deriv{\Gamma\tj A\cand B}{\Gamma\tj B}
\]
\item ($\ctrue$-introduction)
\[ \trule{$\ctrue$-I}\ \deriv{}{\Gamma\tj\ctrue}
\]
\item ($\to$-introduction)
\[ \trule{$\to$-I$_x$}\ \deriv{\Gamma,\typ{x}{A}\tj B}{\Gamma\tj A\to B}
\]
\item ($\to$-elimination)
\[ \trule{$\to$-E}\ \deriv{\Gamma\tj A\to B\sep \Gamma\tj A}{\Gamma\tj B}
\]
\suspendenumerate
\end{enumerate}

\subsection{The Curry-Howard Isomorphism}

There is an obvious one-to-one correspondence between types of the
simply-typed lambda calculus and the formulas of propositional
intuitionistic logic introduced in Section~\ref{subsec-proplogic}
(provided that the set of basic types can be identified with the set
of atomic formulas). We will identify formulas and types from now on,
where it is convenient to do so.

Perhaps less obvious is the fact that derivations are in one-to-one
correspondence with simply-typed lambda terms. To be precise, we
will give a translation from derivations to lambda terms, and a
translation from lambda terms to derivations, which are mutually
inverse up to $\alpha$-equivalence. 

To any derivation of $\typ{x_1}{A_1},\ldots,\typ{x_n}{A_n}\tj B$, we
will associate a lambda term $M$ such that
$\typ{x_1}{A_1},\ldots,\typ{x_n}{A_n}\tj M:B$ is a valid typing
judgment. We define $M$ by recursion on the definition of derivations.
We prove simultaneously, by induction, that
$\typ{x_1}{A_1},\ldots,\typ{x_n}{A_n}\tj M:B$ is indeed a valid typing
judgment.

\begin{enumerate}
\item (Axiom) If the derivation is
\[ \trule{ax$_x$}\ \deriv{}{\Gamma,\typ{x}{A}\tj A},
\]
then the lambda term is $M=x$. Clearly, $\Gamma,\typ{x}{A}\tj x:A$ is
a valid typing judgment by $\tjvar$.
\item ($\cand$-introduction)
If the derivation is
\[ \trule{$\cand$-I}\ \deriv{\ideriv{}{\Gamma\tj A}\sep\ideriv{}{\Gamma\tj B}}{\Gamma\tj A\cand B},
\]
then the lambda term is $M=\pair{P,Q}$, where $P$ and $Q$ are the
terms associated to the two respective subderivations. By induction
hypothesis, $\Gamma\tj P:A$ and $\Gamma\tj Q:B$, thus
$\Gamma\tj\pair{P,Q}:A\times B$ by $\tjpair$.
\item ($\cand$-elimination)
If the derivation is
\[ \trule{$\cand$-E$_1$}\ \deriv{\ideriv{}{\Gamma\tj A\cand B}}{\Gamma\tj A},
\]
then we let $M=\proj1 P$, where $P$ is the term associated to the
subderivation. By induction hypothesis, $\Gamma\tj P:A\times B$, thus
$\Gamma\tj\proj1 P:A$ by $\tjproja$. The case of
{\trule{$\cand$-E$_2$}} is entirely symmetric.
\item ($\ctrue$-introduction)
If the derivation is
\[ \trule{$\ctrue$-I}\ \deriv{}{\Gamma\tj\ctrue},
\]
then let $M=\unit$. We have $\tj\unit:1$ by $\tjunit$.
\item ($\to$-introduction)
If the derivation is
\[ \trule{$\to$-I$_x$}\ \deriv{\ideriv{}{\Gamma,\typ{x}{A}\tj B}}{\Gamma\tj A\to B},
\]
then we let $M=\lamabs{x}{A}.P$, where $P$ is the term associated to
the subderivation. By induction hypothesis, $\Gamma,\typ{x}{A}\tj
P:B$, hence $\Gamma\tj \lamabs{x}{A}.P:A\to B$ by $\tjlam$.
\item ($\to$-elimination)
Finally, if the derivation is
\[ \trule{$\to$-E}\ \deriv{\ideriv{}{\Gamma\tj A\to B}\sep \ideriv{}{\Gamma\tj A}}{\Gamma\tj B},
\]
then we let $M=PQ$, where $P$ and $Q$ are the terms associated to the
two respective subderivations. By induction hypothesis, $\Gamma\tj
P:A\to B$ and $\Gamma\tj Q:A$, thus $\Gamma\tj PQ: B$ by $\tjapp$.
\end{enumerate}

Conversely, given a well-typed lambda term $M$, with associated typing
judgment $\Gamma\tj M:A$, then we can construct a derivation of $A$
from assumptions $\Gamma$. We define this derivation by recursion on
the type derivation of $\Gamma\tj M:A$. The details are too tedious to
spell them out here; we simply go through each of the rules {\tjvar},
{\tjlam}, {\tjapp}, {\tjpair}, {\tjproja}, {\tjprojb}, {\tjunit} and
apply the corresponding rule {\trule{ax}}, {\trule{$\to$-I}},
{\trule{$\to$-E}}, {\trule{$\cand$-I}}, {\trule{$\cand$-E$_1$}},
{\trule{$\cand$-E$_2$}}, {\trule{$\ctrue$-I}}, respectively.

\subsection{Reductions in the simply-typed lambda calculus}
\label{subsec-simply-reductions}

$\beta$- and $\eta$-reductions in the simply-typed lambda calculus are
defined much in the same way as for the untyped lambda calculus,
except that we have introduced some additional terms (such as pairs
and projections), which calls for some additional reduction rules. 
We define the following reductions:
\[ \begin{array}{lllll}\label{page-typed-reductions}
  (\beta_{\to}) & (\lamabs{x}{A}.M)N & \to & \subst{M}{N}{x}, \\
  (\eta_{\to}) &  \lamabs{x}{A}.Mx & \to & M, & \mbox{where $x\not\in\FV{M}$},\\
  (\beta_{\times,1}) & \proj1\pair{M,N} & \to & M,\\
  (\beta_{\times,2}) & \proj2\pair{M,N} & \to & N,\\
  (\eta_{\times}) & \pair{\proj1 M,\proj2 M} & \to & M,\\
  (\eta_{1}) & M & \to & \unit, & \mbox{if $M:1$}.
\end{array}
\]
Then single- and multi-step $\beta$- and $\eta$-reduction are defined
as the usual contextual closure of the above rules, and the
definitions of $\beta$- and $\eta$-equivalence also follow the usual
pattern. In addition to the usual {\trule{cong}} and {\nrule{\xi}}
rules, we now also have congruence rules that apply to pairs and
projections. 

We remark that, to be perfectly precise, we should have defined
reductions between typing judgments, and not between terms. This is
necessary because some of the reduction rules, notably $(\eta_{1})$,
depend on the type of the terms involved. However, this would be
notationally very cumbersome, and we will blur the distinction,
pretending at times that terms appear in some implicit typing context
that we do not write. 

An important property of the reduction is the ``subject reduction''
property, which states that well-typed terms reduce only to well-typed
terms of the same type. This has an immediate application to
programming: subject reduction guarantees that if we write a program
of type ``integer'', then the final result of evaluating the program,
if any, will indeed be an integer, and not, say, a boolean. 

\begin{theorem}[Subject Reduction]
  If $\Gamma\tj M:A$ and $M\redbe M'$, then $\Gamma\tj M':A$. 
\end{theorem}

\begin{proof}
  By induction on the derivation of $M\redbe M'$, and by case
  distinction on the last rule used in the derivation of $\Gamma\tj
  M:A$. For instance, if $M\redbe M'$ by $(\beta_{\to})$, then
  $M=(\lamabs{x}{B}.P)Q$ and $M'=\subst{P}{Q}{x}$. If $\Gamma\tj M:A$,
  then we must have $\Gamma,\typ{x}{B}\tj P:A$ and $\Gamma\tj Q:B$.
  It follows that $\Gamma\tj\subst{P}{Q}{x}:A$; the latter statement
  can be proved separately (as a ``substitution lemma'') by induction
  on $P$ and makes crucial use of the fact that $x$ and $Q$ have the
  same type. 
  
  The other cases are similar, and we leave them as an exercise. Note
  that, in particular, one needs to consider the {\trule{cong}},
  {\nrule{\xi}}, and other congruence rules as well. \eot
\end{proof}

\subsection{A word on Church-Rosser}

One important theorem that does {\em not} hold for
$\beta\eta$-reduction in the simply-typed
$\lam^{\to,\times,1}$-calculus is the Church-Rosser theorem. The
culprit is the rule $(\eta_{1})$. For instance, if $x$ is a variable
of type $A\times 1$, then the term $M=\pair{\proj1 x,\proj2 x}$
reduces to $x$ by $(\eta_{\times})$, but also to $\pair{\proj1
  x,\unit}$ by $(\eta_{1})$. Both these terms are normal forms.
Thus, the Church-Rosser property fails. 
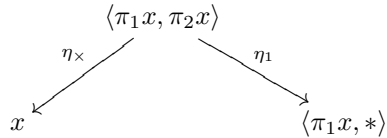
\[ \xymatrix{&\pair{\proj1 x,\proj2 x}\ar[dl]_{\eta_{\times}}\ar[dr]^{\eta_{1}}
  \\x&&\pair{\proj1 x,\unit}}
\]

There are several ways around this problem.  For instance, if we omit
all the $\eta$-reductions and consider only $\beta$-reductions, then
the Church-Rosser property does hold. Eliminating $\eta$-reductions
does not have much of an effect on the lambda calculus from a
computational point of view; already in the untyped lambda calculus,
we noticed that all interesting calculations could in fact be carried
out with $\beta$-reductions alone. We can say that $\beta$-reductions
are the engine for computation, whereas $\eta$-reductions only serve
to clean up the result. In particular, it can never happen that some
$\eta$-reduction inhibits another $\beta$-reduction: if $M\rede M'$,
and if $M'$ has a $\beta$-redex, then it must be the case that $M$
already has a corresponding $\beta$-redex. Also, $\eta$-reductions
always reduce the size of a term. It follows that if $M$ is a
$\beta$-normal form, then $M$ can always be reduced to a
$\beta\eta$-normal form (not necessarily unique) in a finite sequence
of $\eta$-reductions.

\begin{exercise}
  Prove the Church-Rosser theorem for $\beta$-reductions in the
  $\lam^{\to,\times,1}$-calculus. Hint: use the same method that we
  used in the untyped case.
\end{exercise}

Another solution is to omit the type $1$ and the term $\unit$ from the
language. In this case, the Church-Rosser property holds even for
$\beta\eta$-reduction.

\begin{exercise}
  Prove the Church-Rosser theorem for $\beta\eta$-reduction in the
  $\lam^{\to,\times}$-calculus, i.e., the simply-typed lambda calculus
  without $1$ and $\unit$.
\end{exercise}

\subsection{Reduction as proof simplification}

Having made a one-to-one correspondence between simply-typed lambda
terms and derivations in intuitionistic natural deduction, we may now
ask what $\beta$- and $\eta$-reductions correspond to under this
correspondence. It turns out that these reductions can be thought of
as ``proof simplification steps''. 

Consider for example the $\beta$-reduction $\proj1\pair{M,N} \to M$.
If we translate the left-hand side and the right-hand side via the
Curry-Howard isomorphism (here we use the first notation for natural
deduction), we get
\[ \trule{$\cand$-E$_1$}\deriv{\trule{$\cand$-I}\deriv{\ideriv{\Gamma}{A}\sep\ideriv{\Gamma}{B}}{A\cand B}}{A}
\ssep\to\ssep \ideriv{\Gamma}{A}.
\]
We can see that the left derivation contains an introduction rule
immediately followed by an elimination rule. This leads to an obvious
simplification if we replace the left derivation by the right one.

In general, $\beta$-redexes correspond to situations where an
introduction rule is immediately followed by an elimination rule, and
$\eta$-redexes correspond to situations where an elimination rule is
immediately followed by an introduction rule. For example, consider
the $\eta$-reduction $\pair{\proj1 M,\proj2 M} \to M$. This translates
to:
\[
\trule{$\cand$-I}\deriv{\trule{$\cand$-E$_1$}\deriv{\ideriv{\Gamma}{A\cand
      B}}{A}\sep\trule{$\cand$-E$_2$}\deriv{\ideriv{\Gamma}{A\cand
      B}}{B}}{A\cand B}
\ssep\to\ssep \ideriv{\Gamma}{A\cand B}
\]
Again, this is an obvious simplification step, but it has a side
condition: the left and right subderivation must be the same! This
side condition corresponds to the fact that in the redex $\pair{\proj1
  M,\proj2 M}$, the two subterms called $M$ must be equal. It is
another characteristic of $\eta$-reductions that they often carry such
side conditions.

The reduction $M\to\unit$ translates as follows:
\[
\ideriv{\Gamma}{\ctrue} \ssep\to\ssep \trule{$\ctrue$-I}\deriv{\hspace{.5in}}{\ctrue}
\]
In other words, any derivation of $\ctrue$ can be replaced by the
canonical such derivation.

More interesting is the case of the $(\beta_{\to})$ rule. Here, we
have $(\lamabs{x}{A}.M)N \to \subst{M}{N}{x}$, which can be translated
via the Curry-Howard Isomorphism as follows:
\[
\trule{$\to$-E}\ \deriv{\trule{$\to$-I}\ 
  \deriv{\ideriv{\Gamma,[\typ{x}{A}]}{B}}{A\to B} x
  \sep\ideriv{\Gamma}{A}}{B}
\ssep\to \ssep 
\ideriv{\Gamma,\ideriv{\Gamma}{A}}{B}.
\]
What is going on here is that we have a derivation $M$ of $B$ from
assumptions $\Gamma$ and $A$, and we have another derivation $N$ of
$A$ from $\Gamma$. We can directly obtain a derivation of $B$ from
$\Gamma$ by stacking the second derivation on top of the first!

Notice that this last proof ``simplification'' step may not actually
be a simplification. Namely, if the hypothesis labeled $x$ is used
many times in the derivation $M$, then $N$ will have to be copied many
times in the right-hand side term. This corresponds to the fact that
if $x$ occurs several times in $M$, then $\subst{M}{N}{x}$ might be a
longer and more complicated term than $(\lam x.M)N$.

Finally, consider the $(\eta_{\to})$ rule $\lamabs{x}{A}.Mx \to M$,
where $x\not\in\FV{M}$. This translates to derivations as follows:
\[
\trule{$\to$-I}\ \deriv{\trule{$\to$-E}\ \deriv{\ideriv{\Gamma}{A\to
      B}\sep\trule{ax}\ \deriv{[\typ{x}{A}]}{A} x}{B}}{A\to B} x
\ssep\to\ssep \ideriv{\Gamma}{A\to B}
\]

\subsection{Getting mileage out of the Curry-Howard isomorphism}

The Curry-Howard isomorphism makes a connection between the lambda
calculus and logic. We can think of it as a connection between
``programs'' and ``proofs''. What is such a connection good for? Like
any isomorphism, it allows us to switch back and forth and think in
whichever system suits our intuition in a given situation. Moreover,
we can save a lot of work by transferring theorems that were proved
about the lambda calculus to logic, and vice versa. As an example, we
will see in the next section how to add disjunctions to propositional
intuitionistic logic, and then we will explore what we can learn about
the lambda calculus from that.

\subsection{Disjunction and sum types}

To the BNF for formulas of propositional intuitionistic logic from
Section~\ref{subsec-proplogic}, we add the following clauses:
\[ \mbox{Formulas:}\ssep A,B \bnf \ldots \bor A\cor B \bor \cfalse.
\]
Here, $A\cor B$ stands for disjunction, or ``or'', and $\cfalse$
stands for falsity, which we can also think of as zero-ary
disjunction.  The symbol $\cfalse$ is also known by the names of
``bottom'', ``absurdity'', or ``contradiction''.  The rules for
constructing derivations are extended by the following cases:
\begin{enumerate}
\resumeenumerate
\item ($\cor$-introduction)
\[ \trule{$\cor$-I$_1$}\ \deriv{\Gamma\tj A}{\Gamma\tj A\cor B}
\sep
\trule{$\cor$-I$_2$}\ \deriv{\Gamma\tj B}{\Gamma\tj A\cor B}
\]
In other words, if we have proven $A$ or we have proven $B$, then we
may conclude $A\cor B$.
\item ($\cor$-elimination)
\[ \trule{$\cor$-E$_{x,y}$}\ \deriv{\Gamma\tj A\cor B
  \sep \Gamma,\typ{x}{A}\tj C
  \sep \Gamma,\typ{y}{B}\tj C}{\Gamma\tj C}
\]
This is known as the ``principle of case distinction''. If we know
$A\cor B$, and we wish to prove some formula $C$, then we may proceed
by cases. In the first case, we assume $A$ holds and prove $C$. In the
second case, we assume $B$ holds and prove $C$. In either case, we
prove $C$, which therefore holds independently. 

Note that the $\cor$-elimination rule differs from all other rules we
have considered so far, because it involves some arbitrary formula $C$
that is not directly related to the principal formula $A\cor B$ being
eliminated.
\item ($\cfalse$-elimination)
\[ \trule{$\cfalse$-E}\ \deriv{\Gamma\tj\cfalse}{\Gamma\tj C},
\]
for an arbitrary formula $C$. This rule formalizes the familiar
principle ``ex falsum quodlibet'', which means that falsity implies
anything.
\end{enumerate}

There is no $\cfalse$-introduction rule. This is symmetric to the fact
that there is no $\ctrue$-elimination rule.

Having extended our logic with disjunctions, we can now ask what these
disjunctions correspond to under the Curry-Howard isomorphism.
Naturally, we need to extend the lambda calculus by as many new terms
as we have new rules in the logic. It turns out that disjunctions
correspond to a concept that is quite natural in programming: ``sum'' or
``union'' types. 

To the lambda calculus, add type constructors $A+B$ and $0$.
\[ \mbox{Simple types:}\ssep A,B \bnf \ldots\bor A+B\bor 0.
\]
Intuitively, $A+B$ is the disjoint union of $A$ and $B$, as in set
theory: an element of $A+B$ is either an element of $A$ or an element
of $B$, together with an indication of which one is the case. In
particular, if we consider an element of $A+A$, we can still tell
whether it is in the left or right component, even though the two
types are the same. In programming languages, this is sometimes known
as a ``union'' or ``variant'' type. We call it a ``sum'' type here.
The type $0$ is simply the empty type, corresponding to the empty set
in set theory.

What should the lambda terms be that go with these new types?  We know
from our experience with the Curry-Howard isomorphism that we have to
have precisely one term constructor for each introduction or
elimination rule of natural deduction. Moreover, we know that if such
a rule has $n$ subderivations, then our term constructor has to have
$n$ immediate subterms. We also know something about bound variables:
Each time a hypothesis is canceled in a natural deduction rule, there
must be a binder of the corresponding variable in the lambda calculus.
This information more or less uniquely determines what the lambda
terms should be; the only choice that is left is what to call them!

We add four terms to the lambda calculus:
\[ \begin{array}{lll}
  \mbox{Raw terms:}\ssep M,N,P &\bnf&\ldots\bor \inj1 M\bor \inj2 M \\
  && \bor
  \caseof{M}{x}{A}{N}{y}{B}{P} \bor \Box_A M
\end{array}
\]
The typing rules for these new terms are shown in
Table~\ref{tab-sum-typing-rules}.
\begin{table*}[tbp]
\[ \begin{array}{rc}
        \tjinja
&       \deriv{\Gamma\tj M:A}{\Gamma\tj \inj1 M:A+B}
\nl     \tjinjb
&       \deriv{\Gamma\tj M:B}{\Gamma\tj \inj2 M:A+B}
\nl     \tjcase
&       \deriv{\Gamma\tj M: A+B\sep\Gamma,\typ{x}{A}\tj N: C
  \sep\Gamma,\typ{y}{B}\tj P: C}
                {\Gamma\tj (\caseof{M}{x}{A}{N}{y}{B}{P}): C}
\nl     \tjbox
&       \deriv{\Gamma\tj M: 0}
                {\Gamma\tj\Box_A M:A}
\end{array}
\]
\caption{Typing rules for sums}
\label{tab-sum-typing-rules}
\end{table*}
By comparing these rules to \trule{$\cor$-I$_1$},
\trule{$\cor$-I$_2$}, \trule{$\cor$-E}, and \trule{$\cfalse$-E}, you
can see that they are precisely analogous.

But what is the meaning of these new terms? The term $\inj1 M$ is
simply an element of the left component of $A+B$. We can think of
$\inj1$ as the injection function $A\to A+B$. Similar for $\inj2$.
The term $(\caseof{M}{x}{A}{N}{y}{B}{P})$ is a case distinction:
evaluate $M$ of type $A+B$. The answer is either an element of the
left component $A$ or of the right component $B$. In the first case,
assign the answer to the variable $x$ and evaluate $N$. In the second
case, assign the answer to the variable $y$ and evaluate $P$. Since
both $N$ and $P$ are of type $C$, we get a final result of type $C$.
Note that the case statement is very similar to an if-then-else; the
only difference is that the two alternatives also carry a value.
Indeed, the booleans can be defined as $1+1$, in which case
$\truet=\inj1\unit$, $\falset=\inj2\unit$, and $\ifthenelset
MNP=\caseof{M}{x}{1}{N}{y}{1}{P}$, where $x$ and $y$ don't occur in
$N$ and $P$, respectively.

Finally, the term $\Box_A M$ is a simple type cast, corresponding to
the unique function $\Box_A:0\to A$ from the empty set to any set $A$.

\subsection{Classical logic vs.\ intuitionistic logic}

We have mentioned before that the natural deduction calculus we have
presented corresponds to intuitionistic logic, and not classical
logic. But what exactly is the difference? Well, the difference is
that in intuitionistic logic, we have no rule for proof by
contradiction, and we do not have $A\cor\neg A$ as an axiom. 

Let us adopt the following convention for negation: the formula $\neg
A$ (``not $A$'') is regarded as an abbreviation for $A\to\cfalse$.
This way, we do not have to introduce special formulas and rules for
negation; we simply use the existing rules for $\to$ and $\cfalse$.

In intuitionistic logic, there is no derivation of $A\cor\neg A$, for
general $A$. Or equivalently, in the simply-typed lambda calculus,
there is no closed term of type $A+(A\to 0)$. We are not yet in a
position to prove this formally, but informally, the argument goes as
follows: If the type $A$ is empty, then there can be no closed term of
type $A$ (otherwise $A$ would have that term as an element). On the
other hand, if the type $A$ is non-empty, then there can be no closed
term of type $A\to 0$ (or otherwise, if we applied that term to some
element of $A$, we would obtain an element of $0$). But if we were to
write a {\em generic} term of type $A+(A\to 0)$, then this term would
have to work no matter what $A$ is. Thus, the term would have to
decide whether to use the left or right component independently of
$A$. But for any such term, we can get a contradiction by choosing $A$
either empty or non-empty.

Closely related is the fact that in intuitionistic logic, we do not
have a principle of proof by contradiction. The ``proof by
contradiction'' rule is the following:
\[ \trule{contra$_x$}\ \deriv{\Gamma,\typ{x}{\neg A}\tj\cfalse}{\Gamma\tj A}.
\]
This is {\em not} a rule of intuitionistic
propositional logic, but we can explore what would happen if we were
to add such a rule. First, we observe that the contradiction rule is
very similar to the following:
\[ \deriv{\Gamma,\typ{x}{A}\tj\cfalse}{\Gamma\tj\neg A}.
\]
However, since we defined $\neg A$ to be the same as $A\to\cfalse$,
the latter rule is an instance of \trule{$\to$-I}. The contradiction
rule, on the other hand, is not an instance of \trule{$\to$-I}.

If we admit the rule \trule{contra}, then $A\cor\neg A$ can be
derived. The following is such a derivation:
{\footnotesize
\[ \trule{$\to$-E}\ \deriv{\raisebox{-6ex}{$\squishr{\trule{ax$_y$}~}\deriv{}{\typ{y}{\neg(A\cor\neg A)}\tj\neg(A\cor\neg A)}$}\hspace{-24ex}
  \squishr{\trule{$\to$-E}~}\deriv{\squishr{\trule{ax$_y$}~}\deriv{}{\typ{y}{\neg(A\cor\neg A)},\typ{x}{A}\tj\neg(A\cor\neg A)}
    \sep \trule{$\cor$-I$_1$~}\deriv{\squishr{\trule{ax$_x$}~}\deriv{}{\typ{y}{\neg(A\cor\neg A)},\typ{x}{A}\tj A}}{\typ{y}{\neg(A\cor\neg A)},\typ{x}{A}\tj A\cor\neg A}}{
   \trule{$\cor$-I$_2$~}\deriv{ 
  \squishr{\trule{$\to$-I$_x$}~}\deriv{\typ{y}{\neg(A\cor\neg A)},\typ{x}{A}\tj \cfalse}{\typ{y}{\neg(A\cor\neg A)}\tj \neg A}
  }{\typ{y}{\neg(A\cor\neg A)}\tj A\cor \neg A}\hspace{-18ex}
  }}{\trule{contra$_y$~}\deriv{\typ{y}{\neg(A\cor\neg A)}\tj\cfalse}{\tj A\cor\neg A}}
\]}
Conversely, if we added $A\cor\neg A$ as an axiom to intuitionistic
logic, then this already implies the \trule{contra} rule. Namely, from
any derivation of $\Gamma,\typ{x}{\neg A}\tj\cfalse$, we can obtain a
derivation of $\Gamma\tj A$ by using $A\cor\neg A$ as an axiom. Thus,
we can {\em simulate} the \trule{contra} rule, in the presence of
$A\cor\neg A$.
\[ \trule{$\cor$-E$_{x,y}$}\ \deriv{\deriv{\trule{excluded middle}}{\Gamma\tj A\cor \neg A} \sep
 \trule{$\cfalse$-E}\ \deriv{ \Gamma,\typ{x}{\neg A}\tj\cfalse}{\Gamma,\typ{x}{\neg A}\tj A}
\sep \trule{ax$_y$}\ \deriv{}{\Gamma,\typ{y}{A}\tj A}}{\Gamma\tj A}
\]
In this sense, we can say that the rule $\trule{contra}$ and the axiom
$A\cor\neg A$ are equivalent, in the presence of the other axioms and
rules of intuitionistic logic.

It turns out that the system of intuitionistic logic plus
\trule{contra} is equivalent to classical logic as we know it. It is
in this sense that we can say that intuitionistic logic is ``classical
logic without proofs by contradiction''.

\begin{exercise}
  The formula $((A\to B)\to A)\to A$ is called ``Peirce's law''. It is
  valid in classical logic, but not in intuitionistic logic. Give a
  proof of Peirce's law in natural deduction, using the rule
  \trule{contra}.
\end{exercise}

Conversely, Peirce's law, when added to intuitionistic logic for all
$A$ and $B$, implies \trule{contra}. Here is the proof. Recall that
$\neg A$ is an abbreviation for $A\to\cfalse$.
\[ \trule{$\to$-E}~
   \deriv{
   \deriv{\trule{Peirce's law for $B=\cfalse$}}
         {\Gamma\tj((A\to\cfalse)\to A)\to A}
   \sep
   \trule{$\to$-I$_x$}~
   \deriv{
   \squishr{\trule{$\cfalse$-E}~}
   \deriv{\Gamma,\typ{x}{A\to\cfalse}\tj\cfalse}
         {\Gamma,\typ{x}{A\to\cfalse}\tj A}
   }
   {\Gamma\tj(A\to\cfalse)\to A}
   }
   {\Gamma\tj A}
\]

We summarize the results of
this section in terms of a slogan:
\[ \begin{array}{ll}
  &  \mbox{intuitionistic logic + \trule{contra}} \\
  =& \mbox{intuitionistic logic + ``$A\cor\neg A$''} \\
  =& \mbox{intuitionistic logic + Peirce's law} \\
  =& \mbox{classical logic.}
\end{array}
\]

The proof theory of intuitionistic logic is a very interesting subject
in its own right, and an entire course could be taught just on that
subject.

\subsection{Classical logic and the Curry-Howard isomorphism}

To extend the Curry-Howard isomorphism to classical logic, according
to the observations of the previous section, it is sufficient to add
to the lambda calculus a term representing Peirce's law. All we have
to do is to add a term $\fC:((A\to B)\to A)\to A$, for all types $A$
and $B$. 

Such a term is known as {\em Felleisen's $\fC$}, and it has a specific
interpretation in terms of programming languages. It can be understood
as a control operator (similar to ``goto'', ``break'', or exception
handling in some procedural programming languages).

Specifically, Felleisen's interpretation requires a term of the form
\[ M = \fC(\lam k^{A\to B}.N) : A
\]
to be evaluated as follows. To evaluate $M$, first evaluate $N$.  Note
that both $M$ and $N$ have type $A$. If $N$ returns a result, then
this immediately becomes the result of $M$ as well. On the other hand,
if during the evaluation of $N$, the function $k$ is ever called with
some argument $x:A$, then the further evaluation of $N$ is aborted,
and $x$ immediately becomes the result of $M$.

In other words, the final result of $M$ can be calculated anywhere
inside $N$, no matter how deeply nested, by passing it to $k$ as an
argument. The function $k$ is known as a {\em continuation}.

There is a lot more to programming with continuations than can be
explained in these lecture notes. For an interesting application of
continuations to compiling, see e.g. {\cite{App92}} from the
bibliography (Section~\ref{ssec-bibliography}). The above explanation
of what it means to ``evaluate'' the term $M$ glosses over several
details.  In particular, we have not given a reduction rule for $\fC$
in the style of $\beta$-reduction. To do so is rather complicated and
is beyond the scope of these notes.

\section{Weak and strong normalization}

\subsection{Definitions}

As we have seen, computing with lambda terms means reducing lambda
terms to normal form. By the Church-Rosser theorem, such a normal form
is guaranteed to be unique if it exists. But so far, we have paid
little attention to the question whether normal forms exist for a
given term, and if so, how we need to reduce the term to find a normal
form. 

\begin{definition}
  Given a notion of term and a reduction relation, we say that a term
  $M$ is {\em weakly normalizing} if there exists a finite sequence of
  reductions $M\red M_1\red\ldots\red M_n$ such that $M_n$ is a normal
  form. We say that $M$ is {\em strongly normalizing} if there does
  not exist an infinite sequence of reductions starting from $M$, or
  in other words, if {\em every} sequence of reductions starting from
  $M$ is finite. 
\end{definition}

Recall the following consequence of the Church-Rosser theorem, which
we stated as Corollary~\ref{cor-cr-3}: If $M$ has a normal form $N$,
then $M\reds N$. It follows that a term $M$ is weakly normalizing if
and only if it has a normal form. This does not imply that every
possible way of reducing $M$ leads to a normal form. A term is
strongly normalizing if and only if every way of reducing it leads to
a normal form in finitely many steps. 

Consider for example the following terms in the untyped lambda
calculus:
\begin{enumerate}
\item The term $\Omega=(\lam x.xx)(\lam x.xx)$ is neither weakly nor
  strongly normalizing. It does not have a normal form.
\item The term $(\lam x.y)\Omega$ is weakly normalizing, but not
  strongly normalizing. It reduces to the normal form $y$, but it also
  has an infinite reduction sequence.
\item The term $(\lam x.y)((\lam x.x)(\lam x.x))$ is strongly
  normalizing. While there are several different ways to reduce this
  term, they all lead to a normal form in finitely many steps. 
\item The term $\lam x.x$ is strongly normalizing, since it has no
  reductions, much less an infinite reduction sequence. More
  generally, every normal form is strongly normalizing.
\end{enumerate}

We see immediately that strongly normalizing implies weakly
normalizing. However, as the above examples show, the converse is not
true. 

\subsection{Weak and strong normalization in typed lambda 
  calculus}

We found that the term $\Omega=(\lam x.xx)(\lam x.xx)$ is not weakly
or strongly normalizing. On the other hand, we also know that this
term is not typable in the simply-typed lambda calculus. This is not
a coincidence, as the following theorem shows.

\begin{theorem}[Weak normalization theorem]\label{thm-weak-norm}
  In the simply-typed lambda calculus, all terms are weakly
  normalizing.
\end{theorem}

\begin{theorem}[Strong normalization theorem]\label{thm-strong-norm}
  In the simply-typed lambda calculus, all terms are strongly
  normalizing.
\end{theorem}

Clearly, the strong normalization theorem implies the weak
normalization theorem. However, the weak normalization theorem is much
easier to prove, which is the reason we proved both these theorems in
class. In particular, the proof of the weak normalization theorem
gives an explicit measure of the complexity of a term, in terms of the
number of redexes of a certain degree in the term. There is no
corresponding complexity measure in the proof of the strong
normalization theorem.

Please refer to Chapters 4 and 6 of ``Proofs and Types'' by
Girard, Lafont, and Taylor {\cite{GLT89}} for the proofs of
Theorems~\ref{thm-weak-norm} and {\ref{thm-strong-norm}},
respectively. 

\section{Polymorphism}

The polymorphic lambda calculus, also known as ``System F'', is
obtained extending the Curry-Howard isomorphism to the quantifier
$\forall$. For example, consider the identity function $\lambda
x^A.x$.  This function has type $A\to A$. Another identity function is
$\lambda x^B.x$ of type $B\to B$, and so forth for every type. We can
thus think of the identity function as a family of functions, one for
each type. In the polymorphic lambda calculus, there is a dedicated
syntax for such families, and we write $\Lambda\alpha.\lambda
x^\alpha.x$ of type $\forall\alpha.\alpha\to\alpha$.

System F was independently discovered by Jean-Yves Girard and John
Reynolds in the early 1970's.

\subsection{Syntax of System F}

The primary difference between System F and simply-typed lambda
calculus is that System F has a new kind of function that takes a {\em
  type}, rather than a {\em term}, as its argument. We can also think
of such a function as a family of terms that is indexed by a type.

Let $\alpha,\beta,\gamma$ range over a countable set of {\em type
  variables}. The types of System F are given by the grammar
\[ \mbox{Types:}\ssep A,B \bnf \alpha \bor A\to B\bor \forall\alpha.A
\]
A type of the form $A\to B$ is called a {\em function type}, and a
type of the form $\forall\alpha.A$ is called a {\em universal type}.
The type variable $\alpha$ is bound in $\forall\alpha.A$, and we
identify types up to renaming of bound variables; thus,
$\forall\alpha.\alpha\to\alpha$ and $\forall\beta.\beta\to\beta$ are
the same type. We write $\FTV{A}$ for the set of free type variables
of a type $A$, defined inductively by: 
\begin{itemize}
\item $\FTV{\alpha} = \s{\alpha}$,
\item $\FTV{A\to B} = \FTV{A}\cup\FTV{B}$,
\item $\FTV{\forall\alpha.A} = \FTV{A}\setminus\s{\alpha}$.
\end{itemize}
We also write $A[B/\alpha]$ for the result of replacing all free
occurrences of $\alpha$ by $B$ in $A$. Just like the substitution of
terms (see Section~\ref{ssec-substitution}), this type substitution
must be {\em capture-free}, i.e., special care must be taken to rename
any bound variables of $A$ so that their names are different from the
free variables of $B$.

The terms of System F are:
\[ \mbox{Terms:}\ssep M,N \bnf x \bor MN \bor \lamabs{x}{A}.M \bor MA
\bor \Lamabs{\alpha}.M
\]
Of these, variables $x$, applications $MN$, and lambda abstractions
$\lamabs{x}{A}.M$ are exactly as for the simply-typed lambda
calculus. The new terms are {\em type application} $MA$, which is the
application of a type function $M$ to a type $A$, and {\em type
  abstraction} $\Lamabs{\alpha}.M$, which denotes the type function
that maps a type $\alpha$ to a term $M$. The typing rules for System F
are shown in Table~\ref{tab-system-f-typing-rules}.
\begin{table*}[tbp]
\[
\begin{array}{rc}
        \tjvar
&       \deriv{}{\Gamma,\typ{x}{ A}\tj x: A}
\nl     \tjapp
&       \deriv{\Gamma\tj M: A\to B\sep\Gamma\tj N: A}
                {\Gamma\tj MN: B}
\nl     \tjlam
&       \deriv{\Gamma,\typ{x}{ A}\tj M: B}
                {\Gamma\tj\lamabs{x}{A}.M: A\to B}
\nl     \tjApp
&       \deriv{\Gamma\tj M: \forall\alpha.A}
                {\Gamma\tj MB: A[B/\alpha]}
\nl     \tjLam
&       \deriv{\Gamma \tj M: A\sep \alpha\not\in\FTV{\Gamma}}
                {\Gamma\tj\Lamabs{\alpha}.M: \forall\alpha.A}
\end{array}
\]
\caption{Typing rules for System F}
\label{tab-system-f-typing-rules}
\end{table*}

We also write $\FTV{M}$ for the set of free type variables in the term
$M$. We need a final notion of substitution: if $M$ is a term, $B$ a type,
and $\alpha$ a type variable, we write $M[B/\alpha]$ for the
capture-free substitution of $B$ for $\alpha$ in $M$. 

\subsection{Reduction rules}

In System F, there are two rules for $\beta$-reduction. The first one
is the familiar rule for the application of a function to a term. The
second one is an analogous rule for the application of a type function
to a type. 
\[ \begin{array}{lllll}
  (\beta_{\to}) & (\lamabs{x}{A}.M)N & \to & \subst{M}{N}{x}, \\
  (\beta_{\forall}) & (\Lamabs{\alpha}.M)A & \to & \subst{M}{A}{\alpha},
\end{array}
\]
Similarly, there are two rules for $\eta$-reduction.
\[ \begin{array}{lllll}
  (\eta_{\to}) & \lamabs{x}{A}.Mx & \to & M, & \mbox{if $x\not\in\FV{M}$,} \\
  (\eta_{\forall}) & \Lamabs{\alpha}.M\alpha & \to & M, & \mbox{if $\alpha\not\in\FTV{M}$.}
\end{array}
\]
The congruence and $\xi$-rules are as expected:
\[ \deriv{M\to M'}{MN\to M'N}
   \sep
   \deriv{N\to N'}{MN\to MN'}
   \sep
   \deriv{M\to M'}{\lamabs{x}{A}{M}\to\lamabs{x}{A}{M'}}
\]
\[ \deriv{M\to M'}{MA\to M'A}
   \sep
   \deriv{M\to M'}{\Lamabs{\alpha}{M}\to\Lamabs{\alpha}{M'}}
\]

\subsection{Examples}

Just as in the untyped lambda calculus, many interesting data types
and operations can be encoded in System F. 

\subsubsection{Booleans}

Define the System F type $\bool$, and terms $\truet,\falset:\bool$, as follows:
\[ \begin{array}{ccl}
  \bool &=& \forall\alpha.\alpha\to\alpha\to\alpha,\\
  \truet &=& \Lamabs{\alpha}.\lamabs{x}{\alpha}.\lamabs{y}{\alpha}.x,\\
  \falset &=& \Lamabs{\alpha}.\lamabs{x}{\alpha}.\lamabs{y}{\alpha}.y.\\
\end{array}
\]
It is easy to see from the typing rules that $\tj\truet:\bool$ and
$\tj\falset:\bool$ are valid typing judgements. We can define an
if-then-else operation 
\[ \begin{array}{l}
  \ifthenelset : \forall\beta.\bool\to\beta\to\beta\to\beta,
  \\  \ifthenelset = \Lamabs{\beta}.\lamabs{z}{\bool}.z\beta.
\end{array}
\]
It is then easy to see that, for any type $B$ and any pair of terms
$M,N:B$, we have
\[ \begin{array}{rcl}
  \ifthenelset B\, \truet\, MN &\redbs& M,\\
  \ifthenelset B\, \falset\, MN &\redbs& N.
\end{array}
\]
Once we have if-then-else, it is easy to define other boolean
operations, for example
\[ \begin{array}{ccl}
  \andt &=& \lamabs{a}{\bool}.\lamabs{b}{\bool}.\ifthenelset\bool a\,b\,\falset,\\
  \ort &=& \lamabs{a}{\bool}.\lamabs{b}{\bool}.\ifthenelset\bool a\,\truet\,b,\\
  \nott &=& \lamabs{a}{\bool}.\ifthenelset\bool a\,\falset\,\truet.
\end{array}
\]
Later, in Proposition~\ref{prop-unique-bool}, we will show that up to
$\beta\eta$ equality, $\truet$ and and $\falset$ are the {\em only}
closed terms of type $\bool$. This, together with the if-then-else
operation, justifies calling this the type of booleans.

Note that the above encodings of the booleans and their if-then-else
operation in System F is exactly the same as the corresponding
encodings in the untyped lambda calculus from
Section~\ref{ssec-booleans}, provided that one erases all the types
and type abstractions. However, there is an important difference: in
the untyped lambda calculus, the booleans were just two terms among
many, and there was no guarantee that the argument of a boolean
function (such as $\andt$ and $\ort$) was actually a boolean. In
System F, the typing guarantees that all closed boolean terms
eventually reduce to either $\truet$ or $\falset$. 

\subsubsection{Natural numbers}

We can also define a type of Church numerals in System F. We define:
\[ \begin{array}{ccl}
  \nat &=& \forall\alpha.(\alpha\to\alpha)\to\alpha\to\alpha,\\
  \chnum{0} &=& \Lamabs{\alpha}.\lamabs{f}{\alpha\to\alpha}.\lamabs{x}{\alpha}.x,\\
  \chnum{1} &=& \Lamabs{\alpha}.\lamabs{f}{\alpha\to\alpha}.\lamabs{x}{\alpha}.fx,\\
  \chnum{2} &=& \Lamabs{\alpha}.\lamabs{f}{\alpha\to\alpha}.\lamabs{x}{\alpha}.f(fx),\\
  \chnum{3} &=&
  \Lamabs{\alpha}.\lamabs{f}{\alpha\to\alpha}.\lamabs{x}{\alpha}.f(f(fx)),\\
  \ldots
\end{array}
\]
It is then easy to define simple functions, such as successor,
addition, and multiplication:
\[ \begin{array}{ccl}
  \succt &=& \lamabs{n}{\nat}.\Lamabs{\alpha}.\lamabs{f}{\alpha\to\alpha}.\lamabs{x}{\alpha}.f(n\alpha f x),\\
  \addt &=&  \lamabs{n}{\nat}.\lamabs{m}{\nat}.\Lamabs{\alpha}.\lamabs{f}{\alpha\to\alpha}.\lamabs{x}{\alpha}.n\alpha f(m\alpha fx), \\
  \multt &=& \lamabs{n}{\nat}.\lamabs{m}{\nat}.\Lamabs{\alpha}.\lamabs{f}{\alpha\to\alpha}.n\alpha (m\alpha f).
\end{array}
\]
Just as for the booleans, these encodings of the Church numerals and
functions are exactly the same as those of the untyped lambda calculus
from Section~\ref{ssec-natural-numbers}, if one erases all the types
and type abstractions.  We will show in
Proposition~\ref{prop-unique-nat} below that the Church numerals are,
up to $\beta\eta$-equivalence, the only closed terms of type $\nat$.

\subsubsection{Pairs}\label{ssec-pairs}

You will have noticed that we didn't include a cartesian product type
$A\times B$ in the definition of System F. This is because such a type
is definable. Specifically, let
\[ \begin{array}{ccl}
  A\times B &=& \forall\alpha.(A\to B\to\alpha)\to\alpha,\\
  \pair{M,N} &=& \Lamabs{\alpha}.\lamabs{f}{A\to B\to\alpha}.fMN.
\end{array}
\]

Note that when $M:A$ and $N:B$, then $\pair{M,N}:A\times B$. Moreover,
for any pair of types $A,B$, we have projection
functions $\leftt AB : A\times B\to A$ and $\rightt AB : A\times B\to
B$, defined by
\[ \begin{array}{ccl}
  \leftt &=& \Lamabs{\alpha}.\Lamabs{\beta}.\lamabs{p}{\alpha\times
    \beta}.p\alpha(\lamabs{x}{\alpha}.\lamabs{y}{\beta}.x),\\
  \rightt &=& \Lamabs{\alpha}.\Lamabs{\beta}.\lamabs{p}{\alpha\times
    \beta}.p\beta(\lamabs{x}{\alpha}.\lamabs{y}{\beta}.y).
\end{array}
\]
This satisfies the usual laws
\[ \begin{array}{ccl}
  \leftt AB\pair{M,N} &\redbs& M,\\
  \rightt AB\pair{M,N} &\redbs& N.
\end{array}
\]
Once again, these encodings of pairs and projections are exactly the
same as those we used in the untyped lambda calculus, when one erases
all the type-related parts of the terms. You will show in
Exercise~\ref{ex-unique-pair} that every closed term of type $A\times
B$ is $\beta\eta$-equivalent to a term of the form $\pair{M,N}$.

\begin{remark}
  It is also worth noting that the corresponding $\eta$-laws, such as
  \[  \pair{\leftt M,\rightt M}=M,
  \] are {\em not} derivable in System F. These laws hold whenever $M$
  is a closed term, but not necessarily when $M$ contains free
  variables.
\end{remark}

\begin{exercise}\label{ex-encodings}
  Find suitable encodings in System F of the types $1$, $A+B$, and
  $0$, along with the corresponding terms $\unit$, $\inj1$, $\inj2$,
  $\caseof{M}{x}{A}{N}{y}{B}{P}$, and $\Box_A{M}$.
\end{exercise}

\subsection{Church-Rosser property and strong normalization}

\begin{theorem}[Church-Rosser]
  System F satisfies the Church-Rosser property, both for
  $\beta$-reduction and for $\beta\eta$-reduction.
\end{theorem}

\begin{theorem}[Strong normalization]\label{thm-strong-norm-system-f}
  In System F, all terms are strongly normalizing.
\end{theorem}

The proof of the Church-Rosser property is similar to that of the
simply-typed lambda calculus, and is left as an exercise. The proof of
strong normalization is much more complex; it can be found in Chapter
14 of ``Proofs and Types'' {\cite{GLT89}}.

\subsection{The Curry-Howard isomorphism}

From the point of view of the Curry-Howard isomorphism,
$\forall\alpha.A$ is the universally quantified logical statement
``for all $\alpha$, $A$ is true''. Here $\alpha$ ranges over atomic
propositions. For example, the formula
$\forall\alpha.\forall\beta.\alpha\to(\beta\to\alpha)$ expresses the
valid fact that the implication $\alpha\to(\beta\to\alpha)$ is true
for all propositions $\alpha$ and $\beta$. Since this quantifier
ranges over {\em propositions}, it is called a {\em second-order
  quantifier}, and the corresponding logic is {\em second-order
  propositional logic}. 

Under the Curry-Howard isomorphism, the typing rules for System F
become the following logical rules:
\begin{itemize}
\item (Axiom)
\[ \trule{ax$_x$}\ \deriv{}{\Gamma,\typ{x}{A}\tj A}
\]
\item ($\to$-introduction)
\[ \trule{$\to$-I$_x$}\ \deriv{\Gamma,\typ{x}{A}\tj B}{\Gamma\tj A\to B}
\]
\item ($\to$-elimination)
\[ \trule{$\to$-E}\ \deriv{\Gamma\tj A\to B\sep \Gamma\tj A}{\Gamma\tj B}
\]
\item ($\forall$-introduction)
\[ \trule{$\forall$-I}\ \deriv{\Gamma\tj A\sep \alpha\not\in\FTV{\Gamma}}{\Gamma\tj \forall\alpha.A}
\]
\item ($\forall$-elimination)
\[ \trule{$\forall$-E}\ \deriv{\Gamma\tj \forall\alpha.A}{\Gamma\tj A[B/\alpha]}
\]
\end{itemize}
The first three of these rules are familiar from propositional
logic. 

The $\forall$-introduction rule is also known as {\em universal
  generalization}. It corresponds to a well-known logical reasoning
principle: If a statement $A$ has been proven for some {\em arbitrary}
$\alpha$, then it follows that it holds for {\em all} $\alpha$. The
requirement that $\alpha$ is ``arbitrary'' has been formalized in the
logic by requiring that $\alpha$ does not appear in any of the
hypotheses that were used to derive $A$, or in other words, that
$\alpha$ is not among the free type variables of $\Gamma$.

The $\forall$-elimination rule is also known as {\em universal
  specialization}. It is the simple principle that if some statement
is true for all propositions $\alpha$, then the same statement is true for
any particular proposition $B$. Note that, unlike the $\forall$-introduction
rule, this rule does not require a side condition. 

Finally, we note that the side condition in the $\forall$-introduction
rule is of course the same as that of the typing rule $\tjLam$ of
Table~\ref{tab-system-f-typing-rules}. From the point of view of
logic, the side condition is justified because it asserts that
$\alpha$ is ``arbitrary'', i.e., no assumptions have been made about
it. From a lambda calculus view, the side condition also makes sense:
otherwise, the term $\lamabs{x}{\alpha}.\Lamabs{\alpha}.x$ would be
well-typed of type $\alpha\to\forall\alpha.\alpha$, which clearly does
not make any sense: there is no way that an element $x$ of some fixed
type $\alpha$ could suddenly become an element of an arbitrary type. 

\subsection{Supplying the missing logical connectives}

It turns out that a logic with only implication $\to$ and a
second-order universal quantifier $\forall$ is sufficient for
expressing all the other usual logical connectives, for example:
\begin{eqnarray}
  A\cand B &\iff& \forall\alpha.(A\to B\to\alpha)\to\alpha,\label{eqn-A-cand-B}\\
  A\cor B &\iff& \forall\alpha.(A\to\alpha)\to(B\to\alpha)\to\alpha,\label{eqn-A-cor-B}\\
  \cnot A &\iff& \forall\alpha.A\to \alpha,\\
  \ctrue &\iff& \forall\alpha.\alpha\to\alpha,\\
  \cfalse &\iff& \forall\alpha.\alpha,\\
  \exists\beta.A &\iff& \forall\alpha.(\forall\beta.(A\to\alpha))\to\alpha.\label{eqn-exists-beta-A}
\end{eqnarray}

\begin{exercise}
  Using informal intuitionistic reasoning, prove that the left-hand side
  is logically equivalent to the right-hand side for each of
  {\eqref{eqn-A-cand-B}}--{\eqref{eqn-exists-beta-A}}.
\end{exercise}

\begin{remark}
  The definitions {\eqref{eqn-A-cand-B}}--{\eqref{eqn-exists-beta-A}}
  are somewhat reminiscent of De Morgan's laws and double negations.
  Indeed, if we replace the type variable $\alpha$ by the constant
  $\falset$ in {\eqref{eqn-A-cand-B}}, the right-hand side becomes
  $(A\to B\to\falset)\to\falset$, which is intuitionistically
  equivalent to $\cnot\cnot(A\cand B)$. Similarly, the right-hand side
  of {\eqref{eqn-A-cor-B}} becomes
  $(A\to\falset)\to(B\to\falset)\to\falset$, which is
  intuitionistically equivalent to $\cnot(\cnot A\cand\cnot B)$, and
  similarly for the remaining connectives. However, the versions of
  {\eqref{eqn-A-cand-B}}, {\eqref{eqn-A-cor-B}}, and
  {\eqref{eqn-exists-beta-A}} using $\falset$ are only {\em
    classically}, but not {\em intuitionistically} equivalent to
  their respective left-hand sides. On the other hand, it is
  remarkable that by the use of $\forall\alpha$, each right-hand
  side is {\em intuitionistically} equivalent to the left-hand sides.
\end{remark}

\begin{remark}
  Note the resemblance between {\eqref{eqn-A-cand-B}} and the
  definition of $A\times B$ given in Section~\ref{ssec-pairs}.
  Naturally, this is not a coincidence, as logical conjunction $A\cand
  B$ should correspond to cartesian product $A\times B$ under the
  Curry-Howard correspondence. Indeed, by applying the same principle
  to the other logical connectives, one arrives at a good hint for
  Exercise~\ref{ex-encodings}.
\end{remark}

\begin{exercise}
  Extend System F with an existential quantifier $\exists\beta.A$, not
  by using {\eqref{eqn-exists-beta-A}}, but by adding a new type with
  explicit introduction and elimination rules to the language. Justify
  the resulting rules by comparing them with the usual rules of
  mathematical reasoning for ``there exists''. Can you explain the
  meaning of the type $\exists\beta.A$ from a programming language or
  lambda calculus point of view?
\end{exercise}

\subsection{Normal forms and long normal forms}

Recall that a $\beta$-normal form of System F is, by definition, a
term that contains no $\beta$-redex, i.e., no subterm of the form
$(\lamabs{x}{A}.M)N$ or $(\Lamabs{\alpha}.M)A$. The following
proposition gives another useful way to characterize the
$\beta$-normal forms.

\begin{proposition}[Normal forms]
  A term of System F is a $\beta$-normal form if and only if it is of
  the form
  \begin{equation}\label{eqn-nf}
    \Lamabs{a_1}.\Lamabs{a_2}\ldots\Lamabs{a_n}.zQ_1Q_2\ldots Q_k,
  \end{equation}
  where:\rm
  \begin{itemize}
  \item $n\geq 0$ and $k\geq 0$;
  \item Each $\Lamabs{a_i}$ is either a lambda abstraction
    $\lamabs{x_i}{A_i}$ or a type abstraction $\Lamabs{\alpha_i}$;
  \item Each $Q_j$ is either a term $M_j$ or a type $B_j$; and
  \item Each $Q_j$, when it is a term, is recursively in normal form.
  \end{itemize}
\end{proposition}

\begin{proof}
  First, it is clear that every term of the form {\eqref{eqn-nf}} is
  in normal form: the term cannot itself be a redex, and the only
  place where a redex could occur is inside one of the $Q_j$, but
  these are assumed to be normal. 

  For the converse, consider a term $M$ in $\beta$-normal form. We
  show that $M$ is of the form {\eqref{eqn-nf}} by induction on $M$. 
  \begin{itemize}
  \item If $M=z$ is a variable, then it is of the form
    {\eqref{eqn-nf}} with $n=0$ and $k=0$.
  \item If $M=NP$ is normal, then $N$ is normal, so
    by induction hypothesis, $N$ is of the form {\eqref{eqn-nf}}. But
    since $NP$ is normal, $N$ cannot be a lambda abstraction, so we
    must have $n=0$. It follows that $NP=zQ_1Q_2\ldots Q_kP$ is itself
    of the form {\eqref{eqn-nf}}.
  \item If $M=\lamabs{x}{A}.N$ is normal, then $N$ is normal, so by
    induction hypothesis, $N$ is of the form {\eqref{eqn-nf}}. It
    follows immediately that $\lamabs{x}{A}.N$ is also of the form
    {\eqref{eqn-nf}}.
  \item The case for $M=NA$ is like the case for $M=NP$. 
  \item The case for $M=\Lamabs{\alpha}.N$ is like the case for
    $M=\lamabs{x}{A}.N$.\eot
  \end{itemize}
\end{proof}

\begin{definition}
  In a term of the form {\eqref{eqn-nf}}, the variable $z$ is called
  the {\em head variable} of the term.
\end{definition}

Of course, by the Church-Rosser property together with strong
normalization, it follows that every term of System F is
$\beta$-equivalent to a unique $\beta$-normal form, which must then be
of the form {\eqref{eqn-nf}}. On the other hand, the normal forms
{\eqref{eqn-nf}} are not unique up to $\eta$-conversion; for example,
$\lamabs{x}{A\to B}.x$ and $\lamabs{x}{A\to B}.\lamabs{y}{A}.xy$ are
$\eta$-equivalent terms and are both of the form {\eqref{eqn-nf}}. 
In order to achieve uniqueness up to $\beta\eta$-conversion, we
introduce the notion of a {\em long normal form}.

\begin{definition}
  A term of System F is a {\em long normal form} if
  \begin{itemize}
  \item it is of the form {\eqref{eqn-nf}};
  \item the body $zQ_1\ldots Q_k$ is of atomic type (i.e., its type is
    a type variable); and
  \item each $Q_j$, when it is a term, is recursively in long normal form.
  \end{itemize}
\end{definition}

\begin{proposition}\label{prop-unique-lnf}
  Every term of System F is $\beta\eta$-equivalent to a unique long
  normal form.
\end{proposition}

\begin{proof}
  By strong normalization and the Church-Rosser property of
  $\beta$-reduction, we already know that every term is
  $\beta$-equivalent to a unique $\beta$-normal form. It therefore
  suffices to show that every $\beta$-normal form is $\eta$-equivalent
  to a unique long normal form.

  We first show that every $\beta$-normal form is $\eta$-equivalent to
  some long normal form. We prove this by induction. Indeed, consider
  a $\beta$-normal form of the form {\eqref{eqn-nf}}. By induction
  hypothesis, each of $Q_1,\ldots,Q_k$ can be $\eta$-converted to long
  normal form. Now we proceed by induction on the type $A$ of
  $zQ_1\ldots Q_k$. If $A=\alpha$ is atomic, then the normal form is
  already long, and there is nothing to show. If $A=B\to C$, then we
  can $\eta$-expand {\eqref{eqn-nf}} to 
  \[
  \Lamabs{a_1}.\Lamabs{a_2}\ldots\Lamabs{a_n}.\lamabs{w}{B}.zQ_1Q_2\ldots Q_kw
  \]
  and proceed by the inner induction hypothesis. If
  $A=\forall\alpha.B$, then we can $\eta$-expand {\eqref{eqn-nf}} to
  \[
  \Lamabs{a_1}.\Lamabs{a_2}\ldots\Lamabs{a_n}.\Lamabs{\alpha}.zQ_1Q_2\ldots Q_k\alpha
  \]
  and proceed by the inner induction hypothesis.

  For uniqueness, we must show that no two different long normal forms
  can be $\beta\eta$-equivalent to each other. We leave this as an
  exercise. \eot
\end{proof}

\subsection{The structure of closed normal forms}

It is a remarkable fact that if $M$ is in long normal form, then a lot
of the structure of $M$ is completely determined by its type.
Specifically: if the type of $M$ is atomic, then $M$ must start with a
head variable. If the type of $M$ is of the form $B\to C$, then $M$
must be, up to $\alpha$-equivalence, of the form $\lamabs{x}{B}.N$,
where $N$ is a long normal form of type $C$. And if the type of $M$ is
of the form $\forall\alpha.C$, then $M$ must be, up to
$\alpha$-equivalence, of the form $\Lamabs{\alpha}.N$, where $N$ is a
long normal form of type $C$. 

So for example, consider the type 
\[ A=B_1\to B_2\to\forall \alpha_3. B_4\to\forall\alpha_5.\beta.
\]
We say that this type have five {\em prefixes}, where each prefix is
of the form ``$B_i\to$'' or ``$\forall\alpha_i.$''. Therefore, every
long normal form of type $A$ must also start with five prefixes;
specifically, it must start with 
\[ \lamabs{x_1}{B_1}.\lamabs{x_2}{B_2}.\Lamabs{\alpha_3}.\lamabs{x_4}{B_4}.\Lamabs{\alpha_5}.\ldots
\]
The next part of the long normal form is a choice of head variable. If
the term is closed, the head variable must be one of the $x_1$, $x_2$,
or $x_4$. Once the head variable has been chosen, then {\em its} type
determines how many arguments $Q_1,\ldots,Q_k$ the head variable must
be applied to, and the types of these arguments. The structure of each
of $Q_1,\ldots,Q_k$ is then recursively determined by its type, with
its own choice of head variable, which then recursively determines its
subterms, and so on. 

In other words, the degree of freedom in a long normal form is a
choice of head variable at each level. This choice of head variables
completely determines the long normal form.

Perhaps the preceding discussion can be made more comprehensible by
means of some concrete examples. The examples take the form of the
following propositions and their proofs.

\begin{proposition}\label{prop-unique-bool}
  Every closed term of type $\bool$ is $\beta\eta$-equivalent to
  either $\truet$ or $\falset$.
\end{proposition}

\begin{proof}
  Let $M$ be a closed term of type $\bool$. By
  Proposition~\ref{prop-unique-lnf}, we may assume that $M$ is a long
  normal form. Since $\bool = \forall\alpha.\alpha\to\alpha\to\alpha$,
  every long normal form of this type must start, up to
  $\alpha$-equivalence, with
  \[ \Lamabs{\alpha}.\lamabs{x}{\alpha}.\lamabs{y}{\alpha}.\ldots
  \]
  This must be followed by a head variable, which, since $M$ is
  closed, can only be $x$ or $y$. Since both $x$ and $y$ have atomic
  type, neither of them can be applied to further arguments, and
  therefore, the only two possible long normal forms are:
  \[ \begin{array}{l}
    \Lamabs{\alpha}.\lamabs{x}{\alpha}.\lamabs{y}{\alpha}.x \\
    \Lamabs{\alpha}.\lamabs{x}{\alpha}.\lamabs{y}{\alpha}.y,
  \end{array}
  \]
  which are $\truet$ and $\falset$, respectively.\eot
\end{proof}

\begin{proposition}\label{prop-unique-nat}
  Every closed term of type $\nat$ is $\beta\eta$-equivalent to a
  Church numeral $\chnum{n}$, for some $n\in\N$.
\end{proposition}

\begin{proof}
  Let $M$ be a closed term of type $\nat$. By
  Proposition~\ref{prop-unique-lnf}, we may assume that $M$ is a long
  normal form. Since
  $\nat=\forall\alpha.(\alpha\to\alpha)\to\alpha\to\alpha$, every long
  normal form of this type must start, up to $\alpha$-equivalence,
  with
  \[
  \Lamabs{\alpha}.\lamabs{f}{\alpha\to\alpha}.\lamabs{x}{\alpha}.\ldots
  \]
  This must be followed by a head variable, which, since $M$ is
  closed, can only be $x$ or $f$. If the head variable is $x$, then it
  takes no argument, and we have 
  \[ M = \Lamabs{\alpha}.\lamabs{f}{\alpha\to\alpha}.\lamabs{x}{\alpha}.x
  \]
  If the head variable is $f$, then it takes exactly one argument, so
  $M$ is of the form
  \[ M = \Lamabs{\alpha}.\lamabs{f}{\alpha\to\alpha}.\lamabs{x}{\alpha}.fQ_1.
  \]
  Because $Q_1$ has type $\alpha$, its own long normal form has no
  prefix; therefore $Q_1$ must start with a head variable, which must
  again be $x$ or $f$. If $Q_1=x$, we have
  \[ M = \Lamabs{\alpha}.\lamabs{f}{\alpha\to\alpha}.\lamabs{x}{\alpha}.fx.
  \]
  If $Q_1$ has head variable $f$, then we have $Q_1=fQ_2$, and
  proceeding in this manner, we find that $M$ has to be of the form
  \[ M =
  \Lamabs{\alpha}.\lamabs{f}{\alpha\to\alpha}.\lamabs{x}{\alpha}.f(f(\ldots(fx)\ldots)),
  \]
  i.e., a Church numeral.\eot
\end{proof}

\begin{exercise}\label{ex-unique-pair}
  Prove that every closed term of type $A\times B$ is
  $\beta\eta$-equivalent to a term of the form $\pair{M,N}$, where
  $M:A$ and $N:B$.
\end{exercise}

\subsection{Application: representation of arbitrary data in System F}

Let us consider the definition of a long normal form one more time. By
definition, every long normal form is of the form 
\begin{equation}\label{eqn-nf2}
  \Lamabs{a_1}.\Lamabs{a_2}\ldots\Lamabs{a_n}.zQ_1Q_2\ldots Q_k,
\end{equation}
where $zQ_1Q_2\ldots Q_k$ has atomic type and $Q_1,\ldots,Q_k$ are,
recursively, long normal forms.  Instead of writing the long normal
form on a single line as in {\eqref{eqn-nf2}}, let us write it in tree
form instead:
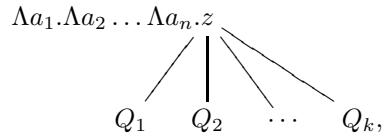
\[
  \xymatrix@C-3ex{
    &\makebox[0in][r]{$\Lamabs{a_1}.\Lamabs{a_2}\ldots\Lamabs{a_n}.$}z
    \ar@{-}[dl]
    \ar@{-}[d]
    \ar@{-}[dr]
    \ar@{-}[drr]
    \\
    Q_1 & Q_2 & \cdots & Q_k,
  }
\]
where the long normal forms $Q_1,\ldots,Q_k$ are recursively also
written as trees. For example, with this notation, the Church numeral
$\chnum{2}$ becomes
\begin{equation}\label{eqn-chnum}
  \m{\xymatrix@R-1.5em@C-2ex{
    \makebox[0in][r]{$\Lamabs{\alpha}.\lamabs{f}{\alpha\to\alpha}.\lamabs{x}{\alpha}.$}f
    \ar@{-}[d]
    \\
    f\ar@{-}[d]
    \\
    x,
  }}
\end{equation}
and the pair $\pair{M,N}$ becomes
\[ 
  \xymatrix@R-1.5em@C-5ex{
    &\makebox[0in][r]{$\Lamabs{\alpha}.\lamabs{f}{A\to B\to\alpha}.$}f
    \ar@{-}[dl]\ar@{-}[dr]
    \\
    M&&N.
  }
\]
We can use this very idea to encode (almost) arbitrary data
structures. For example, suppose that the data structure we wish to
encode is a binary tree whose leaves are labelled by natural numbers. 
Let's call such a thing a {\em leaf-labelled binary tree}.
Here is an example:
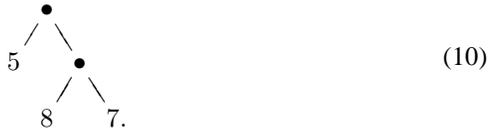
\begin{equation}\label{eqn-tree}
  \m{\xymatrix@R-1.5em@C-5ex{
    &&\bullet\ar@{-}[dl]\ar@{-}[dr]
    \\
    &5 &&\bullet\ar@{-}[dl]\ar@{-}[dr]
    \\
    && 8 && 7.
  }}
\end{equation}
In general, every leaf-labelled binary tree is either a {\em leaf},
which is labelled by a natural number, or else a {\em branch} that has
exactly two {\em children} (a left one and a right one), each of which
is a leaf-labelled binary tree. Written as a BNF, we have the
following grammar for leaf-labelled binary trees:
\[ \mbox{Tree:}\ssep T,S \bnf \leaft(n) \bor \brancht(T,S).
\]
When translating this as a System F type, we think along the lines of
long normal forms. We need a type variable $\alpha$ to represent
leaf-labelled binary trees. We need two head variables whose type ends
in $\alpha$: The first head variable, let's call it $\ell$, represents
a leaf, and takes a single argument that is a natural number. Thus
$\ell:\nat\to\alpha$. The second head variable, let's call it $b$,
represents a branch, and takes two arguments that are leaf-labelled
binary trees. Thus $b:\alpha\to\alpha\to\alpha$. We end up with the
following System F type:
\[ \treet =
\forall\alpha.(\nat\to\alpha)\to(\alpha\to\alpha\to\alpha)\to\alpha.
\]
A typical long normal form of this type is:
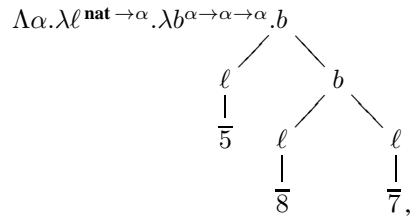
\[ \xymatrix@R-1.5em@C-3ex{
  &&\makebox[0in][r]{$\Lamabs{\alpha}.\lamabs{\ell}{\,\nat\to\alpha}.\lamabs{b}{\alpha\to\alpha\to\alpha}.$}b\ar@{-}[dl]\ar@{-}[dr]
  \\
  &\ell\ar@{-}[d] &&b\ar@{-}[dl]\ar@{-}[dr]
  \\
  &\chnum{5}& \ell\ar@{-}[d] && \ell\ar@{-}[d]
  \\
  && \chnum{8} && \chnum{7}\makebox[0in][l]{,}
}
\]
where $\chnum{5}$, $\chnum{7}$, and $\chnum{8}$ denote Church numerals
as in {\eqref{eqn-chnum}}, here not expanded into long normal form for
brevity. Notice how closely this long normal form follows
{\eqref{eqn-tree}}. Here is the same term written on a single line:
\[ \Lamabs{\alpha}.\lamabs{\ell}{\,\nat\to\alpha}.\lamabs{b}{\alpha\to\alpha\to\alpha}.b(\ell\chnum{5})(b(\ell\chnum{8})(\ell\chnum{7}))
\]

\begin{exercise}
  Prove that the closed long normal forms of type $\treet$ are in
  one-to-one correspondence with leaf-labelled binary trees.
\end{exercise}

\section{Type inference}

In Section~\ref{sec-simply-typed-lc}, we introduced the simply-typed lambda
calculus, and we discussed what it means for a term to be well-typed.
We have also asked the question, for a given term, whether it is
typable or not.

In this section, we will discuss an algorithm that decides, given a
term, whether it is typable or not, and if the answer is yes, it also
outputs a type for the term. Such an algorithm is known as a {\em type
  inference algorithm}.

A weaker kind of algorithm is a {\em type checking algorithm}. A type
checking algorithm takes as its input a term with full type
annotations, as well as the types of any free variables, and it
decides whether the term is well-typed or not. Thus, a type checking
algorithm does not infer any types; the type must be given to it as an
input and the algorithm merely checks whether the type is legal. 

Many compilers of programming languages include a type checker, and
programs that are not well-typed are typically refused.  The compilers
of some programming languages, such as ML or Haskell, go one step
further and include a type inference algorithm. This allows
programmers to write programs with no or very few type annotations,
and the compiler will figure out the types automatically. This makes
the programmer's life much easier, especially in the case of
higher-order languages, where types such as $((A\to B)\to C)\to D$ are
not uncommon and would be very cumbersome to write down. However, in
the event that type inference {\em fails}, it is not always easy for
the compiler to issue a meaningful error message that can help the
human programmer fix the problem. Often, at least a basic
understanding of how the type inference algorithm works is necessary
for programmers to understand these error messages.

\subsection{Principal types}

A simply-typed lambda term can have more than one possible type.
Suppose that we have three basic types $\iota_1,\iota_2,\iota_3$ in
our type system. Then the following are all valid typing judgments for
the term $\lam x.\lam y.yx$:
\[ \begin{array}{l}
  \tj \lamabs{x}{\iota_1}.\lamabs{y}{\iota_1\to\iota_1}.yx : 
  \iota_1\to(\iota_1\to\iota_1)\to\iota_1, \\
  \tj \lamabs{x}{\iota_2\to\iota_3}.\lamabs{y}{(\iota_2\to\iota_3)\to\iota_3}.yx : 
  (\iota_2\to\iota_3)\to((\iota_2\to\iota_3)\to\iota_3)\to\iota_3, \\
  \tj \lamabs{x}{\iota_1}.\lamabs{y}{\iota_1\to\iota_3}.yx : 
  \iota_1\to(\iota_1\to\iota_3)\to\iota_3, \\
  \tj \lamabs{x}{\iota_1}.\lamabs{y}{\iota_1\to\iota_3\to\iota_2}.yx : 
  \iota_1\to(\iota_1\to\iota_3\to\iota_2)\to\iota_3\to\iota_2, \\
  \tj \lamabs{x}{\iota_1}.\lamabs{y}{\iota_1\to\iota_1\to\iota_1}.yx : 
  \iota_1\to(\iota_1\to\iota_1\to\iota_1)\to\iota_1\to\iota_1.
\end{array}
\]
What all these typing judgments have in common is that they are of the
form
\[ \begin{array}{l}
  \tj \lamabs{x}{A}.\lamabs{y}{A\to B}.yx : A\to(A\to B)\to B,
\end{array}
\]
for certain types $A$ and $B$. In fact, as we will see, {\em every}
possible type of the term $\lam x.\lam y.yx$ is of this form. We also
say that $A\to(A\to B)\to B$ is the {\em most general type} or the
{\em principal type} of this term, where $A$ and $B$ are placeholders
for arbitrary types.

The existence of a most general type is not a peculiarity of the term
$\lam xy.yx$, but it is true of the simply-typed lambda calculus in
general: every typable term has a most general type. This statement
is known as the {\em principal type property}. 

We will see that our type inference algorithm not only calculates a
possible type for a term, but in fact it calculates the most general
type, if any type exists at all. In fact, we will prove the principal
type property by closely examining the type inference algorithm.

\subsection{Type templates and type substitutions}

In order to formalize the notion of a most general type, we need to be
able to speak of types with placeholders. 

\begin{definition}
  Suppose we are given an
  infinite set of {\em type variables}, which we denote by upper case
  letters $X,Y,Z$ etc. A {\em type template} is a simple type, built
  from type variables and possibly basic types. Formally, type templates
  are given by the BNF
  \[ \mbox{Type templates:}\ssep A,B \bnf X \bor \iota \bor A\to B\bor 
  A\times B\bor 1
  \]
\end{definition}

Note that we use the same letters $A,B$ to denote type templates that
we previously used to denote types. In fact, from now on, we will
simply regard types as special type templates that happen to contain
no type variables.

The point of type variables is that they are placeholders (just like
any other kind of variables). This means, we can replace type
variables by arbitrary types, or even by type templates. A type
substitution is just such a replacement.

\begin{definition}
  A {\em type substitution} $\sigma$ is a function from type variables
  to type templates. We often write $\tsubst{X_1\mapsto A_1,\ldots,X_n\mapsto
  A_n}$ for the substitution defined by $\sigma(X_i)=A_i$ for
  $i=1\ldots n$, and $\sigma(Y)=Y$ if $Y\not\in\s{X_1,\ldots,X_n}$. 
  If $\sigma$ is a type substitution, and $A$ is a type template, then
  we define $\sigmabar A$, the {\em application} of $\sigma$ to $A$,
  as follows by recursion on $A$:
  \[ \begin{array}{rcl}
    \sigmabar X &=& \sigma X, \\
    \sigmabar \iota &=& \iota, \\
    \sigmabar (A\to B) &=& \sigmabar A \to \sigmabar B, \\
    \sigmabar (A\times B) &=& \sigmabar A \times \sigmabar B, \\
    \sigmabar 1 &=& 1.
  \end{array}
  \]
\end{definition}

In words, $\sigmabar A$ is simply the same as $A$, except that all the
type variables have been replaced according to $\sigma$. We are now in
a position to formalize what it means for one type template to be more
general than another.

\begin{definition}
  Suppose $A$ and $B$ are type templates. We say that $A$ is {\em more
    general} than $B$ if there exists a type substitution $\sigma$
  such that $\sigmabar A=B$.
\end{definition}

In other words, we consider $A$ to be more general than $B$ if $B$ can
be obtained from $A$ by a substitution. We also say that $B$ is an
{\em instance} of $A$. Examples:

\begin{itemize}
\item $X\to Y$ is more general than $X\to X$.
\item $X\to X$ is more general than $\iota\to\iota$. 
\item $X\to X$ is more general than $(\iota\to\iota)\to(\iota\to\iota)$.
\item Neither of $\iota\to\iota$ and
  $(\iota\to\iota)\to(\iota\to\iota)$ is more general than the
  other. We say that these types are {\em incomparable}.
\item $X\to Y$ is more general than $W\to Z$, and vice versa. We say
  that $X\to Y$ and $W\to Z$ are {\em equally general}.
\end{itemize}

We can also speak of one substitution being more general than another:
\begin{definition}
  If $\tau$ and $\rho$ are type substitutions, we say that $\tau$
  is more general than $\rho$ if there exists a type substitution
  $\sigma$ such that $\sigmabar\cp\tau=\rho$. 
\end{definition}

\subsection{Unifiers}

We will be concerned with solving equations between type templates. 
The basic question is not very different from solving equations in
arithmetic: given an equation between expressions, for instance
$x+y=x^2$, is it possible to find values for $x$ and $y$ that make
the equation true? The answer is yes in this case, for instance
$x=2,y=2$ is one solution, and $x=1,y=0$ is another possible
solution. We can even give the most general solution, which is
$x=\mbox{arbitrary}, y=x^2-x$. 

Similarly, for type templates, we might ask whether an equation such
as 
\[ X\to(X\to Y)=(Y\to Z)\to W
\]
has any solutions. The answer is yes, and one solution, for instance,
is $X=\iota\to\iota$, $Y=\iota$, $Z=\iota$,
$W=(\iota\to\iota)\to\iota$. But this is not the most general
solution; the most general solution, in this case, is
$Y=\mbox{arbitrary}$, $Z=\mbox{arbitrary}$, $X=Y\to Z$, $W=(Y\to Z)\to
Y$.

We use substitutions to represent the solutions to such equations. For
instance, the most general solution to the sample equation from the
last paragraph is represented by the substitution
\[ \sigma=\tsubst{X\mapsto Y\to Z, W\mapsto (Y\to Z)\to Y}.
\]
If a substitution $\sigma$ solves the equation $A=B$ in this way, then
we also say that $\sigma$ is a {\em unifier} of $A$ and $B$.

To give another example, consider the equation
\[ X\times(X\to Z) = (Z\to Y)\times Y.
\]
This equation does not have any solution, because we would have to
have both $X=Z\to Y$ and $Y=X\to Z$, which implies $X=Z\to(X\to Z)$,
which is impossible to solve in simple types. We also say that
$X\times(X\to Z)$ and $(Z\to Y)\times Y$ cannot be unified. 

In general, we will be concerned with solving not just single
equations, but systems of several equations. The formal definition of
unifiers and most general unifiers is as follows:

\begin{definition}
  Given two sequences of type templates $\bar{A}=A_1,\ldots,A_n$ and
  $\bar{B}=B_1,\ldots,B_n$, we say that a type substitution $\sigma$
  is a {\em unifier} of $\bar{A}$ and $\bar{B}$ if $\sigmabar
  A_i=\sigmabar B_i$, for all $i=1\ldots n$. Moreover, we say that
  $\sigma$ is a {\em most general unifier} of $\bar{A}$ and $\bar{B}$
  if it is a unifier, and if it is more general than any other
  unifier of $\bar{A}$ and $\bar{B}$.
\end{definition}

\subsection{The unification algorithm}

Unification is the process of determining a most general unifier. More
specifically, unification is an algorithm whose input are two
sequences of type templates $\bar{A}=A_1,\ldots,A_n$ and
$\bar{B}=B_1,\ldots,B_n$, and whose output is either ``failure'', if
no unifier exists, or else a most general unifier $\sigma$. We call
this algorithm $\mgu$ for ``most general unifier'', and we write
$\mgu(\bar{A};\bar{B})$ for the result of applying the algorithm to
$\bar{A}$ and $\bar{B}$. 

Before we state the algorithm, let us note that we only use finitely
many type variables, namely, the ones that occur in $\bar{A}$ and
$\bar{B}$. In particular, the substitutions generated by this
algorithm are finite objects that can be represented and manipulated
by a computer. 

The algorithm for calculating $\mgu(\bar{A};\bar{B})$ is as follows.
By convention, the algorithm chooses the first applicable clause in the
following list. Note that the algorithm is recursive.

\begin{enumerate}
\item $\mgu(X;X)=\id$, the identity substitution.
\item\label{item-mgu-2} $\mgu(X;B)=\tsubst{X\mapsto B}$, if $X$ does
  not occur in $B$.
\item $\mgu(X;B)$ fails, if $X$ occurs in $B$ and $B\neq X$.
\item $\mgu(A;Y)=\tsubst{Y\mapsto A}$, if $Y$ does not occur in $A$.
\item $\mgu(A;Y)$ fails, if $Y$ occurs in $A$ and $A\neq Y$.
\item $\mgu(\iota;\iota)=\id$.
\item $\mgu(A_1\to A_2; B_1\to B_2) = \mgu(A_1,A_2; B_1,B_2)$.
\item $\mgu(A_1\times A_2; B_1\times B_2) = \mgu(A_1,A_2; B_1,B_2)$.
\item $\mgu(1; 1) = \id$.
\item\label{item-mgu-10} $\mgu(A; B)$ fails, in all other cases.
\item\label{item-mgu-11} $\mgu(A,\bar{A};B,\bar{B}) = \taubar \cp
  \rho$, where $\rho=\mgu(\bar{A};\bar{B})$ and $\tau=\mgu(\rhobar
  A;\rhobar B)$.
\end{enumerate}

Note that clauses 1--\ref{item-mgu-10} calculate the most general
unifier of two type templates, whereas clause {\ref{item-mgu-11}}
deals with lists of type templates. Clause {\ref{item-mgu-10}} is a
catch-all clause that fails if none of the earlier clauses apply.  In
particular, this clause causes the following to fail: $\mgu(A_1\to
A_2; B_1\times B_2)$, $\mgu(A_1\to A_2; \iota)$, etc.

\begin{proposition}\label{pro-unification}
  If $\mgu(\bar{A};\bar{B})=\sigma$, then $\sigma$ is a most general
  unifier of $\bar{A}$ and $\bar{B}$. If $\mgu(\bar{A};\bar{B})$
  fails, then $\bar{A}$ and $\bar{B}$ have no unifier. 
\end{proposition}

\begin{proof}
  First, it is easy to prove by induction on the definition of $\mgu$
  that if $\mgu(\bar{A};\bar{B})=\sigma$, then $\sigma$ is a unifier
  of $\bar{A}$ and $\bar{B}$. This is evident in all cases except
  perhaps clause \ref{item-mgu-11}: but here, by induction hypothesis,
  $\rhobar\bar{A}=\rhobar\bar{B}$ and $\taubar(\rhobar
  A)=\taubar(\rhobar B)$, hence also
  $\taubar(\rhobar(A,\bar{A}))=\taubar(\rhobar(B,\bar{B}))$. Here we
  have used the evident notation of applying a substitution to a list
  of type templates.
  
  Second, we prove that if $\bar{A}$ and $\bar{B}$ can be unified,
  then $\mgu(\bar{A};\bar{B})$ returns a most general unifier.
  This is again proved by induction. For example, in clause
  \ref{item-mgu-2}, we have $\sigma=\tsubst{X\mapsto B}$. Suppose
  $\tau$ is another unifier of $X$ and $B$.  Then $\taubar X=\taubar
  B$. We claim that $\taubar\cp\sigma=\tau$. But $\taubar(\sigma(X)) =
  \taubar(B) = \taubar(X) = \tau(X)$, whereas if $Y\neq X$, then
  $\taubar(\sigma(Y))=\taubar(Y)=\tau(Y)$. Hence
  $\taubar\cp\sigma=\tau$, and it follows that $\sigma$ is more
  general than $\tau$. The clauses 1--\ref{item-mgu-10} all follow by
  similar arguments.  For clause {\ref{item-mgu-11}}, suppose that
  $A,\bar{A}$ and $B,\bar{B}$ have some unifier $\sigma'$. Then
  $\sigma'$ is also a unifier for $\bar{A}$ and $\bar{B}$, and thus
  the recursive call return a most general unifier $\rho$ of $\bar{A}$
  and $\bar{B}$. Since $\rho$ is more general than $\sigma'$, we have
  $\kappabar\cp\rho=\sigma'$ for some substitution $\kappa$. But then
  $\kappabar(\rhobar A)=\sigmabar' A=\sigmabar' B=\kappabar(\rhobar
  B)$, hence $\kappabar$ is a unifier for $\rhobar A$ and $\rhobar B$.
  By induction hypothesis, $\tau=\mgu(\rhobar A;\rhobar B)$ exists and
  is a most general unifier for $\rhobar A$ and $\rhobar B$. It
  follows that $\tau$ is more general than $\kappabar$, thus
  $\kappabar'\cp\tau = \kappabar$, for some substitution $\kappa'$.
  Finally we need to show that $\sigma=\taubar \cp \rho$ is more
  general than $\sigma'$. But this follows because
  $\kappabar'\cp\sigma = \kappabar'\cp\taubar\cp\rho =
  \kappabar\cp\rho=\sigma'$. \eot
\end{proof}

\begin{remark}
  Proving that the algorithm $\mgu$ terminates is tricky. In
  particular, termination can't be proved by induction on the size of
  the arguments, because in the second recursive call in clause
  \ref{item-mgu-11}, the application of $\rhobar$ may well increase
  the size of the arguments. To prove termination, note that each
  substitution $\sigma$ generated by the algorithm is either the
  identity, or else it eliminates at least one variable. We can use
  this to prove termination by nested induction on the number of
  variables and on the size of the arguments. We leave the details for
  another time.
\end{remark}

\subsection{The type inference algorithm}

Given the unification algorithm, type inference is now relatively
easy. We formulate another algorithm, $\typeinfer$, which takes a
typing judgment $\Gamma\tj M:B$ as its input (using templates instead
of types, and not necessarily a {\em valid} typing judgment). The
algorithm either outputs a most general substitution $\sigma$ such
that $\sigmabar\Gamma\tj M:\sigmabar B$ is a valid typing judgment, or
if no such $\sigma$ exists, the algorithm fails.

In other words, the algorithm calculates the most general substitution
that makes the given typing judgment valid. It is defined as follows:

\begin{enumerate}
\item $\typeinfer(\typ{x_1}{A_1},\ldots,\typ{x_n}{A_n}\tj x_i:B) =
  \mgu(A_i;B)$.
\item $\typeinfer(\Gamma\tj MN:B)= \taubar\cp\sigma$, where
  $\sigma=\typeinfer(\Gamma\tj M:X\to B)$,
  $\tau=\typeinfer(\sigmabar\Gamma\tj N:\sigmabar X)$, for a fresh
  type variable $X$.
\item $\typeinfer(\Gamma\tj\lamabs{x}{A}.M:B)=\taubar\cp\sigma$, where
  $\sigma=\mgu(B;A\to X)$ and
  $\tau=\typeinfer(\sigmabar\Gamma,\typ{x}{\sigmabar A}\tj M:\sigmabar
  X)$, for a fresh type variable $X$.
\item $\typeinfer(\Gamma\tj\pair{M,N}:A) =
  \rhobar\cp\taubar\cp\sigma$, where $\sigma=\mgu(A;X\times Y)$,
  $\tau=\typeinfer(\sigmabar\Gamma\tj M:\sigmabar X)$, and
  $\rho=\typeinfer(\taubar\sigmabar\Gamma\tj N:\taubar\sigmabar Y)$,
  for fresh type variables $X$ and $Y$.
\item $\typeinfer(\Gamma\tj\proj1 M:A) = \typeinfer(\Gamma\tj
  M:A\times Y)$, for a fresh type variable $Y$.
\item $\typeinfer(\Gamma\tj\proj2 M:B) = \typeinfer(\Gamma\tj
  M:X\times B)$, for a fresh type variable $X$.
\item $\typeinfer(\Gamma\tj\unit:A) = \mgu(A;1)$.
\end{enumerate}

Strictly speaking, the algorithm is non-deterministic, because some of
the clauses involve choosing one or more fresh type variables, and the
choice is arbitrary. However, the choice is not essential, since we
may regard all fresh type variables are equivalent. Here, a type
variable is called ``fresh'' if it has never been used.

Note that the algorithm $\typeinfer$ can fail; this happens if and
only if the call to $\mgu$ fails in steps 1, 3, 4, or 7.

Also note that the algorithm obviously always terminates; this follows
by induction on $M$, since each recursive call only uses a smaller
term $M$. 

\begin{proposition}
  If there exists a substitution $\sigma$ such that
  $\sigmabar\Gamma\tj M:\sigmabar B$ is a valid typing judgment, then
  $\typeinfer (\Gamma\tj M: B)$ will return a most general such
  substitution. Otherwise, the algorithm will fail. 
\end{proposition}

\begin{proof}
  The proof is similar to that of Proposition~\ref{pro-unification}.
  \eot
\end{proof}

Finally, the question ``is $M$ typable'' can be answered by choosing
distinct type variables $X_1,\ldots,X_n,Y$ and applying the algorithm
$\typeinfer$ to the typing judgment
$\typ{x_1}{X_1},\ldots,\typ{x_n}{X_n}\tj M:Y$. Note that if the
algorithm succeeds and returns a substitution $\sigma$, then $\sigma
Y$ is the most general type of $M$, and the free variables have types
$\typ{x_1}{\sigma X_1},\ldots,\typ{x_n}{\sigma X_n}$.

\section{Denotational semantics}\label{sec-set-semantics}

We introduced the lambda calculus as the ``theory of functions''. But
so far, we have only spoken of functions in abstract terms. Do lambda
terms correspond to any {\em actual} functions, such as, functions in
set theory? And what about the notions of $\beta$- and
$\eta$-equivalence?  We intuitively accepted these concepts as
expressing truths about the equality of functions. But do these
properties really hold of real functions? Are there other properties
that functions have that that are not captured by $\beta\eta$-equivalence?

The word ``semantics'' comes from the Greek word for ``meaning''.
{\em Denotational semantics} means to give meaning to a language by
interpreting its terms as mathematical objects. This is done by
describing a function that maps syntactic objects (e.g., types,
terms) to semantic objects (e.g., sets, elements). This function is
called an {\em interpretation} or {\em meaning function}, and we
usually denote it by $\semm{-}$. Thus, if $M$ is a term, we will
usually write $\semm{M}$ for the meaning of $M$ under a given
interpretation.

Any good denotational semantics should be {\em compositional}, which
means, the interpretation of a term should be given in terms of the
interpretations of its subterms. Thus, for example, $\semm{MN}$
should be a function of $\semm{M}$ and $\semm{N}$. 

Suppose that we have an axiomatic notion of equality $\simeq$ on terms
(for instance, $\beta\eta$-equivalence in the case of the lambda
calculus). With respect to a particular class of interpretations, {\em
  soundness} is the property
\[      M\simeq N \sep\imp\sep \semm{M}=\semm{N} \mbox{ for all
  interpretations in the class}.
\]
{\em Completeness} is the property
\[      \semm{M}=\semm{N} \mbox{ for all
  interpretations in the class} \sep\imp\sep M\simeq N.
\]
Depending on our viewpoint, we will either say the axioms are sound
(with respect to a given interpretation), or the interpretation is
sound (with respect to a given set of axioms). Similarly for
completeness. Soundness expresses the fact that our axioms (e.g.,
$\beta$ or $\eta$) are true with respect to the given interpretation. 
Completeness expresses the fact that our axioms are sufficient.

\subsection{Set-theoretic interpretation}

The simply-typed lambda calculus can be given a straightforward
set-theoretic interpretation as follows. We map types to sets and
typing judgments to functions. For each basic type $\iota$, assume
that we have chosen a non-empty set $S_{\iota}$. We can then associate
a set $\semm{A}$ to each type $A$ recursively:
\[ \begin{array}{lll}
  \semm{\iota} &=& S_{\iota} \\
  \semm{A\to B} &=& \semm{B}^{\semm{A}} \\
  \semm{A\times B} &=& \semm{A}\times\semm{B} \\
  \semm{1} &=& \s{*}
\end{array}
\]
Here, for two sets $X,Y$, we write $Y^X$ for the set of all functions
from $X$ to $Y$, i.e., $Y^X=\s{f\such f:X\to Y}$. Of course, $X\times
Y$ denotes the usual cartesian product of sets, and $\s{*}$ is some
singleton set. 

We can now interpret lambda terms, or more precisely, typing
judgments, as certain functions. Intuitively, we already know which
function a typing judgment corresponds to. For instance, the typing
judgment $\typ{x}{A},\typ{f}{A\to B}\tj fx:B$ corresponds to the
function that takes an element $x\in\semm{A}$ and an element
$f\in\semm{B}^{\semm{A}}$, and that returns $f(x)\in\semm{B}$. In
general, the interpretation of a typing judgment
\[     \typ{x_1}{A_1},\ldots,\typ{x_n}{A_n} \tj M:B
\]
will be a function
\[     \semm{A_1}\times\ldots\times\semm{A_n} \to \semm{B}.
\]
Which particular function it is depends of course on the term $M$. 
For convenience, if $\Gamma=\typ{x_1}{A_1},\ldots,\typ{x_n}{A_n}$ is a
context, let us write $\semm{\Gamma}=\semm{A_1}\times\ldots\times\semm{A_n}$. 
We now define $\semm{\Gamma\tj M:B}$ by recursion on $M$.

\begin{itemize}
\item If $M$ is a variable, we define 
  \[ \semm{\typ{x_1}{A_1},\ldots,\typ{x_n}{A_n}\tj x_i:A_i} = \proj{i}
  : \semm{A_1}\times\ldots\times\semm{A_n}\to\semm{A_i},
  \]
  where $\proj{i}(a_1,\ldots,a_n) = a_i$.
\item If $M=NP$ is an application, we recursively calculate
  \[ \begin{array}{lll}
    f &=& \semm{\Gamma\tj N:A\to B}
    : \semm{\Gamma}\to\semm{B}^{\semm{A}}, \\
    g &=& \semm{\Gamma\tj P:A} : \semm{\Gamma}\to\semm{A}.
  \end{array}
  \]
  We then define
  \[ \semm{\Gamma\tj NP:B} = h : \semm{\Gamma}\to\semm{B}
  \]
  by $h(\abar)=f(\abar)(g(\abar))$, for all $\abar\in\semm{\Gamma}$.
\item If $M=\lamabs{x}{A}.N$ is an abstraction, we recursively calculate
  \[ \begin{array}{lll}
    f &=& \semm{\Gamma,\typ{x}{A}\tj N:B}
    : \semm{\Gamma}\times\semm{A}\to\semm{B}.
  \end{array}
  \]
  We then define
  \[ \semm{\Gamma\tj\lamabs{x}{A}.N:A\to B} = h 
  : \semm{\Gamma}\to\semm{B}^{\semm{A}}
  \]
  by $h(\abar)(a) = f(\abar,a)$, for all $\abar\in\semm{\Gamma}$ and
  $a\in\semm{A}$.
\item If $M=\pair{N,P}$ is an pair, we recursively calculate
  \[ \begin{array}{lll}
    f &=& \semm{\Gamma\tj N:A} : \semm{\Gamma}\to\semm{A}, \\
    g &=& \semm{\Gamma\tj P:B} : \semm{\Gamma}\to\semm{B}.
  \end{array}
  \]
  We then define
  \[ \semm{\Gamma\tj \pair{N,P}:A\times B} = h 
  : \semm{\Gamma}\to\semm{A}\times\semm{B}
  \]
  by $h(\abar)=\rpair{f(\abar),g(\abar)}$, for all $\abar\in\semm{\Gamma}$.
\item If $M=\proj{i}N$ is a projection (for $i=1,2$), we recursively
  calculate
  \[ \begin{array}{lll}
    f &=& \semm{\Gamma\tj N:B_1\times B_2} 
    : \semm{\Gamma}\to\semm{B_1}\times\semm{B_2}.
  \end{array}
  \]
  We then define
  \[ \semm{\Gamma\tj \proj{i}N:B_i} = h 
  : \semm{\Gamma}\to\semm{B_i}
  \]
  by $h(\abar)=\proj{i}(f(\abar))$, for all $\abar\in\semm{\Gamma}$.
  Here $\proj{i}$ in the meta-language denotes the set-theoretic
  function $\proj{i}:\semm{B_1}\times\semm{B_2}\to\semm{B_i}$ given by
  $\proj{i}\rpair{b_1,b_2} = b_i$.
\item If $M=\unit$, we define
  \[ \semm{\Gamma\tj\unit:1} = h : \semm{\Gamma}\to\s{*}
  \]
  by $h(\abar)=*$, for all $\abar\in\semm{\Gamma}$.
\end{itemize}

To minimize notational inconvenience, we will occasionally abuse the
notation and write $\semm{M}$ instead of $\semm{\Gamma\tj M:B}$, thus
pretending that terms are typing judgments. However, this is only an
abbreviation, and it will be understood that the interpretation really
depends on the typing judgment, and not just the term, even if we use
the abbreviated notation.

We also refer to an interpretation as a {\em model}.

\subsection{Soundness}

\begin{lemma}[Context change]\label{lem-context-set}
  The interpretation behaves as expected under reordering of contexts
  and under the addition of dummy variables to contexts. More
  precisely, if $\sigma:\s{1,\ldots,n}\to\s{1,\ldots,m}$ is an
  injective map, and if the free variables of $M$ are among
  $x_{\sigma 1},\ldots,x_\sigma n$, then the interpretations of the
  two typing judgments,
  \[ \begin{array}{l}
    f = \semm{\typ{x_1}{A_1},\ldots,\typ{x_m}{A_m}\tj M:B}
    : \semm{A_1}\times\ldots\times\semm{A_m} \to \semm{B}, \\
    g = \semm{\typ{x_{\sigma 1}}{A_{\sigma 1}},\ldots,
      \typ{x_{\sigma n}}{A_{\sigma n}} \tj M:B} 
    : \semm{A_{\sigma 1}}\times\ldots\times\semm{A_{\sigma n}} \to \semm{B}
  \end{array}
  \]
  are related as follows:
  \[  f(a_1,\ldots,a_m) = g(a_{\sigma 1},\ldots,a_{\sigma n}),
  \]
  for all $a_1\in\semm{A_1},\ldots,a_m\in\semm{A_m}$.
\end{lemma}

\begin{proof}
  Easy, but tedious, induction on $M$.\eot
\end{proof}

The significance of this lemma is that, to a certain extent, the
context does not matter. Thus, if the free variables of $M$ and $N$
are contained in $\Gamma$ as well as $\Gamma'$, then we have 
\[ \semm{\Gamma\tj M:B}=\semm{\Gamma\tj N:B} \sep\mbox{iff}\sep
 \semm{\Gamma'\tj M:B}=\semm{\Gamma'\tj N:B}.
\]
Thus, whether $M$ and $N$ have equal denotations only depends on $M$
and $N$, and not on $\Gamma$.

\begin{lemma}[Substitution Lemma]
  If 
  \[ \begin{array}{l}
    \semm{\Gamma,\typ{x}{A}\tj M:B} = f 
    : \semm{\Gamma}\times\semm{A}\to\semm{B}\sep\mbox{and}\\
    \semm{\Gamma\tj N:A} = g
    : \semm{\Gamma}\to\semm{A},
  \end{array}
  \]
  then
  \[ \semm{\Gamma\tj \subst{M}{N}{x}:B} = h
  : \semm{\Gamma}\to\semm{B},
  \]
  where $h(\abar) = f(\abar,g(\abar))$, for all
  $\abar\in\semm{\Gamma}$.
\end{lemma}

\begin{proof}
  Very easy, but very tedious, induction on $M$.\eot
\end{proof}

\begin{proposition}[Soundness]
  The set-theoretic interpretation is sound for
  $\beta\eta$-reasoning. In other words,
  \[ M\eqbe N \sep\imp\sep \semm{\Gamma\tj M:B}=\semm{\Gamma\tj N:B}. 
  \]
\end{proposition}

\begin{proof}
  Let us write $M\sim N$ if $\semm{\Gamma\tj M:B}=\semm{\Gamma\tj
    N:B}$. By the remark after Lemma~\ref{lem-context-set}, this
  notion is independent of $\Gamma$, and thus a well-defined relation
  on terms (as opposed to typing judgments). To prove soundness, we
  must show that $M\eqbe N$ implies $M\sim N$, for all $M$ and $N$.
  It suffices to show that $\sim$ satisfies all the axioms of
  $\beta\eta$-equivalence.
  
  The axioms $\trule{refl}$, $\trule{symm}$, and $\trule{trans}$ hold
  trivially. Similarly, all the $\trule{cong}$ and $\nrule{\xi}$ rules
  hold, due to the fact that the meaning of composite terms was
  defined solely in terms of the meaning of their subterms. It remains
  to prove that each of the various $\nrule{\beta}$ and $\nrule{\eta}$
  laws is satisfied (see page~\pageref{page-typed-reductions}). We prove the rule
  $\nrule{\beta_{\to}}$ as an example; the remaining rules are left as
  an exercise.

  Assume $\Gamma$ is a context such that $\Gamma,\typ{x}{A}\tj M:B$
  and $\Gamma\tj N:A$. Let
  \[ \begin{array}{l}
    f = \semm{\Gamma,\typ{x}{A}\tj M:B}
    : \semm{\Gamma}\times\semm{A}\to\semm{B},\\
    g = \semm{\Gamma\tj N:A}
    : \semm{\Gamma}\to\semm{A}, \\
    h = \semm{\Gamma\tj (\lamabs{x}{A}.M):A\to B}
    : \semm{\Gamma}\to\semm{B}^{\semm{A}}, \\
    k = \semm{\Gamma\tj (\lamabs{x}{A}.M)N:B} :
    \semm{\Gamma}\to\semm{B},\\
    l = \semm{\Gamma\tj \subst{M}{N}{x} : B} :
    \semm{\Gamma}\to\semm{B}.
  \end{array}
  \]
  We must show $k=h$. By definition, we have $k(\abar)=h(\abar)(g(\abar))=
  f(\abar,g(\abar))$. On the other hand, $l(\abar)=f(\abar,g(\abar))$
  by the substitution lemma. \eot
\end{proof}

Note that the proof of soundness amounts to a simple calculation;
while there are many details to attend to, no particularly interesting
new idea is required. This is typical of soundness proofs in general. 
Completeness, on the other hand, is usually much more difficult to
prove and often requires clever ideas.

\subsection{Completeness}

We cite two completeness theorems for the set-theoretic
interpretation. The first one is for the class of all models with
finite base type. The second one is for the single model with one
countably infinite base type.

\begin{theorem}[Completeness, Plotkin, 1973]
  The class of set-theoretic models with finite base types is complete
  for the lambda-$\beta\eta$ calculus. 
\end{theorem}

Recall that completeness for a class of models means that if
$\semm{M}=\semm{N}$ holds in {\em all} models of the given class, then
$M\eqbe N$. This is not the same as completeness for each individual
model in the class. 

Note that, for each {\em fixed} choice of finite sets as the
interpretations of the base types, there are some lambda terms such
that $\semm{M}=\semm{N}$ but $M\not\eqbe N$. For instance, consider
terms of type $(\iota\to\iota)\to\iota\to\iota$. There are infinitely
many $\beta\eta$-distinct terms of this type, namely, the Church
numerals. On the other hand, if $S_{\iota}$ is a finite set, then
$\semm{(\iota\to\iota)\to\iota\to\iota}$ is also a finite set. Since a
finite set cannot have infinitely many distinct elements, there must
necessarily be two distinct Church numerals $M,N$ such that
$\semm{M}=\semm{N}$.

Plotkin's completeness theorem, on the other hand, shows that whenever
$M$ and $N$ are distinct lambda terms, then there exist {\em some}
set-theoretic model with finite base types in which $M$ and $N$ are
different.

The second completeness theorem is for a {\em single} model, namely
the one where $S_{\iota}$ is a countably infinite set.

\begin{theorem}[Completeness, Friedman, 1975]
  The set-theoretic model with base type equal to $\N$, the set of
  natural numbers, is complete for the lambda-$\beta\eta$ calculus.
\end{theorem}

We omit the proofs.

\section{The language PCF}

PCF stands for ``programming with computable functions''. The language
PCF is an extension of the simply-typed lambda calculus with booleans,
natural numbers, and recursion. It was first introduced by Dana Scott
as a simple programming language on which to try out techniques for
reasoning about programs. Although PCF is not intended as a ``real
world'' programming language, many real programming languages can be
regarded as (syntactic variants of) extensions of PCF, and many of the
reasoning techniques developed for PCF also apply to more complicated
languages.

PCF is a ``programming language'', not just a ``calculus''. By this we
mean, PCF is equipped with a specific evaluation order, or rules that
determine precisely how terms are to be evaluated. We follow the
slogan:
\begin{center}
  Programming language = syntax + evaluation rules.
\end{center}

After introducing the syntax of PCF, we will look at three different
equivalence relations on terms.
\begin{itemize}
\item {\em Axiomatic equivalence} $\eqax$ will be given by axioms in
  the spirit of $\beta\eta$-equivalence.
\item {\em Operational equivalence} $\eqop$ will be defined in terms
  of the operational behavior of terms. Two terms are operationally
  equivalent if one can be substituted for the other in any context
  without changing the behavior of a program.
\item {\em Denotational equivalence} $\eqden$ is defined via a
  denotational semantics.
\end{itemize}

We will develop methods for reasoning about these equivalences, and
thus for reasoning about programs. We will also investigate how the
three equivalences are related to each other.

\subsection{Syntax and typing rules}

PCF types are simple types over two base types $\boolt$ and $\natt$.
\[ A,B \bnf \boolt \bor \natt \bor A\to B\bor A\times
B\bor 1
\]
The raw terms of PCF are those of the simply-typed lambda calculus,
together with some additional constructs that deal with booleans,
natural numbers, and recursion.
\[ \begin{array}{@{}lll}
  M,N,P &\bnf& 
  x \bor MN \bor \lamabs{x}{A}.M
  \bor \pair{M,N} \bor \proj1 M \bor \proj2 M \bor \unit \\
  &&\bor \truet \bor \falset \bor \zerot \bor \succt(M) \bor \predt(M) \\
  &&\bor \iszerot(M) \bor \ite{M}{N}{P} \bor \Y(M)
\end{array}
\]
The intended meaning of these terms is the same as that of the
corresponding terms we used to program in the untyped lambda calculus:
$\truet$ and $\falset$ are the boolean constants, $\zerot$ is the
constant zero, $\succt$ and $\predt$ are the successor and predecessor
functions, $\iszerot$ tests whether a given number is equal to zero,
$\ite{M}{N}{P}$ is a conditional, and $\Y(M)$ is a fixed point of $M$.

The typing rules for PCF are the same as the typing rules for the
simply-typed lambda calculus, shown in
Table~\ref{tab-simple-typing-rules}, plus the additional typing rules
shown in Table~\ref{tab-pcf-typing-rules}.
\begin{table*}[tbp]
\[
\begin{array}{rc}
  \trule{true}  & \deriv{}{\Gamma\tj\truet: \boolt}
  \nl  \trule{false}  & \deriv{}{\Gamma\tj\falset: \boolt}
  \nl  \trule{zero}  & \deriv{}{\Gamma\tj\zerot: \natt}
  \nl  \trule{succ}  & \deriv{\Gamma\tj M: \natt}{\Gamma\tj\succt(M): \natt}
\end{array}
\sep
\begin{array}{rc}
       \trule{pred}  & \deriv{\Gamma\tj M: \natt}{\Gamma\tj\predt(M): \natt}
  \nl  \trule{iszero} &\deriv{\Gamma\tj M: \natt}{\Gamma\tj\iszerot(M): \boolt}
  \nl  \trule{fix} & \deriv{\Gamma\tj M:A\to A}{\Gamma\tj \Y(M):A}
\end{array}
\]
\[
\begin{array}{rc}
  \trule{if} &
  \deriv{\Gamma\tj M: \boolt\sep\Gamma\tj N:A\sep\Gamma\tj P:A}
  {\Gamma\tj \ite{M}{N}{P}:A}
\end{array}
\]
\caption{Typing rules for PCF}
\label{tab-pcf-typing-rules}
\end{table*}

\subsection{Axiomatic equivalence}

The axiomatic equivalence of PCF is based on the
$\beta\eta$-equivalence of the simply-typed lambda calculus. The
relation $\eqax$ is the least relation given by the following:
\begin{itemize}
\item All the $\beta$- and $\eta$-axioms of the simply-typed lambda
  calculus, as shown on page~\pageref{page-typed-reductions}.
\item One congruence or $\xi$-rule for each term constructor. This
  means, for instance
  \[ \deriv {M\eqax M' \sep N\eqax N' \sep P\eqax P'}
  {\ite{M}{N}{P} \eqax \ite{M'}{N'}{P'}},
  \]
  and similar for all the other term constructors.
\item The additional axioms shown in Table~\ref{tab-pcf-axioms}. Here,
  $\numn$ stands for a {\em numeral}, i.e., a term of the form
  $\succt(\ldots(\succt(\zerot))\ldots)$.
\end{itemize}
\begin{table*}[tbp]
\[ \begin{array}{rcl}
  \predt(\zerot) &=& \zerot \\
  \predt(\succt(\numn)) &=& \numn \\
  \iszerot(\zerot) &=& \truet \\
  \iszerot(\succt(\numn)) &=& \falset \\
  \ite{\truet}{N}{P} &=& N \\
  \ite{\falset}{N}{P} &=& P \\
  \Y(M) &=& M(\Y(M))
\end{array}
\]
\caption{Axiomatic equivalence for PCF}
\label{tab-pcf-axioms}
\end{table*}

\subsection{Operational semantics}

The operational semantics of PCF is commonly given in two different
styles: the {\em small-step} or {\em shallow} style, and the {\em
  big-step} or {\em deep} style. We give the small-step semantics
first, because it is closer to the notion of $\beta$-reduction that
we considered for the simply-typed lambda calculus.

There are some important differences between an operational semantics,
as we are going to give it here, and the notion of $\beta$-reduction
in the simply-typed lambda calculus. Most importantly, the operational
semantics is going to be {\em deterministic}, which means, each
term can be reduced in at most one way. Thus, there will never be a
choice between more than one redex. Or in other words, it will always
be uniquely specified which redex to reduce next.

As a consequence of the previous paragraph, we will abandon many of
the congruence rules, as well as the {\nrule{\xi}}-rule. We adopt the
following informal conventions:
\begin{itemize}
\item never reduce the body of a lambda abstraction,
\item never reduce the argument of a function (except for primitive
  functions such as $\succt$ and $\predt$),
\item never reduce the ``then'' or ``else'' part of an if-then-else
  statement,
\item never reduce a term inside a pair.
\end{itemize}

Of course, the terms that these rules prevent from being reduced can
nevertheless become subject to reduction later: the body of a lambda
abstraction and the argument of a function can be reduced after a
$\beta$-reduction causes the $\lam$ to disappear and the argument
to be substituted in the body. The ``then'' or ``else'' parts of an
if-then-else term can be reduced after the ``if'' part evaluates to
true or false. And the terms inside a pair can be reduced after the
pair has been broken up by a projection.

An important technical notion is that of a {\em value}, which is a
term that represents the result of a computation and cannot be reduced
further. Values are given as follows:
\[ \mbox{Values:}\ssep V,W \bnf \truet \bor \falset \bor \zerot \bor
   \succt(V) \bor \unit \bor \pair{M,N} \bor \lamabs{x}{A}.M
\]
The transition rules for the small-step operational semantics of PCF
are shown in Table~\ref{tab-pcf-small}.
\begin{table*}[t]\def\nl{\\[1.8ex]}
\[
\begin{array}{c}
  \deriv{M\to N}{\predt(M)\to\predt(N)} \nl
  \deriv{}{\predt(\zerot)\to\zerot} \nl
  \deriv{}{\predt(\succt(V))\to V} \nl
  \deriv{M\to N}{\iszerot(M)\to\iszerot(N)} \nl
  \deriv{}{\iszerot(\zerot)\to\truet} \nl
  \deriv{}{\iszerot(\succt(V))\to\falset} \nl
  \deriv{M\to N}{\succt(M)\to\succt(N)} \nl
  \deriv{M\to N}{MP\to NP} \nl
  \deriv{}{(\lamabs{x}{A}.M)N\to \subst{M}{N}{x}}
\end{array}
\sep
\begin{array}{c}
  \deriv{M\to M'}{\proj{i} M\to \proj{i}M'} \nl
  \deriv{}{\proj1\pair{M,N}\to M} \nl
  \deriv{}{\proj2\pair{M,N}\to N} \nl
  \deriv{M:1,\ssep M\neq\unit}{M\to\unit} \nl
  \deriv{M\to M'}{\ite{M}{N}{P}\to\ite{M'}{N}{P}} \nl
  \deriv{}{\ite{\truet}{N}{P}\to N} \nl
  \deriv{}{\ite{\falset}{N}{P}\to P} \nl
  \deriv{}{\Y(M)\to M(\Y(M))} \nl
\end{array}
\]
\caption{Small-step operational semantics of PCF}
\label{tab-pcf-small}
\end{table*}

We write $M\to N$ if $M$ reduces to $N$ by these rules. We write
$M\not\to$ if there does not exist $N$ such that $M\to N$. The first
two important technical properties of small-step reduction are
summarized in the following lemma.

\begin{lemma}\label{lem-pcf-lemma1}
  \begin{enumerate}
  \item {\em Values are normal forms.} If $V$ is a value, then $V\not\to$.
  \item {\em Evaluation is deterministic.} If $M\to N$ and $M\to N'$,
    then $N\syntaxeq N'$.
  \end{enumerate}
\end{lemma}

Another important property is subject reduction: a well-typed term
reduces only to another well-typed term of the same type.

\begin{lemma}[Subject Reduction]
  If $\Gamma\tj M:A$ and $M\to N$, then $\Gamma\tj N:A$. 
\end{lemma}

Next, we want to prove that the evaluation of a well-typed term does
not get ``stuck''. If $M$ is some term such that $M\not\to$, but $M$
is not a value, then we regard this as an error, and we also write
$M\to\errort$. Examples of such terms are $\proj1(\lam x.M)$ and
$\pair{M,N}P$. The following lemma shows that well-typed closed terms
cannot lead to such errors.

\begin{lemma}[Progress]\label{lem-pcf-progress}
  If $M$ is a closed, well-typed term, then either $M$ is a value, or
  else there exists $N$ such that $M\to N$.
\end{lemma}

The Progress Lemma is very important, because it implies that a
well-typed term cannot ``go wrong''. It guarantees that a well-typed
term will either evaluate to a value in finitely many steps, or else
it will reduce infinitely and thus not terminate. But a well-typed
term can never generate an error. In programming language terms, a
term that type-checks at {\em compile-time} cannot generate an error
at {\em run-time}.

To express this idea formally, let us write $M\to^* N$ in the usual
way if $M$ reduces to $N$ in zero or more steps, and let us write
$M\to^*\errort$ if $M$ reduces in zero or more steps to an error.

\begin{proposition}[Safety]\label{prop-pcf-safety}
  If $M$ is a closed, well-typed term, then $M\not\to^*\errort$.
\end{proposition}

\begin{exercise}
  Prove Lemmas~\ref{lem-pcf-lemma1}--\ref{lem-pcf-progress} and
  Proposition~\ref{prop-pcf-safety}.
\end{exercise}

\subsection{Big-step semantics}

In the small-step semantics, if $M\to^* V$, we say that $M$ {\em
  evaluates to} $V$. Note that by determinacy, for every $M$, there
exists at most one $V$ such that $M\to^* V$. 

It is also possible to axiomatize the relation ``$M$ evaluates to
$V$'' directly. This is known as the big-step semantics. Here, we
write $M\evto V$ if $M$ evaluates to $V$. The axioms for the big-step
semantics are shown in Table~\ref{tab-pcf-big}.
\begin{table*}[t]
\[
\begin{array}{c}
  \deriv{}{\truet\evto\truet} \nl
  \deriv{}{\falset\evto\falset} \nl
  \deriv{}{\zerot\evto\zerot} \nl
  \deriv{}{\pair{M,N}\evto\pair{M,N}} \nl
  \deriv{}{\lamabs{x}{A}.M\evto\lamabs{x}{A}.M} \nl
  \deriv{M\evto\zerot}{\predt(M)\evto\zerot} \nl
  \deriv{M\evto\succt(V)}{\predt(M)\evto V} \nl
  \deriv{M\evto\zerot}{\iszerot(M)\evto\truet} \nl
  \deriv{M\evto\succt(V)}{\iszerot(M)\evto\falset} \nl
\end{array}
\sep
\begin{array}{c}
  \deriv{M\evto V}{\succt(M)\evto\succt(V)} \nl
  \deriv{M\evto\lamabs{x}{A}.M'\sep\subst{M'}{N}{x}\evto V}{MN\evto V} \nl
  \deriv{M\evto\pair{M_1,M_2}\sep M_1\evto V}{\proj1 M\evto V} \nl
  \deriv{M\evto\pair{M_1,M_2}\sep M_2\evto V}{\proj2 M\evto V} \nl
  \deriv{M:1}{M\evto\unit} \nl
  \deriv{M\evto\truet\sep N\evto V}{\ite{M}{N}{P}\evto V} \nl
  \deriv{M\evto\falset\sep P\evto V}{\ite{M}{N}{P}\evto V} \nl
  \deriv{M(\Y(M))\evto V}{\Y(M)\evto V}
\end{array}
\]
\caption{Big-step operational semantics of PCF}
\label{tab-pcf-big}
\end{table*}

The big-step semantics satisfies properties similar to those of the
small-step semantics.

\begin{lemma}
  \begin{enumerate}
  \item {\em Values.} For all values $V$, we have $V\evto V$.
  \item {\em Determinacy.} If $M\evto V$ and $M\evto V'$, then
    $V\syntaxeq V'$.
  \item {\em Subject Reduction.} If $\Gamma\tj M:A$ and $M\evto V$,
    then $\Gamma\tj V:A$.
  \end{enumerate}
\end{lemma}

The analogues of the Progress and Safety properties cannot be as
easily stated for big-step reduction, because we cannot easily talk
about a single reduction step or about infinite reduction sequences. 
However, some comfort can be taken in the fact that the big-step
semantics and small-step semantics coincide:

\begin{proposition}\label{prop-big-small}
  $M\to^* V$ iff $M\evto V$.
\end{proposition}
 
\subsection{Operational equivalence}

Informally, two terms $M$ and $N$ will be called operationally
equivalent if $M$ and $N$ are interchangeable as part of any larger
program, without changing the observable behavior of the program. This
notion of equivalence is also often called observational equivalence,
to emphasize the fact that it concentrates on observable properties of
terms.

What is an observable behavior of a program? Normally, what we observe
about a program is its output, such as the characters it prints to a
terminal. Since any such characters can be converted in principle to
natural numbers, we take the point of view that the observable
behavior of a program is a natural number that it evaluates to. 
Similarly, if a program computes a boolean, we regard the boolean
value as observable. However, we do not regard abstract values, such
as functions, as being directly observable, on the grounds that a
function cannot be observed until we supply it some arguments and
observe the result.

\begin{definition}
  An {\em observable type} is either $\boolt$ or $\natt$. A {\em
  result} is a closed value of observable type. Thus, a result is
  either $\truet$, $\falset$, or $\numn$.  A {\em program} is a closed
  term of observable type. 

  A {\em context} is a term with a hole, written $C[-]$. Formally, the
  class of contexts is defined by a BNF:
  \[ \begin{array}{@{}lll}
    C[-] &\bnf&
    [-] \bor x \bor C[-]N \bor MC[-] \bor \lamabs{x}{A}.C[-]
    \bor \ldots
  \end{array}
  \]
  and so on, extending through all the cases in the definition of a
  PCF term. 
\end{definition}

Well-typed contexts are defined in the same way as well-typed terms,
where it is understood that the hole also has a type. The free
variables of a context are defined in the same way as for terms.
Moreover, we define the {\em captured variables} of a context to be
those bound variables whose scope includes the hole.  So for instance,
in the context $(\lam x.[-])(\lam y.z)$, the variable $x$ is captured,
the variable $z$ is free, and $y$ is neither free nor captured. 

If $C[-]$ is a context and $M$ is a term of the appropriate type, we
write $C[M]$ for the result of replacing the hole in the context
$C[-]$ by $M$. Here, we do not $\alpha$-rename any bound variables, so
that we allow free variables of $M$ to be captured by $C[-]$.

We are now ready to state the definition of operational equivalence.

\begin{definition}
  Two terms $M,N$ are {\em operationally equivalent}, in symbols
  $M\eqop N$, if for all closed and closing context $C[-]$ of
  observable type and all values $V$,
  \[ C[M]\evto V \iff C[N]\evto V.
  \]
\end{definition}

Here, by a {\em closing} context we mean that $C[-]$ should capture
all the free variables of $M$ and $N$. This is equivalent to requiring
that $C[M]$ and $C[N]$ are closed terms of observable types, i.e.,
programs. Thus, two terms are equivalent if they can be used
interchangeably in any program.

\subsection{Operational approximation}

As a refinement of operational equivalence, we can also define a
notion of operational approximation: We say that $M$ {\em
  operationally approximates} $N$, in symbols $M\sqleqop N$, if for
all closed and closing contexts $C[-]$ of observable type and all
values $V$,
\[ C[M]\evto V \imp C[N]\evto V.
\]
Note that this definition includes the case where $C[M]$ diverges, but
$C[N]$ converges, for some $N$. This formalizes the notion that $N$ is
``more defined'' than $M$. Clearly, we have $M\eqop N$ iff $M\sqleqop
N$ and $N\sqleqop M$. Thus, we get a partial order $\sqleqop$ on the
set of all terms of a given type, modulo operational equivalence.
Also, this partial order has a least element, namely if we let
$\Omega=\Y(\lam x.x)$, then $\Omega\sqleqop N$ for any term $N$ of the
appropriate type.

Note that, in general, $\sqleqop$ is not a complete partial order, due
to missing limits of $\omega$-chains.

\subsection{Discussion of operational equivalence}

Operational equivalence is a very useful concept for reasoning about
programs, and particularly for reasoning about program fragments. If
$M$ and $N$ are operationally equivalent, then we know that we can
replace $M$ by $N$ in any program without affecting its behavior.  For
example, $M$ could be a slow, but simple subroutine for sorting a
list. The term $N$ could be a replacement that runs much faster. If we can
prove $M$ and $N$ to be operationally equivalent, then this means we
can safely use the faster routine instead of the slower one. 

Another example are compiler optimizations. Many compilers will try to
optimize the code that they produce, to eliminate useless
instructions, to avoid duplicate calculations, etc. Such an
optimization often means replacing a piece of code $M$ by another
piece of code $N$, without necessarily knowing much about the context
in which $M$ is used. Such a replacement is safe if $M$ and $N$ are
operationally equivalent. 

On the other hand, operational equivalence is a somewhat problematic
notion. The problem is that the concept is not stable under adding new
language features. It can happen that two terms, $M$ and $N$, are
operationally equivalent, but when a new feature is added to the
language, they become nonequivalent, {\em even if $M$ and $N$ do not
  use the new feature}. The reason is the operational equivalence is
defined in terms of contexts. Adding new features to a language also
means that there will be new contexts, and these new contexts might be
able to distinguish $M$ and $N$. 

This can be a problem in practice. Certain compiler optimizations
might be sound for a sequential language, but might become unsound if
new language features are added. Code that used to be correct might
suddenly become incorrect if used in a richer environment. For
example, many programs and library functions in C assume that they are
executed in a single-threaded environment. If this code is ported to a
multi-threaded environment, it often turns out to be no longer
correct, and in many cases it must be re-written from scratch. 

\subsection{Operational equivalence and parallel or}

Let us now look at a concrete example in PCF. We say that a term $\POR$
implements the {\em parallel or} function if it has the following
behavior:
\[ \begin{array} {llll}
  \POR\truet P &\to& \truet, &\mbox{for all $P$} \\
  \POR N\truet &\to& \truet, &\mbox{for all $N$} \\
  \POR\falset\falset &\to& \falset.
\end{array}
\]
Note that this in particular implies $\POR\truet\Omega=\truet$ and
$\POR\Omega\truet=\truet$, where $\Omega$ is some divergent term. It should
be clear why $\POR$ is called the ``parallel'' or: the only way to
achieve such behavior is to evaluate both its arguments in parallel,
and to stop as soon as one argument evaluates to $\truet$ or both
evaluate to $\falset$.

\begin{proposition}
  $\POR$ is not definable in PCF.
\end{proposition}

We do not give the proof of this fact, but the idea is relatively
simple: one proves by induction that every PCF context $C[-,-]$ with
two holes has the following property: either, there exists a term $N$
such that $C[M,M']=N$ for all $M,M'$ (i.e., the context does not look
at $M,M'$ at all), or else, either $C[\Omega,M]$ diverges for all $M$,
or $C[M,\Omega]$ diverges for all $M$. Here, again, $\Omega$ is some
divergent term such as $\Y(\lam x.x)$. 

Although $\POR$ is not definable in PCF, we can define the following
term, called the {\em POR-tester}:
\[ \begin{array}{l@{}l}
  \nm{POR-test} = \lam x. 
  &            \nm{if} x \truet\Omega \nm{then} \\
  &\sep            \nm{if} x\Omega\truet\nm{then}\\
  &\sep\sep            \nm{if} x\falset\falset\nm{then}\Omega\\
  &\sep\sep            \nm{else}\truet\\
  &\sep            \nm{else}\Omega \\
  &            \nm{else}\Omega
\end{array}
\]

The POR-tester has the property that $\nm{POR-test} M=\truet$ if $M$
implements the parallel or function, and in all other cases
$\nm{POR-test} M$ diverges. In particular, since parallel or is not
definable in PCF, we have that $\nm{POR-test}M$ diverges, for all PCF
terms $M$. Thus, when applied to any PCF term, $\nm{POR-test}$ behaves
precisely as the function $\lam x.\Omega$ does. One can make this into
a rigorous argument that shows that $\nm{POR-test}$ and $\lam
x.\Omega$ are operationally equivalent:
\[        \nm{POR-test}  \eqop \lam x.\Omega \sep\mbox{(in PCF)}.
\]

Now, suppose we want to define an extension of PCF called {\em
  parallel PCF}. It is defined in exactly the same way as PCF, except
that we add a new primitive function $\POR$, and small-step
reduction rules
\[ \begin{array}{c}
  \deriv{M\to M'\sep N\to N'}{\POR MN\to\POR M'N'} \nl
  \deriv{}{\POR\truet N\to\truet} \nl
  \deriv{}{\POR M\truet\to\truet} \nl
  \deriv{}{\POR\falset\falset\to\falset}
\end{array}
\]
Parallel PCF enjoys many of the same properties as PCF, for instance,
Lemmas~\ref{lem-pcf-lemma1}--\ref{lem-pcf-progress} and
Proposition~\ref{prop-pcf-safety} continue to hold for it.

But notice that 
\[   \nm{POR-test}  \not\eqop \lam x.\Omega \sep\mbox{(in parallel PCF)}.
\]
This is because the context $C[-]=[-]\POR$ distinguishes the two
terms: clearly, $C[\nm{POR-test}]\evto\truet$, whereas $C[\lam
x.\Omega]$ diverges.

\section{Complete partial orders}

\subsection{Why are sets not enough, in general?}

As we have seen in Section~\ref{sec-set-semantics}, the interpretation
of types as plain sets is quite sufficient for the simply-typed lambda
calculus. However, it is insufficient for a language such as PCF.
Specifically, the problem is the fixed point operator $\Y:(A\to A)\to A$.
It is clear that there are many functions $f:A\to A$ from a set $A$ to
itself that do not have a fixed point; thus, there is no chance we are
going to find an interpretation for a fixed point operator in the simple
set-theoretic model.

On the other hand, if $A$ and $B$ are types, there are generally many
functions $f:\semm{A}\to\semm{B}$ in the set-theoretic model that are
not definable by lambda terms. For instance, if $\semm{A}$ and
$\semm{B}$ are infinite sets, then there are uncountably many
functions $f:\semm{A}\to\semm{B}$; however, there are only countably
many lambda terms, and thus there are necessarily going to be
functions that are not the denotation of any lambda term.

The idea is to put additional structure on the sets that interpret
types, and to require functions to preserve that structure. This is
going to cut down the size of the function spaces, decreasing the
``slack'' between the functions definable in the lambda calculus and
the functions that exist in the model, and simultaneously increasing
the chances that additional structure, such as fixed point operators,
might exist in the model.

Complete partial orders are one such structure that is commonly used
for this purpose. The method is originally due to Dana Scott.

\subsection{Complete partial orders}

\begin{definition}
  A {\em partially ordered set} or {\em poset} is a set $X$ together
  with a binary relation $\sqleq$ satisfying
  \begin{itemize}
  \item {\em reflexivity:} for all $x\in X$, $x\sqleq x$,
  \item {\em antisymmetry:} for all $x,y\in X$, $x\sqleq y$ and $y\sqleq
    x$ implies $x=y$,
  \item {\em transitivity:} for all $x,y,z\in X$, $x\sqleq y$ and
    $y\sqleq z$ implies $x\sqleq z$.
  \end{itemize}
\end{definition}

The concept of a partial order differs from a total order in that we
do not require that for any $x$ and $y$, either $x\sqleq y$ or $y\sqleq
x$. Thus, in a partially ordered set it is permissible to have
incomparable elements.

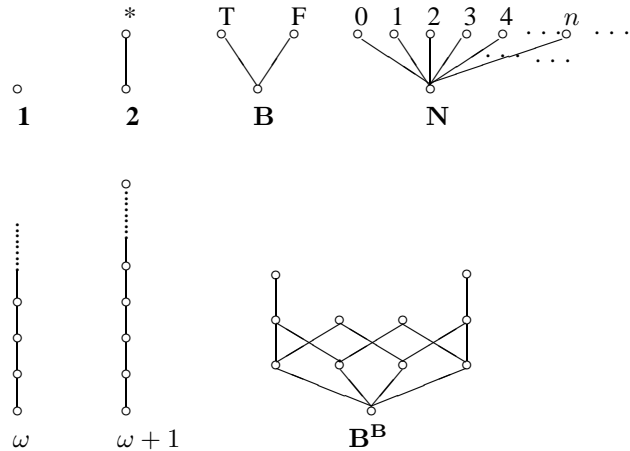
\begin{figure}
\begin{center}
\input{posets}
\end{center}
\caption{Some posets}\label{fig-posets}
\end{figure}

We can often visualize posets, particularly finite ones, by drawing
their line diagrams as in Figure~\ref{fig-posets}. In these diagrams,
we put one circle for each element of $X$, and we draw an edge from
$x$ upward to $y$ if $x\sqleq y$ and there is no $z$ with $x\sqleq
z\sqleq y$. Such line diagrams are also known as {\em Hasse diagrams}.

The idea behind using a partial order to denote computational values
is that $x\sqleq y$ means that $x$ is {\em less defined than} $y$. For
instance, if a certain term diverges, then its denotation will be less
defined than, or below that of a term that has a definite value.
Similarly, a function is more defined than another if it converges on
more inputs.

Another important idea in using posets for modeling computational
value is that of {\em approximation}. We can think of some infinite
computational object (such as, an infinite stream), to be a limit of
successive finite approximations (such as, longer and longer finite
streams). Thus we also read $x\sqleq y$ as $x$ {\em approximates} $y$.
A complete partial order is a poset in which every countable chain of
increasing elements approximates something.

\begin{definition}
  Let $X$ be a poset and let $A\seq X$ be a subset. We say that $x\in
  X$ is an {\em upper bound} for $A$ if $a\sqleq x$ for all $a\in A$.
  We say that $x$ is a {\em least upper bound} for $A$ if $x$ is an
  upper bound, and whenever $y$ is also an upper bound, then $x\sqleq
  y$.
\end{definition}

\begin{definition}
  An {\em $\omega$-chain} in a poset $X$ is a sequence of elements
  $x_0,x_1,x_2,\ldots$ such that
  \[ x_0 \sqleq x_1 \sqleq x_2 \sqleq \ldots \]
\end{definition}

\begin{definition}
  A {\em complete partial order (cpo)} is a poset such that every
  $\omega$-chain of elements has a least upper bound.
\end{definition}

If $x_0,x_1,x_2,\ldots$ is an $\omega$-chain of elements in a cpo, we
write $\dirsup_{i\in\N}x_i$ for the least upper bound. We also call
the least upper bound the {\em limit} of the $\omega$-chain.

Not every poset is a cpo. In Figure~\ref{fig-posets}, the poset
labeled $\omega$ is not a cpo, because the evident $\omega$-chain does
not have a least upper bound (in fact, it has no upper bound at all).
The other posets shown in Figure~\ref{fig-posets} are cpo's. 

\subsection{Properties of limits}

\begin{proposition}\label{prop-limit-properties}
  \begin{enumerate}
  \item {\em Monotonicity.}  Suppose $\s{x_i}_i$ and $\s{y_i}_i$ are
    $\omega$-chains in a cpo $C$, such that $x_i\sqleq y_i$ for all
    $i$. Then
    \[ \dirsup_i x_i \sqleq\dirsup_i y_i.
    \]
  \item {\em Exchange.} Suppose $\s{x_{ij}}_{i,j\in\N}$ is a doubly
    monotone double sequence of elements of a cpo $C$, i.e., whenever
    $i\leq i'$ and $j\leq j'$, then $x_{ij}\sqleq x_{i'j'}$. Then
    \[ \dirsup_{i\in\N}\dirsup_{j\in\N} x_{ij} = 
    \dirsup_{j\in\N}\dirsup_{i\in\N} x_{ij} =
    \dirsup_{k\in\N} x_{kk}.
    \]
    In particular, all limits shown are well-defined.
  \end{enumerate}
\end{proposition}

\begin{exercise}
  Prove Proposition~\ref{prop-limit-properties}.
\end{exercise}

\subsection{Continuous functions}

If we model data types as cpo's, it is natural to model algorithms as
functions from cpo's to cpo's. These functions are subject to two
constraints: they have to be monotone and continuous.

\begin{definition}
  A function $f:C\to D$ between posets $C$ and $D$ is said to be {\em
    monotone} if for all $x,y\in C$, 
  \[ x\sqleq y \sep\imp\sep f(x)\sqleq f(y).
  \]
  A function $f:C\to D$ between cpo's $C$ and $D$ is said to be {\em
    continuous} if it is monotone and it preserves least upper bounds
  of $\omega$-chains, i.e., for all $\omega$-chains $\s{x_i}_{i\in\N}$
  in $C$,
  \[ f(\dirsup_{i\in\N}x_i) = \dirsup_{i\in\N}f(x_i).
  \]
\end{definition}

The intuitive explanation for the monotonicity requirement is that
information is ``positive'': more information in the input cannot lead
to less information in the output of an algorithm. The intuitive
explanation for the continuity requirement is that any particular
output of an algorithm can only depend on a finite amount of input.

\subsection{Pointed cpo's and strict functions}

\begin{definition}
  A cpo is said to be {\em pointed} if it has a least element. The
  least element is usually denoted $\bot$ and pronounced ``bottom''.
  All cpo's shown in Figure~\ref{fig-posets} are pointed.

  A continuous function between pointed cpo's is said to be {\em
  strict} if it preserves the bottom element.
\end{definition}

\subsection{Products and function spaces}

If $C$ and $D$ are cpo's, then their {\em cartesian product} $C\times
D$ is also a cpo, with the pointwise order given by
$\rpair{x,y}\sqleq\rpair{x',y'}$ iff $x\sqleq x'$ and $y\sqleq y'$.
Least upper bounds are also given pointwise, thus
\[ \dirsup_i\rpair{x_i,y_i} = \rpair{\dirsup_i x_i,\dirsup_i y_i}.
\]

\begin{proposition}\label{prop-cpo-products}
  The first and second projections, $\proj1:C\times D\to C$ and
  $\proj2:C\times D\to D$, are continuous functions. Moreover, if
  $f:E\to C$ and $g:E\to D$ are continuous functions, then so is the
  function $h:E\to C\times D$ given by $h(z)=\rpair{f(z),g(z)}$.
\end{proposition}

If $C$ and $D$ are cpo's, then the set of continuous functions $f:C\to
D$ forms a cpo, denoted $D^C$. The order is given pointwise: given two
functions $f,g:C\to D$, we say that
\[ f\sqleq g \sep\mbox{iff}\sep \mbox{for all $x\in C$, $f(x)\sqleq
  g(x)$}.
\]

\begin{proposition}
  The set $D^C$ of continuous functions from $C$ to $D$, together with
  the order just defined, is a complete partial order.
\end{proposition}

\begin{proof}
  Clearly the set $D^C$ is partially ordered. What we must show is
  that least upper bounds of $\omega$-chains exist.  Given an
  $\omega$-chain $f_0,f_1,\ldots$ in $D^C$, we define $g\in D^C$
  to be the pointwise limit, i.e.,
  \[               g(x) = \dirsup_{i\in\N}f_i(x),
  \]
  for all $x\in C$. Note that $\s{f_i(x)}_i$ does indeed form an
  $\omega$-chain in $C$, so that $g$ is a well-defined function. We
  claim that $g$ is the least upper bound of $\s{f_i}_i$. First we
  need to show that $g$ is indeed an element of $D^C$. To see that $g$
  is monotone, we use Proposition~\ref{prop-limit-properties}(1) and
  calculate, for any $x\sqleq y\in C$,
  \[   g(x) = \dirsup_{i\in\N}f_i(x) \sqleq \dirsup_{i\in\N}f_i(y)
  = g(y). \] To see that $g$ is continuous, we use
  Proposition~\ref{prop-limit-properties}(2) and calculate, for any
  $\omega$-chain $x_0,x_1,\ldots$ in $C$,
  \[   g(\dirsup_j x_j) = \dirsup_i \dirsup_j f_i(x_j) 
  = \dirsup_j \dirsup_i f_i(x_j) = \dirsup_j g(x_j).
  \]
  Finally, we must show that $g$ is the least upper bound of the
  $\s{f_i}_i$. Clearly, $f_i\sqleq g$ for all $i$, so that $g$ is an
  upper bound. Now suppose $h\in D^C$ is any other upper bound of
  $\s{f_i}$. Then for all $x$, $f_i(x)\sqleq h(x)$. Since $g(x)$ was
  defined to be the least upper bound of $\s{f_i(x)}_i$, we then have
  $g(x)\sqleq h(x)$. Since this holds for all $x$, we have $g\sqleq
  h$. Thus $g$ is indeed the least upper bound.\eot
\end{proof}

\begin{exercise}
  Recall the cpo $\Bb$ from Figure~\ref{fig-posets}.  The cpo
  $\Bb^{\Bb}$ is also shown in Figure~\ref{fig-posets}. Its 11
  elements correspond to the 11 continuous functions from $\Bb$ to
  $\Bb$. Label the elements of $\Bb^{\Bb}$ with the functions they
  correspond to.
\end{exercise}

\begin{proposition}\label{prop-cpo-app}
  The application function $D^C\times C\to D$, which maps
  $\rpair{f,x}$ to $f(x)$, is continuous.
\end{proposition}

\begin{proposition}\label{prop-cpo-curry}
  Continuous functions can be continuously curried and uncurried. In
  other words, if $f:C\times D\to E$ is a continuous function, then
  $f^*:C\to E^D$, defined by $f^*(x)(y) = f(x,y)$, is well-defined and
  continuous. Conversely, if $g:C\to E^D$ is a continuous function, then
  $g_*:C\times D\to E$, defined by $g_*(x,y) = g(x)(y)$, is well-defined
  and continuous. Moreover, $(f^*)_*=f$ and $(g_*)^*=g$. 
\end{proposition}

\subsection{The interpretation of the simply-typed lambda calculus in
  complete partial orders}\label{subsec-cpo-interp}

The interpretation of the simply-typed lambda calculus in cpo's
resembles the set-theoretic interpretation, except that types are
interpreted by cpo's instead of sets, and typing judgments are
interpreted as continuous functions.

For each basic type $\iota$, assume that we have chosen a pointed cpo
$S_{\iota}$. We can then associate a pointed cpo $\semm{A}$ to each
type $A$ recursively:
\[ \begin{array}{lll}
  \semm{\iota} &=& S_{\iota} \\
  \semm{A\to B} &=& \semm{B}^{\semm{A}} \\
  \semm{A\times B} &=& \semm{A}\times\semm{B} \\
  \semm{1} &=& {\bf 1}
\end{array}
\]
Typing judgments are now interpreted as continuous functions
\[     \semm{A_1}\times\ldots\times\semm{A_n} \to \semm{B}
\]
in precisely the same way as they were defined for the set-theoretic
interpretation. The only thing we need to check, at every step, is
that the function defined is indeed continuous. For variables, this
follows from the fact that projections of cartesian products are
continuous (Proposition~\ref{prop-cpo-products}). For applications, we
use the fact that the application function of cpo's is continuous
(Proposition~\ref{prop-cpo-app}), and for lambda-abstractions, we use
the fact that currying is a well-defined, continuous operation
(Proposition~\ref{prop-cpo-curry}). Finally, the continuity of the
maps associated with products and projections follows from
Proposition~\ref{prop-cpo-products}.

\begin{proposition}[Soundness and Completeness]
  The interpretation of the simply-typed lambda calculus in pointed
  cpo's is sound and complete with respect to the lambda-$\beta\eta$
  calculus.
\end{proposition}

\subsection{Cpo's and fixed points}

One of the reasons, mentioned in the introduction to this section, for
using cpo's instead of sets for the interpretation of the simply-typed
lambda calculus is that cpo's admit fixed point, and thus they can be
used to interpret an extension of the lambda calculus with a fixed point
operator. 

\begin{proposition}
  Let $C$ be a pointed cpo and let $f:C\to C$ be a continuous
  function. Then $f$ has a least fixed point.
\end{proposition}

\begin{proof}
  Define $x_0=\bot$ and $x_{i+1} = f(x_i)$, for all $i\in\N$. The
  resulting sequence $\s{x_i}_i$ is an $\omega$-chain, because clearly
  $x_0\sqleq x_1$ (since $x_0$ is the least element), and if
  $x_{i}\sqleq x_{i+1}$, then $f(x_{i})\sqleq f(x_{i+1})$ by
  monotonicity, hence $x_{i+1}\sqleq x_{i+2}$. It follows by induction
  that $x_{i}\sqleq x_{i+1}$. Let $x=\dirsup_i x_i$ be the limit of
  this $\omega$-chain. Then using continuity of $f$, we have
  \[ f(x) = f(\dirsup_i x_i) = \dirsup_i f(x_i) = \dirsup_i x_{i+1} =
  x.
  \]
  To prove that it is the least fixed point, let $y$ be any other
  fixed point, i.e., let $f(y)=y$. We prove by induction that for all
  $i$, $x_i\sqleq y$. For $i=0$ this is trivial because $x_0=\bot$.
  Assume $x_i\sqleq y$, then $x_{i+1}=f(x_{i})\sqleq f(y)=y$. It
  follows that $y$ is an upper bound for $\s{x_i}_i$. Since $x$ is, by
  definition, the least upper bound, we have $x\sqleq y$. Since $y$
  was arbitrary, $x$ is below any fixed point, hence $x$ is the least
  fixed point of $f$.\eot
\end{proof}

If $f:C\to C$ is any continuous function, let us write $\fix{f}$ for
its least fixed point. We claim that $\fix{f}$ depends continuously on
$f$, i.e., that $\dagger:C^C\to C$ defines a continuous function. 

\begin{proposition}\label{prop-dagger}
  The function $\dagger:C^C\to C$, which assigns to each continuous
  function $f\in C^C$ its least fixed point $\fix{f}\in C$, is
  continuous. 
\end{proposition}

\begin{exercise}
  Prove Proposition~\ref{prop-dagger}.
\end{exercise}

Thus, if we add to the simply-typed lambda calculus a family of
fixed point operators $Y_A:(A\to A)\to A$, the resulting extended lambda
calculus can then be interpreted in cpo's by letting
\[  \semm{Y_A} = \dagger:\semm{A}^{\semm{A}} \to \semm{A}.
\]

\subsection{Example: Streams}

Consider streams of characters from some alphabet $A$. Let
$A^{\leq\omega}$ be the set of finite or infinite sequences of
characters. We order $A$ by the {\em prefix ordering}: if $s$ and $t$
are (finite or infinite) sequences, we say $s\sqleq t$ if $s$ is a
prefix of $t$, i.e., if there exists a sequence $s'$ such that
$t=ss'$. Note that if $s\sqleq t$ and $s$ is an infinite sequence,
then necessarily $s=t$, i.e., the infinite sequences are the maximal
elements with respect to this order.

\begin{exercise}
  Prove that the set $A^{\leq\omega}$ forms a cpo under the prefix
  ordering.
\end{exercise}

\begin{exercise}
  Consider an automaton that reads characters from an input stream
  and writes characters to an output stream. For each input character
  read, it can write zero, one, or more output characters. Discuss how
  such an automaton gives rise to a continuous function from
  $A^{\leq\omega}\to A^{\leq\omega}$. In particular, explain the
  meaning of monotonicity and continuity in this context. Give some
  examples. 
\end{exercise}

\section{Denotational semantics of PCF}

The denotational semantics of PCF is defined in terms of cpo's. It
extends the cpo semantics of the simply-typed lambda calculus. Again,
we assign a cpo $\semm{A}$ to each PCF type $A$, and a continuous
function 
\[ \semm{\Gamma\tj M:B} : \semm{\Gamma} \to \semm{B}
\]
to every PCF typing judgment. The interpretation is defined in
precisely the same way as for the simply-typed lambda calculus. The
interpretation for the PCF-specific terms is shown in
Table~\ref{tab-pcf-denot}. Recall that $\Bb$ and $\Nn$ are the cpo's of
lifted booleans and lifted natural numbers, respectively, as shown in
Figure~\ref{fig-posets}.
\begin{table}
\[ \begin{array}{llll}
  \mbox{Types:} 
  &  \semm{\boolt} &=& \Bb \\
  &  \semm{\natt}  &=& \Nn  \nl
  \mbox{Terms:} 
  & \semm{\truet} &=& T\in\Bb \\
  & \semm{\falset} &=& F\in\Bb \\
  & \semm{\zerot} &=& 0\in\Nn \nl
  & \semm{\succt(M)} &=& \left\{\begin{array}{ll}
      \bot & \mbox{if $\semm{M} = \bot$,} \\
      n+1  & \mbox{if $\semm{M} = n$}
    \end{array}\right. \\\\
  & \semm{\predt(M)} &=& \left\{\begin{array}{ll}
      \bot & \mbox{if $\semm{M} = \bot$,} \\
      0    & \mbox{if $\semm{M} = 0$,} \\
      n    & \mbox{if $\semm{M} = n+1$}
    \end{array}\right. \\\\
  & \semm{\iszerot(M)} &=& \left\{\begin{array}{ll}
      \bot     & \mbox{if $\semm{M} = \bot$,} \\
      \truet   & \mbox{if $\semm{M} = 0$,} \\
      \falset  & \mbox{if $\semm{M} = n+1$}
    \end{array}\right. \\\\
  & \semm{\ite{M}{N}{P}} &=& \left\{\begin{array}{ll}
      \bot        & \mbox{if $\semm{M} = \bot$,} \\
      \semm{N}    & \mbox{if $\semm{M} = \falset$,} \\
      \semm{P}    & \mbox{if $\semm{M} = \truet$,} \\
    \end{array}\right. \\\\
  & \semm{\Y(M)} &=& \semm{M}^{\dagger}
\end{array}
\]
\caption{Cpo semantics of PCF}\label{tab-pcf-denot}
\end{table}

\begin{definition}
  Two PCF terms $M$ and $N$ of equal types are denotationally
  equivalent, in symbols $M\eqden N$, if $\semm{M}=\semm{N}$. 
  We also write $M\sqleqden N$ if $\semm{M}\sqleq\semm{N}$.
\end{definition}

\subsection{Soundness and adequacy}

We have now defined the three notions of equivalence on terms:
$\eqax$, $\eqop$, and $\eqden$. In general, one does not expect the
three equivalences to coincide. For example, any two divergent terms
are operationally equivalent, but there is no reason why they should
be axiomatically equivalent. Also, the POR-tester and the term $\lam
x.\Omega$ are operationally equivalent in PCF, but they are not
denotationally equivalent (since a function representing POR clearly
exists in the cpo semantics). For general terms $M$ and $N$, one has
the following property:

\begin{theorem}[Soundness]
  For PCF terms $M$ and $N$, the following implications hold:
  \[ M\eqax N \sep\imp\sep M\eqden N  \sep\imp\sep  M\eqop N.
  \]
\end{theorem}

Soundness is a very useful property, because $M\eqax N$ is in general
easier to prove than $M\eqden N$, and $M\eqden N$ is in turns easier
to prove than $M\eqop N$. Thus, soundness gives us a powerful proof
method: to prove that two terms are operationally equivalent, it
suffices to show that they are equivalent in the cpo semantics (if
they are), or even that they are axiomatically equivalent. 

As the above examples show, the converse implications are not in
general true. However, the converse implications hold if the terms $M$
and $N$ are closed and of observable type, and if $N$ is a value. This
property is called computational adequacy. Recall that a program is a
closed term of observable type, and a result is a closed value of
observable type. 

\begin{theorem}[Computational Adequacy]
  If $M$ is a program and $V$ is a result, then
  \[ M\eqax V \sep\iff\sep M\eqden V  \sep\iff\sep  M\eqop V.
  \]
\end{theorem}

\begin{proof}
  First note that the small-step semantics is contained in the
  axiomatic semantics, i.e., if $M\to N$, then $M\eqax N$. This is
  easily shown by induction on derivations of $M\to N$. 
  
  To prove the theorem, by soundness, it suffices to show that $M\eqop
  V$ implies $M\eqax V$. So assume $M\eqop V$. Since $V\evto V$ and
  $V$ is of observable type, it follows that $M\evto V$. Therefore
  $M\to^* V$ by Proposition~\ref{prop-big-small}. But this already
  implies $M\eqax V$, and we are done.\eot
\end{proof}

\subsection{Full abstraction}

We have already seen that the operational and denotational semantics
do not coincide for PCF, i.e., there are some terms such that $M\eqop
N$ but $M\not\eqden N$. Examples of such terms are $\nm{POR-test}$ and
$\lam x.\Omega$. 

But of course, the particular denotational semantics that we gave to
PCF is not the only possible denotational semantics. One can ask
whether there is a better one. For instance, instead of cpo's, we
could have used some other kind of mathematical space, such as a cpo
with additional structure or properties, or some other kind of object
altogether. The search for good denotational semantics is a subject of
much research. The following terminology helps in defining precisely
what is a ``good'' denotational semantics.

\begin{definition}
  A denotational semantics is called {\em fully abstract} if for all
  terms $M$ and $N$,
  \[        M\eqden N  \sep\iff\sep   M\eqop N.
  \]
  If the denotational semantics involves a partial order (such as a
  cpo semantics), it is also called {\em order fully abstract} if
  \[        M\sqleqden N   \sep\iff\sep    M\sqleqop N.
  \]
\end{definition}

\vspace{-.1ex}

The search for a fully abstract denotational semantics for PCF was an
open problem for a very long time. Milner proved that there could be
at most one such fully abstract model in a certain sense. This model
has a syntactic description (essentially the elements of the model are
PCF terms), but for a long time, no satisfactory semantic description
was known. The problem has to do with sequentiality: a fully abstract
model for PCF must be able to account for the fact that certain
parallel constructs, such as parallel or, are not definable in PCF.
Thus, the model should consist only of ``sequential'' functions. Berry
and others developed a theory of ``stable domain theory'', which is
based on cpo's with a additional properties intended to capture
sequentiality. This research led to many interesting results, but the
model still failed to be fully abstract. 

Finally, in 1992, two competing teams of researchers, Abramsky,
Jagadeesan and Malacaria, and Hyland and Ong, succeeded in giving a
fully abstract semantics for PCF in terms of games and strategies.
Games capture the interaction between a player and an opponent, or
between a program and its environment. By considering certain kinds of
``history-free'' strategies, it is possible to capture the notion of
sequentiality in just the right way to match PCF. In the last decade,
game semantics has been extended to give fully abstract semantics to a
variety of other programming languages, including, for instance,
Algol-like languages.

Finally, it is interesting to note that the problem with ``parallel or''
is essentially the {\em only} obstacle to full abstraction for the cpo
semantics. As soon as one adds ``parallel or'' to the language, the
semantics becomes fully abstract.

\begin{theorem}
  The cpo semantics is fully abstract for parallel PCF.
\end{theorem}

\section{Acknowledgements}

Thanks to Field Cady, Brendan Gillon, and Francisco Rios for reporting
typos.

\section{Bibliography}\label{ssec-bibliography}

Here are some textbooks and other books on the lambda calculus.\void{None
of them are required reading for the course, but you may nevertheless
find it interesting or helpful to browse them.  I will try to put them
on reserve in the library, to the extent that they are available.}
{\cite{Bar84}} is a standard reference handbook on the lambda
calculus.  {\cite{GLT89}}--{\cite{Rev88}} are textbooks on the lambda
calculus. {\cite{Win93}}--{\cite{Hen90}} are textbooks on the
semantics of programming languages. Finally,
{\cite{Pey87}--\cite{App92}} are textbooks on writing compilers for
functional programming languages. They show how the lambda calculus
can be useful in a more practical context.

\renewcommand{\refname}{\vspace{-5ex}}

\end{document}

%% file: posets.tex
\setlength{\unitlength}{1500sp}%
\begin{picture}(9608,6110)(1643,-6336)
\put(7906,-361){\makebox(0,0)[lb]{\smash{{1}%
}}}
\put(8506,-361){\makebox(0,0)[lb]{\smash{{2}%
}}}
\put(9106,-361){\makebox(0,0)[lb]{\smash{{3}%
}}}
\put(9706,-361){\makebox(0,0)[lb]{\smash{{4}%
}}}
\put(10756,-361){\makebox(0,0)[lb]{\smash{{$n$}%
}}}
\put(5056,-361){\makebox(0,0)[lb]{\smash{{T}%
}}}
\put(6256,-361){\makebox(0,0)[lb]{\smash{{F}%
}}}
\put(3481,-361){\makebox(0,0)[lb]{\smash{{*}%
}}}
\put(7306,-361){\makebox(0,0)[lb]{\smash{{0}%
}}}
{\thinlines
\put(3526,-511){\circle{150}}
}%
{\put(5101,-511){\circle{150}}
}%
{\put(6301,-511){\circle{150}}
}%
{\put(5701,-1411){\circle{150}}
}%
{\put(7351,-511){\circle{150}}
}%
{\put(8551,-1411){\circle{150}}
}%
{\put(7951,-511){\circle{150}}
}%
{\put(8551,-511){\circle{150}}
}%
{\put(9151,-511){\circle{150}}
}%
{\put(9751,-511){\circle{150}}
}%
{\put(10801,-511){\circle{150}}
}%
{\put(1726,-1411){\circle{150}}
}%
{\put(1726,-6736){\circle{150}}
}%
{\put(1726,-6136){\circle{150}}
}%
{\put(1726,-5536){\circle{150}}
}%
{\put(1726,-4936){\circle{150}}
}%
{\put(3526,-6136){\circle{150}}
}%
{\put(3526,-5536){\circle{150}}
}%
{\put(3526,-4936){\circle{150}}
}%
{\put(3526,-4336){\circle{150}}
}%
{\put(3526,-2986){\circle{150}}
}%
{\put(3526,-6736){\circle{150}}
}%
{\put(3526,-1411){\circle{150}}
}%
{\put(6001,-4486){\circle{150}}
}%
{\put(6001,-5236){\circle{150}}
}%
{\put(6001,-5986){\circle{150}}
}%
{\put(7051,-5986){\circle{150}}
}%
{\put(7051,-5236){\circle{150}}
}%
{\put(8101,-5236){\circle{150}}
}%
{\put(9151,-5236){\circle{150}}
}%
{\put(9151,-4486){\circle{150}}
}%
{\put(9151,-5986){\circle{150}}
}%
{\put(7576,-6736){\circle{150}}
}%
{\put(8101,-5986){\circle{150}}
}%
{\put(3526,-586){\line( 0,-1){750}}
}%
{\put(5176,-586){\line( 2,-3){507.692}}
}%
{\put(6226,-586){\line(-2,-3){507.692}}
}%
{\put(7426,-586){\line( 3,-2){1125}}
}%
{\put(8026,-586){\line( 2,-3){507.692}}
}%
{\put(8551,-586){\line( 0,-1){750}}
}%
{\put(8551,-1336){\line( 2, 3){507.692}}
}%
{\put(8551,-1336){\line( 3, 2){1125}}
}%
{\put(10726,-586){\line(-3,-1){2182.500}}
}%
{\put(1726,-6661){\line( 0, 1){450}}
}%
{\put(1726,-6061){\line( 0, 1){450}}
}%
{\put(1726,-5461){\line( 0, 1){450}}
}%
{\put(1726,-4861){\line( 0, 1){450}}
}%
{\put(3526,-6661){\line( 0, 1){450}}
}%
{\put(3526,-6061){\line( 0, 1){450}}
}%
{\put(3526,-5461){\line( 0, 1){450}}
}%
{\put(3526,-4861){\line( 0, 1){450}}
}%
{\put(3526,-4261){\line( 0, 1){375}}
}%
{\multiput(1726,-4411)(0.00000,93){9}{\makebox(1.6667,11.6667){.}}
}%
{\multiput(3526,-3886)(0.00000,93){9}{\makebox(1.6667,11.6667){.}}
}%
{\put(6001,-4561){\line( 0,-1){600}}
}%
{\put(9151,-4561){\line( 0,-1){600}}
}%
{\put(9151,-5311){\line( 0,-1){600}}
}%
{\put(6001,-5311){\line( 0,-1){600}}
}%
{\put(7051,-5311){\line(-5,-3){1036.765}}
}%
{\put(8101,-5311){\line(-5,-3){1036.765}}
}%
{\put(8101,-5311){\line( 5,-3){1036.765}}
}%
{\put(8101,-5911){\line(-5, 3){1036.765}}
}%
{\put(7051,-5911){\makebox(1.6667,11.6667){.}}
}%
{\put(7051,-5911){\line(-5, 3){1036.765}}
}%
{\put(8101,-5911){\line( 5, 3){1036.765}}
}%
{\put(7576,-6661){\line(-5, 2){1564.655}}
}%
{\put(7576,-6661){\line(-5, 6){510.246}}
}%
{\put(7576,-6661){\line( 5, 6){510.246}}
}%
{\put(7576,-6661){\line( 5, 2){1564.655}}
}%
\put(10126,-511){\makebox(0,0)[lb]{\smash{{. . .}%
}}}
\put(11251,-511){\makebox(0,0)[lb]{\smash{{. . .}%
}}}
\put(9451,-886){\makebox(0,0)[lb]{\smash{{. . .}%
}}}
\put(10276,-961){\makebox(0,0)[lb]{\smash{{. . .}%
}}}
\put(1726,-2011){\makebox(0,0)[lb]{\smash{{\bf 1}%
}}}
\put(5626,-2011){\makebox(0,0)[lb]{\smash{{$\Bb$}%
}}}
\put(3526,-2011){\makebox(0,0)[lb]{\smash{{\bf 2}%
}}}
\put(8476,-2011){\makebox(0,0)[lb]{\smash{{$\Nn$}%
}}}
\put(1651,-7336){\makebox(0,0)[lb]{\smash{{$\omega$}%
}}}
\put(3376,-7336){\makebox(0,0)[lb]{\smash{{$\omega+1$}%
}}}
\put(7201,-7336){\makebox(0,0)[lb]{\smash{{$\Bb^{\Bb}$}%
}}}
\end{picture}